\documentclass[12pt]{book}
\usepackage{latexsym}
\usepackage{makeidx}
\usepackage{hyphenat}
\usepackage{graphicx}
\usepackage{hyperref}
\usepackage{graphics}
\makeindex

\newcommand{\bea}{\begin{eqnarray}}
\newcommand{\eea}{\end{eqnarray}}
\renewcommand{\a}{\alpha}

\renewcommand{\b}{\beta}

\newcommand{\pa}{\partial}

\newcommand{\sgn}{{\rm sgn}}

\def\As{A\!\!\!/}
\def\Bs{B\!\!\!/}
\def\ks{k\!\!\!/}
\def\ls{l\!\!\!/}
\def\ps{p\!\!\!/}

\def\as{a\!\!\!/}
\def\bs{b\!\!\!/}

\def\Ds{D\!\!\!\!/}
\def\ds{\partial\!\!\!/}
\def\os{\omega\!\!\!/}
\newcommand{\Slash}[1]{{#1}\!\!/}
\newcommand{\SLASH}[1]{{#1}\!\!\!/}
\newcommand{\SSLASH}[1]{{#1}\!\!\!\!/}

\makeindex
\begin{document}

\begin{center}
{\LARGE\bf Lorentz symmetry breaking -- classical and quantum aspects}

\vspace*{1cm}

T. Mariz, J. R. Nascimento, A. Yu. Petrov

\end{center}

\vfill

This is a preprint of the following work: Tiago Mariz, Jose Roberto Nascimento, Albert Petrov, Lorentz Symmetry Breaking -- Classical and Quantum Aspects, 2022, Springer, reproduced with permission of Springer Nature Switzerland AG 2022. The final authenticated version is available online at: https://doi.org/10.1007/978-3-031-20120-2.

\newpage

\thispagestyle{empty}


\tableofcontents

\newpage

\chapter*{\centering Abstract}

In this book, we review various aspects of the Lorentz symmetry breaking,
both classical and quantum ones, with the special interest to perturbative
generation of Lorentz-breaking terms. We present impacts of Lorentz symmetry
breaking in noncommutative and supersymmetric theories. Also, we discuss the
problem of Lorentz symmetry breaking in a curved space-time. The book is closed
with a review of experimental studies of Lorentz symmetry breaking.

\chapter{Introduction}

The Lorentz symmetry, being certainly one of the most important symmetries of the quantum field theory, naturally requires detailed studies. One of the most important related issues is the determination of its limits of applicability. While validity of this symmetry continues to be verified with a high degree of precision \cite{datatables}, some profound theoretical motivations for its breaking must be discussed. Already in 1989, in \cite{KosSam}, it has been supposed that, in the low-energy limit of the bosonic string, some vector, or, in general, tensor fields can acquire non-zero expectation values, which, in their turn, introduce some privileged directions in the space-time, and, therefore, break the Lorentz symmetry. As a next step, the first Lorentz-violating (LV) generalization of a known field theory model was proposed, that is, the modification of the usual four-dimensional QED by the following additive term called the Carroll-Field-Jackiw (CFJ) term \cite{CFJ}: \index{Carroll-Field-Jackiw (CFJ) term}
\bea
\label{CFJ}
{\cal L}_{CFJ}=\epsilon^{\alpha\beta\gamma\delta}k_{\alpha}A_{\beta}\partial_{\gamma}A_{\delta}.
\eea
Here, the $k_{\alpha}$ is a constant vector, whose components are assumed to be small,  introducing the anisotropy in the space-time and thus breaking the Lorentz symmetry\footnote{Within this book we discuss only small LV modifications of known field theory models and do not address the theories characterized by strong space-time asymmetry like the Horava-Lifshitz-like ones \cite{Horava:2009uw}. \index{Horava-Lifshitz-like theories}}. This term can be treated as a natural four-dimensional extension of the well-known Chern-Simons term, being gauge invariant up to a total derivative. Evidently, the introduction of the CFJ term called an interest to investigation of its implications both at the classical level, especially dispersion relations and issues related with unitarity and causality, and at the quantum level, that is, possibilities for its perturbative generation, studies of renormalization and renormalizability, relation with possible anomalies, etc., and some important results obtained within these studies are reviewed in this book. Further, other LV extensions have been proposed, not only for the electromagnetic field, but for spinor and scalar ones as well, which allowed to formulate the LV standard model extension (LV SME) \cite{ColKost1,ColKost2}, making thus a Lorentz symmetry breaking to be an experimentally testable phenomenon. It is important to say here that the CFJ term breaks the CPT symmetry as well, due to the presence of the constant vector. In general, the LV terms involving constant tensors of odd ranks break also the CPT symmetry, while the presence of constant tensors of even ranks maintains the CPT invariance, see e.g. \cite{Cha2002}.

Clearly, it is necessary here to discuss the reasons allowing for the applicability of the usual relativistic formalism within LV theories. Really, in a LV theory there can be two types of Lorentz transformations \cite{ColKost1,ColKost2,Kost2001,Mattingly}: the observer ones and the particle ones. The background fields, that is, just those ones represented by constant vectors or tensors, transform under the observer Lorentz transformations \index{observer Lorentz transformations} but not under the particle Lorentz transformations, while the dynamical fields transform both under observer and particle Lorentz transformations. \index{particle Lorentz transformations} Therefore, the Lagrangians of LV theories continue to be invariant under observer Lorentz transformations but not invariant under particle Lorentz transformations (the last ones are sometimes also denominated as active Lorentz transformations) \cite{ColKost1,ColKost2}. This allows to apply the relativistic methodology within the framework of the observer Lorentz transformations.

Besides of the string theory, there are other reasons for the Lorentz symmetry breaking. The most important of them are, first, the possible variability of the fundamental constants, especially, the speed of light, which began to be discussed in order to explain the cosmic acceleration (see e.g. \cite{Mag}), second, the space-time noncommutativity, which in the most used, Moyal-type version involves a constant second-rank antisymmetric tensor $\Theta^{\mu\nu}$, which clearly is not Lorentz-invariant \cite{SW}, third, the loop quantum gravity proposed as an attempt to quantize the geometry (see for a review \cite{GamPul}), fourth, the concept of the space-time foam originally proposed in \cite{Hawk} and further discussed, e.g., in \cite{BernKlink,Klinkfoam} (three last  scenarios evidently  require existence of some minimal distance scale), and, probably, other reasons.

The main implications of the LV extensions of field theory models, especially, of the electrodynamics, at the tree level consist in nontrivial behavior of the possible classical solutions, which can display in a vacuum the phenomena characteristic for propagation of electromagnetic waves in nontrivial media, such as birefringence  \cite{JaNC,SchreckBir} and rotation of the polarization plane \cite{JaNC}, which can allow to use LV models for an effective description for certain condensed matter phenomena, some examples of such studies are given in \cite{Grushin,SchreckCM,KostSchreck}. Besides of this, one of the important issues at the tree level is also related with the problems of unitarity and causality, which, in general, are not guaranteed for LV theories, so that, typically, to achieve unitarity and causality, the LV parameters should satisfy special conditions which vary for different theories, see e.g. \cite{KostLehn,KlinkSchreck,KlinkSchreck1}. However, it turns out to be that for many examples of the LV theories neither unitarity nor causality are broken for some  characteristic, usually small, values of LV parameters. 

At the perturbative level, the main idea consists in a possibility to generate certain LV additive terms which can be employed within extensions of known field theory models, as quantum corrections, through possible couplings of these fields to some matter, typically spinor one. This method is a reminiscence of an old concept of emergent dynamics \cite{Bjorken}, and its first application in the context of Lorentz symmetry breaking has been carried out to generate the CFJ term \cite{JaKo}. This term turned out to be finite, although it is formally superficially divergent, and ambiguous, which called a discussion about possible origins of these ambiguities and their relation with the chiral anomaly \index{anomaly} \cite{JackAmb}. Further, the ambiguity was shown to occur not only for the CFJ term but also for some other LV terms, being also related with anomalies -- straightforward generalizations of the chiral anomaly and gravitational anomalies \cite{AlvWitt}. The possibility to generate different LV terms, especially the modifications for the electrodynamics, has been studied in different dimensions of the space-time, from two to five, and will be one of the main issues considered in this book.

We already noted that one of the reasons for interest to LV theories is motivated by development of the concept of space-time noncommutativity \cite{KostCarr} (for a discussion of a relation between Lorentz symmetry breaking and noncommutativity, see also \cite{SchreckNC}; generalization of this correspondence to a curved space-time case is considered in \cite{BaileyLane}). In this book we review the LV impacts of most popular formulations of noncommutative field theories, that is, the Moyal product formulation and the Seiberg-Witten map. We note that unlike other LV theories, the noncommutative field theories are non-polynomial with respect to the LV parameters even at the classical level. We also describe the noncommutative field approach which is more similar to the usual LV methodology.

One more important point in studying the LV theories consists in possibility of development of their supersymmetric extensions. There was a common belief that breaking of the Lorentz symmetry would necessarily imply breaking of the supersymmetry as well. However, the prescription allowing to construct the LV deformation of the supersymmetry algebra was proposed already in 2001  \cite{BK}. Further, two other manners to construct the supersymmetric LV theories were proposed: first, introducing of the extra superfield(s) whose component(s) are related with some constant vectors (tensors), second, direct adding of LV terms where derivatives acting on superfields are contracted with some LV vectors and tensors. We note that in all these cases, all powerful machinery of supergraphs can be applied to obtain quantum corrections.

An important issue of studying the Lorentz symmetry breaking is devoted to the possibility of constructing LV extensions of gravity. The main difficulty in this study consists in the fact that a straightforward LV modification of gravity models would break the general coordinate invariance as well, while it would be probably preferable to maintain the general coordinate invariance, playing the role of the gauge symmetry for the gravitational field. The only known Lorentz-CPT breaking additive term in gravity possessing this property is the four-dimensional gravitational Chern-Simons term \cite{JaPi}. Many issues related to this term have been studied in great details, including possibilities of its perturbative generation, arising of ambiguities, which, in this case, are linked with gravitational anomalies, and verification of consistency of known exact solutions of the usual general relativity within the modified gravity model whose action is given by a sum of this term and  the usual Einstein-Hilbert action, that is, the Chern-Simons modified gravity (for a review of various results obtained in this theory, see also \cite{AlexYunes}). In the book, we discuss this term and some other possible Lorentz and/or CPT breaking terms in gravity.

Our book has the following structure. In the Chapter 2, we review the tree-level aspects of LV theories; in the Chapter 3, we discuss perturbative generation of LV quantum corrections depending on scalar and vector fields, in space-times of different dimensions from 2 to 5; the Chapter 4 will be devoted to discussing of the space-time noncommutativity within the context of the Lorentz symmetry breaking; in the Chapter 5, we discuss supersymmetric LV theories;  in the Chapter 6, we discuss LV extensions of gravity, especially the four-dimensional gravitational Chern-Simons term;  and, in the Chapter 7, we present a review of experimental studies of the Lorentz symmetry breaking.

\newpage

\chapter[Classical aspects]{Classical aspects of the Lorentz symmetry breaking}

The aim of this chapter consists in discussing of typical phenomena characteristic for LV theories at the classical level. There is a variety of manners to construct LV extensions for different theories describing scalar, spinor, electromagnetic and gravitational fields (a very extensive review of corresponding additive terms can be found in \cite{KosGra}), however, the possible impacts of all these extensions are very similar. One of the most important consequences of the Lorentz symmetry breaking consists in a deformation of dispersion relations, which leads to the concept of variable speed of light (see e.g. \cite{Mag}). Another its important implications consists in unusual behavior of plane wave solutions which display the phenomena typical for propagation of electromagnetic waves in nontrivial media, such as birefringence of waves and rotation of their plane of polarization. In this chapter, we, first, review the most important LV modifications of different field theory models and review their tree-level properties, second, describe the characteristic features of LV theories, third, consider their duality aspects, where the dispersion relations play an important role, fourth, discuss the spontaneous Lorentz symmetry breaking. 

\section[Review of LV extensions]{Review of most important LV extensions of field theory models}

Here, we present the most studied terms among those ones proposed in \cite{KosGra}, and review some other interesting terms.
We start our consideration with the more generic renormalizable LV extension of the four-dimensional QED without higher derivatives, discussed in \cite{Kost2001a}:
\bea
\label{genrenmod}
{\cal L}=\bar{\psi}(i\Gamma^{\nu}D_{\nu}-M)\psi-\frac{1}{4}F_{\mu\nu}F^{\mu\nu}-\frac{1}{4}\kappa_{\mu\nu\lambda\rho}F^{\mu\nu}F^{\lambda\rho}+\epsilon^{\mu\nu\lambda\rho}k_{\mu}A_{\nu}\partial_{\lambda}A_{\rho},
\eea
where
\bea
\Gamma^{\nu}&=&\gamma^{\nu}+c^{\mu\nu}\gamma_{\mu}+d^{\mu\nu}\gamma_{\mu}\gamma_5+e^{\nu}+if^{\nu}\gamma_5+\frac{1}{2}g^{\lambda\mu\nu}\sigma_{\lambda\mu},\nonumber\\
M&=&m+a_{\mu}\gamma^{\mu}+b_{\mu}\gamma^{\mu}\gamma_5+\frac{1}{2}H^{\mu\nu}\sigma_{\mu\nu},
\eea
$D_{\mu}=\pa_{\mu}-ieA_{\mu}$ is the simplest covariant derivative, and $\kappa_{\mu\nu\lambda\rho}$, $k^{\mu}$, $a^{\mu}$, $b^{\mu}$ $c^{\mu\nu}$, $d^{\mu\nu}$, $e^{\mu}$, $f^{\mu}$, $g^{\lambda\mu\nu}$, $H^{\mu\nu}$ are the constant (pseudo)tensors of corresponding ranks. In what follows, we discuss different terms contributing to this Lagrangian, and further, we consider as well non-minimal operators \index{non-minimal interactions} whose dimension is higher than four, so that they yield non-renormalizable couplings or higher-derivative extensions of quadratic actions. Also, we will introduce the LV modifications for scalar field theories and some other models.

The first consistent example of a LV theory is the electrodynamics with the additive CFJ term (\ref{CFJ}). It should be noted that, besides of the Lorentz symmetry, the CPT symmetry is also broken in this case, as it occurs for any term involving a constant tensor of an odd rank \cite{Cha2002}, while for an even-rank constant tensor the CPT invariance is maintained. The important aspect of this theory consists also in its gauge invariance. So, the Lagrangian of the simplest LV extension of the spinor QED, being the particular case of the generic theory (\ref{genrenmod}),  can be written as (see e.g. \cite{JaKo}):
\bea
\label{lvqed}
{\cal L}_{QED,LV}=-\frac{1}{4}F_{\mu\nu}F^{\mu\nu}+\epsilon^{\alpha\beta\gamma\delta}k_{\alpha}A_{\beta}\partial_{\gamma}A_{\delta}+
\bar{\psi}(i\ds-e\As-m+\bs\gamma_5)\psi.
\eea 
The dispersion relation in the purely gauge sector of this theory were found already in \cite{CFJ} and looks like \index{dispersion relations}
\bea
\label{disprelCFJ}
p^4+p^2k^2=(k\cdot p)^2,
\eea
where the $p^{\alpha}$ is a 4-momentum. Anticipating the further discussions, we can say that in principle, there are two solutions for it. However, there is no actual birefringence in this case since one of solutions is non-causal. 

The CFJ term can be naturally generalized to a non-Abelian \index{non-Abelian CFJ term} case. Indeed, it is well known that the non-Abelian gauge invariant generalization of the Chern-Simons (CS) term, for the Lie-algebra valued vector field $A^{\mu}=A^{\mu a}T^a$, where $T^a$ are the generators of the algebra, has the form \cite{DJT}:
\bea
{\cal L}_{CS}=\epsilon^{\alpha\beta\gamma}{\rm tr}(A_{\alpha}\pa_{\beta}A_{\gamma}+\frac{2}{3}A_{\alpha}A_{\beta}A_{\gamma}).
\eea
Promoting this expression to four dimensions, in the same manner in which one can treat the CFJ term as an extension of the usual CS term, one arrives at
\bea
\label{nab}
{\cal L}_{na}=\epsilon^{\alpha\beta\gamma\delta}k_{\delta}{\rm tr}(A_{\alpha}\pa_{\beta}A_{\gamma}+\frac{2}{3}A_{\alpha}A_{\beta}A_{\gamma}).
\eea
This term has been originally proposed in \cite{ColMac}, in order to obtain the most generic renormalizable LV extension of the spinor QED. And, in \cite{ourYM}, this term was shown to arise as a one-loop quantum correction.

One more important LV extension of electrodynamics is CPT-even. Its standard form looks like (cf. \cite{KosGra}):
\bea
\label{keven}
{\cal L}_{even}=-\frac{1}{4}\kappa^{\alpha\beta\gamma\delta}F_{\alpha\beta}F_{\gamma\delta}.
\eea
The $\kappa^{\alpha\beta\gamma\delta}$ is a constant fourth-rank tensor whose symmetry is the same as of the curvature tensor. In one of the most typical cases, that is, the aether one \index{aether} \cite{Carroll}, this tensor is chosen to be expressed in terms of a single constant vector $u_{\alpha}$:
\bea
\label{kappavect}
\kappa_{\alpha\beta\gamma\delta}=-(u_{\alpha}u_{\gamma}\eta_{\beta\delta}-u_{\alpha}u_{\delta}\eta_{\beta\gamma}+u_{\beta}u_{\delta}\eta_{\alpha\gamma}-u_{\beta}u_{\gamma}\eta_{\alpha\delta}), 
\eea  
which allows to write the ${\cal L}_{even}$ as \cite{aether}:
\bea
\label{aether}
{\cal L}_{even}=u^{\alpha}u_{\gamma}F_{\alpha\beta}F^{\gamma\beta}.
\eea
We note that the form (\ref{kappavect}) is actiually a particular case of the ansatz proposed in \cite{Alt}, where, instead of $u^{\mu}u^{\nu}$, the generic second-rank tensor $k^{\mu\nu}$ is used.
In the scalar sector, one could define an aether-like term for the scalar field \cite{Carroll}, given by
\bea
\label{scalaether}
{\cal L}_{sc,aether}=\frac{1}{2}\phi(u\cdot\pa)^2\phi.
\eea
In the spinor sector one can introduce, first, the CPT-Lorentz breaking term $\bar{\psi}\bs\gamma_5\psi$ which already has been considered within the model (\ref{lvqed}), second, the aether-like contribution to the Lagrangiian of the spinor field \cite{Carroll}, that is, 
\bea
{\cal L}_{sp,aether}=iu^{\alpha}u^{\beta}\bar{\psi}\gamma_{\alpha}\partial_{\beta}\psi.
\eea
In the gravity sector, the most important term is certainly the four-dimensional gravitational Chern-Simons term \index{gravitational Chern-Simons term} \cite{JaPi}:
\bea
I_{CS}=\epsilon^{\alpha\beta\gamma\delta}k_{\delta}(\Gamma_{\alpha\mu}^{\nu}\pa_{\beta}\Gamma_{\gamma\nu}^{\mu}+\frac{2}{3}\Gamma_{\alpha\mu}^{\nu}\Gamma_{\beta\rho}^{\mu}\Gamma_{\gamma\nu}^{\rho}).
\eea
Besides of these terms, it is worth to mention the higher-derivative LV terms, characterized by dimensions 5 and more, with the most important among them is the Myers-Pospelov (MP) term \index{Myers-Pospelov term} \cite{MP}, the simplest dimension-5 one, looking like 
\bea
\label{MP}
{\cal L}_{MP}=\frac{\xi}{M}n^{\alpha}F_{\alpha\beta}(n\cdot\partial)n_{\lambda}\epsilon^{\mu\nu\lambda\beta}F_{\mu\nu}.
\eea
Here $\xi$ is a dimensionless constant, $M$ is some mass scale, usually the Planck mass, and $n^{\mu}$ is a dimensionless LV vector. It is important here that this vector can be chosen  in a manner allowing to avoid higher time derivatives in (\ref{MP}) (e.g. with $n^0=0$) and thus eliminating the ghost states, and this approach can be generalized to other higher-derivative LV terms, both in QED and other field theory models, which further will be referred as Myers-Pospelov-like ones (a detailed discussion of unitarity in Myers-Pospelov-like theories can be found, e.g. \cite{ReyesUnit,ReyesUnit2,ReyesUnit3,ReyesUrr,ReyesSarr}, for many issues related to dispersion relations in these theories see also \cite{SchreUn} and references therein). This possibility opens the way to construct consistent higher-derivative field theories without use of the Horava-Lifshitz methodology.

The terms listed above are well defined either in four dimensions (those ones involving the Levi-Civita symbol), or in any dimensions (aether-like ones). At the same time, there is one more LV term which is well defined only in three dimensions \cite{JT0}, arising in the context of the Julia-Toulouse mechanism (see  \cite{JT1} and references therein), and, therefore, sometimes called the \index{Julia-Toulouse term} Julia-Toulouse (JT) term, namely,
\bea
\label{JT}
{\cal L}_{JT}=\epsilon^{\alpha\beta\gamma}v_{\alpha}F_{\beta\gamma}\phi,
\eea 
introducing the ``mixing'' between dynamics of scalar and gauge fields, with $v_{\alpha}$ being a constant vector, and even the two-dimensional LV term \cite{BM}, which is written as
\bea
\label{scal2d}
{\cal L}_{2d}=\epsilon^{\alpha\beta}k_{\alpha}\phi\pa_{\beta}\bar{\phi}.
\eea
In the next chapter we discuss the manners to generate these quadratic terms. Typically, it can be done with use of appropriate LV couplings.

We already noted that the most popular of these couplings, used in the paradigmatic Lagrangian (\ref{lvqed}), is given by the term $\bar{\psi}\bs\gamma_5\psi$. Its importance is related with the fact that its presence allows to generate the Levi-Civita symbol, and hence, the CFJ term and its generalizations. However, recently the non-minimal (i.e. non-renormalizable) interactions, allowing for simpler ways to generate many new terms,  also began to attract the interest. The first one is the magnetic coupling \index{magnetic coupling} involving the coupling constant $g_1$:
\bea
\label{magn}
{\cal L}_{magn}=g_1\epsilon^{\mu\nu\lambda\rho}b_{\mu}j_{\nu}F_{\lambda\rho},
\eea
where $j_{\rho}=\bar{\psi}\gamma_{\rho}\psi$ is the usual Dirac current. Namely, this interaction has been used in \cite{aether} to generate the aether term. The non-renormalizability of this vertex, as well as of other non-minimal vertices, in principle can be controlled through restriction of the study to the fermionic determinant, that is, to the one-loop order, with the external lines are purely gauge (or scalar, in an extension of the Yukawa model) ones. 

One can introduce as well the analogue of this coupling involving the usual vector $a_{\mu}$ rather than the axial one $b_{\mu}$, and a coupling constant $g_2$,  given by the Lagrangian
\bea
\label{vecc}
{\cal L}_{v}=g_2a^{\mu}F_{\mu\nu}j^{\nu},
\eea
which was used to generate the axion term \index{axion} in \cite{axion}.

Another vertex, playing an essential role in the scalar sector, is the Yukawa-like LV coupling with the coupling constant $h$ \cite{aether}:
\bea
\label{yulv}
{\cal L}_{Yukawa}=h\bar{\psi}\as\psi\phi.
\eea
This vertex is renormalizable in four dimensions and allows us to generate the aether term for the scalar field. 

There can be other LV couplings as well. A list of all possible LV terms, contributing either to quadratic action or to interaction vertices,  with dimensions up to 6, and, in the purely gauge sector, with dimensions up to 8, is given in \cite{KosLi}. The typical examples of such terms are, first, 
	${\cal L}^{(4+n)}_A=-\frac{1}{2}k^{(4+n)\kappa\lambda\alpha_1\ldots\alpha_n\mu\nu}F_{\kappa\lambda}D_{(\alpha_1}\ldots D_{\alpha_n)}F_{\mu\nu}$ in the gauge sector, being the natural higher-derivative extensions of the CPT-even term $-\frac{1}{4}\kappa^{\mu\nu\alpha\beta}F_{\mu\nu}F_{\alpha\beta}$, and second, ${\cal L}^{(3+n)}_{\psi D}=\frac{1}{2}c^{(3+n)\mu\alpha_1\ldots\alpha_n}\bar{\psi}\gamma_{\mu}iD_{(\alpha_1}\ldots iD_{\alpha_n)}\psi$ in the spinor sector, that is, higher-derivative extensions for the term $\bar{\psi}c^{\mu\nu}i\gamma_{\mu}D_{\nu}\psi$ contributing to the theory (\ref{genrenmod}). The generalizations of other terms from the spinor sector of (\ref{genrenmod}) can be constructed in a similar manner. In the Chapter 3, we use the simplest examples of LV couplings to generate the quadratic LV terms in scalar, spinor and vector sectors.

\section{Wave propagation in LV theories}

Already in \cite{Mag}, the variability of the speed of light has been proposed as one of possible explanation of a cosmic acceleration. It is easy to see that it can occur only if the dispersion relation for the electromagnetic field  differs from the usual one $E=p$). The deformation of dispersion relations can be implemented also in massive theories \cite{AmCam}, implying the form $E^2=\vec{p}^2+m^2+f(E,p,m,\Lambda)$, where $f(E,p,m,\Lambda)$ is some function suggested to be suppressed in the limit $\Lambda\to\infty$, with $\Lambda$ is a characteristic energy  scale. This function characterizes the deviation from the usual relativistic scenario. Sometimes it is supposed that $\Lambda$ is the Planck mass, 
For example, in \cite{AmCam}, the relation
\bea
E^2=m^2+p^2-L_P p^2E
\eea
was considered, with $L_P$  being a Planck length. In this case the usual phase speed is found to be $v\simeq 1+L_P|p|/2$, so, even in the massless case, it is not a constant more but a parameter depending on the reference frame. Other interesting examples of dispersion relations, discussed in \cite{AmCam}, are $E^2=p^2-L^2_Pp^2E^2$ and $E^2=p^2-L^2_PE^4$. Also, it is worth to mention the following generalized dispersion relations used in \cite{Mag1} for cosmological studies:
\bea
E^2-p^2f^2(E)=0.
\eea
Different forms of $f(E)$ were discussed in \cite{Mag1}, e.g., $f_1(E)=(1+\lambda E)^{\gamma}$, with $2/3\leq \gamma \leq 1$, and $f_2(E)=\frac{2\lambda E}{1-e^{-2\lambda E}}$. 
Within these contexts, the key motivation for the Lorentz symmetry breaking stems from the idea of the minimal length (and presence of the corresponding energy scale) whose existence clearly imposes natural limits of validity on the Lorentz symmetry. The most important concepts based on this idea are the loop quantum gravity and the space-time foam. Many examples of modified dispersion relations constructed in similar manners were discussed also in \cite{MagSmo}, forming a base of so-called generalized Lorentz invariance with an invariant energy scale, further referred as double special relativity. \index{double special relativity}

While in the papers mentioned above, the deformed dispersion relations were discussed on their own base, the natural question consists in possibility for their arising in some consistent field theory models. As an example, we study the relation (\ref{disprelCFJ}), which  describes the LV electrodynamics with the CFJ term.
There are the following characteristic situations.

1. The vector $k_{\mu}=(\mu,0,0,0)$ is time-like. In this case we have $E^2=\vec{p}^2\pm |\vec{p}||\mu|$ \cite{CFJ}, where the solution with the minus sign is unstable at $|\vec{p}|<|\mu|$, making the wave vector to be space-like and implying instability of the solution, and should be disregarded by imposing appropriate boundary conditions.

2. The vector $k_{\mu}=(0,\mu',0,0)$ is space-like. In this case, if we, for the sake of the simplicity, suggest $\vec{p}$ to be directed along $x$ axis, we have $E^2=\vec{p}^2+\frac{{\mu'}^2}{2}\pm\sqrt{\vec{p}^2{\mu'}^2+\frac{{\mu'}^4}{4}}$. However, in this case again the solution with the minus sign implies $E^2<\vec{p}^2$, so the corresponding 4-momentum is space-like, and this solution is unstable.

The possibility for the unstable solutions allowing a 4-momentum to be space-like, which can be eliminated through boundary conditions, frequently takes place in LV theories. Moreover, although in these cases the birefringence \index{birefringence} is ruled out with the requirements of stability, in certain situations two solutions consistent with stability conditions can exist. For example, it was proved in \cite{ColKost2}, that in an extended LV QED with the quadratic Lagrangian involving both CFJ and CPT-even terms:
\bea
{\cal L}=-\frac{1}{4}F_{\alpha\beta}F^{\alpha\beta}+\epsilon_{\alpha\beta\gamma\delta}k^{\alpha}A^{\beta}\partial^{\gamma}A^{\delta}-\frac{1}{4}\kappa_{\alpha\beta\gamma\delta}F^{\alpha\beta}F^{\gamma\delta},
\eea
and with $\vec{p}=(0,0,p)$, one finds the dispersion relations
\bea
E=p(1+\rho)\pm\sqrt{\sigma^2p^2+\tau^2},
\eea
where $\rho$ and $\sigma$ are constants constructed from components of $\kappa_{\alpha\beta\gamma\delta}$, and $\tau$ is a constant constructed of components of $k^{\alpha}$, their explicit form is given in \cite{ColKost2}. If $\rho,\sigma$, and $\tau$ are small enough, the stability of both solutions is guaranteed (at the same time, we note that it was shown in \cite{Casana2010} that, for $k^{\alpha}=0$ with the special form of $\kappa_{\alpha\beta\gamma\delta}$, the birefringence does not occur). 

Another important effect related to the wave propagation in LV theories is related with the \index{rotation of plane of polarization} rotation of a plane of light polarization. Its possibility was noted in \cite{JaNC}, and here we demonstrate this situation by the example of the electrodynamics with an additive Myers-Pospelov (MP) term \cite{MP}. Indeed, if we consider the sum of the usual Maxwell term and the MP term (\ref{MP}), with again $\vec{p}=(0,0,p)$ and $n^{\alpha}=(1,0,0,0)$, one will have the following dispersion relations, with $\vec{\epsilon}_{x,y}$ being the polarization vectors along $x$ and $y$ axes \cite{MP}:
\bea
(E^2-p^2\pm\frac{2\xi}{M}p^3)(\vec{\epsilon}_x\pm i\vec{\epsilon}_y)=0.
\eea
We see that the polarizations along the axes $x$ and $y$ satisfy different dispersion relations and thus have different velocities, so, the polarization plane will rotate. 

In principle, these effects justify the possibility to apply the Lorentz symmetry breaking as an effective analogical description of condensed matter phenomena. Some successful examples of such applications are presented in \cite{Grushin}. 

\section{Duality issues in LV theories}

An important issue in the context of the Lorentz symmetry breaking is related with discussions of the possible LV generalizations of the known duality between self-dual and Maxwell-Chern-Simons theories. Originally, this duality was established in \cite{DJ}. The essence of the idea of the duality is as follows: if one consider the Lagrangians of self-dual \index{self-dual model} and Maxwell-Chern-Simons theories, given respectively by the expressions:
\bea
{\cal L}_{SD}&=&\frac{m^2}{2}f^{\alpha}f_{\alpha}-\frac{m}{2}\epsilon^{\alpha\beta\gamma}f_{\alpha}\partial_{\beta}f_{\gamma};
\nonumber\\
{\cal L}_{MCS}&=&-\frac{1}{4}F_{\alpha\beta}F^{\alpha\beta}+\frac{m}{2}\epsilon^{\alpha\beta\gamma}A_{\alpha}\partial_{\beta}A_{\gamma},
\eea
the dynamics of these theories is related through the mapping
\bea
f^{\alpha}\leftrightarrow \frac{1}{m}F^{\alpha}\equiv\frac{1}{m}\epsilon^{\alpha\beta\gamma}\partial_{\beta}A_{\gamma}.
\eea 
Moreover, if one introduces to both these Lagrangians the interaction terms, $f^{\alpha}j_{\alpha}$ and $A^{\alpha}G_{\alpha}$ respectively, one will arrive at the equations of motion:
\bea
&& -m\epsilon^{\alpha\beta\gamma}\pa_{\beta}f_{\gamma}+m^2f^{\alpha}=j^{\alpha}; \nonumber\\
&& -m\epsilon^{\alpha\beta\gamma}\pa_{\beta}F_{\gamma}+m^2F^{\alpha}=G^{\alpha},
\eea 
which, besides of the duality between $f^{\alpha}$ and $\frac{1}{m}F^{\alpha}$, establishes also the mapping between the currents $j^{\alpha}$ and $G^{\alpha}$. This duality has been discussed in great details in \cite{Anacleto}, where it was confirmed to occur not only for the case of the usual spinor coupling characterized by the current $j^{\alpha}=\bar{\psi}\gamma^{\alpha}\psi$, but also for coupling to scalar matter whose contribution to a Lagrangian cannot be represented as $f^{\alpha}j_{\alpha}$. The natural question is the possibility to construct the LV extensions of these results, so that new couplings for fields and currents turn out to be implied by the duality. The simplest example where such a coupling arises, without Lorentz-breaking effects yet, is given in \cite{Anacleto}, where it was shown that the dual term to the usual coupling $f^{\alpha}j_{\alpha}$ is the magnetic coupling $\frac{1}{m}F^{\mu}j_{\mu}$.

We start with the following simple generalization of the self-dual theory \cite{Anacleto2}:
\bea
{\cal L}_{SD}=-\frac{m^2}{2}f^{\alpha}f_{\alpha}+\frac{m}{2}\epsilon^{\alpha\beta\gamma}f_{\alpha}\partial_{\beta}f_{\gamma}+\frac{1}{2}\pa^{\mu}\phi\pa_{\mu}\phi+f^{\mu}(2m\phi v_{\mu}+j_{\mu}),
\eea
where we have introduced the LV quadratic term, $2m\phi v^{\mu}f_{\mu}$. Using the gauge embedding method (which is one of most efficient methods to construct a dual to a given theory, see e.g. \cite{Anacleto}), we arrive at the following dual Lagrangian:
\bea
{\cal L}_{MCS}&=&\frac{1}{2}F_{\alpha}F^{\alpha}-\frac{m}{2}\epsilon^{\alpha\beta\gamma}A_{\alpha}\partial_{\beta}A_{\gamma}
+\frac{1}{2}\pa^{\mu}\phi\pa_{\mu}\phi+\phi\epsilon^{\mu\nu\lambda}v_{\mu}F_{\nu\lambda}+2\phi^2v_{\mu}v^{\mu}+
\nonumber\\&+& \frac{1}{2m^2}j^{\mu}j_{\mu}+\frac{1}{m}j^{\mu}F_{\mu}+\frac{2}{m}\phi v^{\mu}j_{\mu}.
\eea
First, the duality of these two theories is confirmed with use of the mapping
\bea
f^{\alpha}\leftrightarrow \frac{1}{m}F^{\alpha}+\frac{2\phi}{m}v^{\mu}+\frac{1}{m^2}j^{\mu},
\eea
which means that two theories are dynamically equivalent. Second, our important result consists in generation of new terms, not only the Thirring current-current interaction and the magnetic coupling, but also the remarkable mixed quadratic term (\ref{JT}), which links the dynamics of gauge and scalar fields and has a key role within the Julia-Toulouse mechanism \cite{JT0}.

Third, one can discuss as well the dispersion relations in these theories. For the self-dual theory, one has three types of dispersion relations, namely, 
\begin{eqnarray}
\label{direl1}
&& ({\rm i}): E^2=\vec{p}^2;\quad\,
({\rm ii}): E^2=\vec{p}^2+m^2;\nonumber\\
&& ({\rm iii}): (E^2-\vec{p}^2)^2-(E^2-\vec{p}^2)m^2+4m^2v^2=0;
\end{eqnarray}
 while for the dual MCS-like theory, one has 
\begin{eqnarray}
\label{direl2}
&& ({\rm i}): E^2=\vec{p}^2;\quad\, 
({\rm ii}): E^2=\vec{p}^2+m^2; \quad\, ({\rm iii}): E^2-\vec{p}^2=4v^2;
\\
&& 
({\rm iv}): (E^2-\vec{p}^2-m^2)(E^2-\vec{p}^2-4v^2)+v^2(E^2-\vec{p}^2)-\nonumber\\
&&\:\:\:\:\:-(\vec{v}\cdot\vec{p}-v_0E)^2=0.
\nonumber
\end{eqnarray} 
Formally, some of these dispersion relations are different, however, only those ones corresponding to physical degrees of freedom should coincide for both theories \cite{Anacleto}. In this case,  only the relations (i) and (ii) are common for these theories, and they correspond to the scalar field and the only physical degree of freedom of the three-dimensional vector field  respectively.

In the paper \cite{Scarpdual}, the similar analysis has been carried out also for CPT-even LV extensions of the self-dual and Maxwell-Chern-Simons theories, whose Lagrangians look like
\bea
{\cal L}_{SD}&=&\frac{m^2}{2}f^{\alpha}h_{\alpha\beta}f^{\beta}-\frac{m}{2}\epsilon^{\alpha\beta\gamma}f_{\alpha}\partial_{\beta}f_{\gamma}+f^{\mu}j_{\mu};\nonumber\\
{\cal L}_{MCS}&=&-\frac{1}{4}F_{\alpha\beta}F^{\alpha\beta}+\frac{m}{2}\epsilon^{\alpha\beta\gamma}A_{\alpha}\partial_{\beta}A_{\gamma}-\frac{\alpha}{8}(\epsilon_{\mu\nu\lambda}b^{\mu}F^{\nu\lambda})^2
\nonumber\\&-& \frac{1}{2m^2}j^{\mu}(h^{-1})_{\mu\nu}j^{\nu}+\frac{1}{m}j^{\mu}(h^{-1})_{\mu\nu}F^{\mu}.
\eea
Here, $h_{\alpha\beta}=\eta_{\alpha\beta}-\beta b_{\alpha}b_{\beta}$ and $(h^{-1})_{\alpha\beta}=\eta_{\alpha\beta}+\alpha b_{\alpha}b_{\beta}$, with $\alpha=\frac{\beta}{1-\beta b^2}$, where we assumed smallness of Lorentz-breaking effects requiring $|\beta|\ll 1$, and $b_{\beta}$ is a constant vector implementing the Lorentz symmetry breaking. In this case we obtain, among all, the aether-like term (\ref{aether}), arising from the $(\epsilon_{\mu\nu\lambda}b^{\mu}F^{\nu\lambda})^2$ term. The equivalence of the equations of motion in these theories can be shown as well \cite{Scarpdual}.
In this case there is only one physical degree of freedom, and only one physical dispersion relation, which is common in both theories. It looks like
\bea
E^2-\vec{p}^2-m^2+\alpha[b^2(E^2-\vec{p}^2)-(\vec{b}\cdot\vec{p}-b_0E)^2]=0.
\eea
The detailed analysis of stability and causality issues can be done in this case, it is presented in \cite{Scarpdual}. We note again that if the Lorentz-breaking parameters are enough small we have no problems with unitarity and causality.

Moreover, the Lorentz symmetry breaking opens the way to generalize the duality for the four-dimensional case.
In this case we start with the self-dual model extension
\begin{eqnarray}
\label{sd4}
L_{SD}=\frac{m^2}{2}f^{\alpha}f_{\alpha}-\frac{1}{2}\epsilon^{\alpha\beta\gamma\delta}k_{\alpha}f_{\beta}\partial_{\gamma} f_{\delta}+f^{\alpha}j_{\alpha},
\end{eqnarray}
where we have replaced  the Chern-Simons term by its four-dimensional analogue, that is, the CFJ term. Under the gauge embedding procedure \cite{Anacleto}, we arrive at the following dual action:
\begin{eqnarray}
\label{new}
L_{new}&=&\frac{1}{2}\epsilon^{\alpha\beta\gamma\delta}k_{\alpha}f_{\beta}\partial_{\gamma} f_{\delta}+\frac{1}{2m^2}j^{\alpha}j_{\alpha}+
\frac{1}{2m^2}\epsilon^{\alpha\beta\gamma\delta}k_{\alpha}j_{\beta} F_{\gamma\delta}+\frac{k^2}{4m^2}F^{\alpha\beta}F_{\alpha\beta}-\nonumber\\&-&\frac{1}{2m^2}k^{\alpha}k_{\beta}F^{\beta\gamma}F_{\alpha\gamma}.
\end{eqnarray}
In this case, the dual action turns out to involve the new terms, that is, the aether term (\ref{aether}) and the LV magnetic coupling (\ref{magn}). Thus, we see that, first, the aether term arises together with the coupling used for its generation \cite{aether}, second, the dual of the CPT-odd theory is CPT-even. The dispersion relations in these theories \cite{Clovis}, with the first of them valid only in the Maxwell-like theory (\ref{new}), are given by
\begin{eqnarray}
\label{disprel1}
&&{\rm(i)} \,\,\, k^2p^2-(k\cdot p)^2=0,\nonumber\\
&&{\rm (ii)}\, k^2p^2-(k\cdot p)^2+m^4=0.
\end{eqnarray}
The second of these relations is common for both models and thus corresponds to the physical degree of freedom. This duality can be verified in different manners \cite{Clovis}. To close this section, we emphasize that the dual immersion of Lorentz-breaking theories indeed allows to generate new LV terms.

\section{Spontaneous breaking of Lorentz symmetry}

In principle, there are two main manners to break the Lorentz symmetry, that is, explicit one, where a new term proportional to a constant vector or tensor is added to a Lorentz invariant action, and the spontaneous one, when the original action is Lorentz invariant, while its minimum introduces a privileged space-time direction. While in most part of this book we suggest that the Lorentz symmetry breaking is explicit, it is important to consider the spontaneous one as well, because, first, it provides a natural mechanism for originating the LV theories, second, it is deeply related with string theory, and thus possesses profound motivations \cite{KosSam}, third, as we will see in the Chapter 6, it plays a fundamental role within the curved space-time context. Following this paper, namely the spontaneous breaking of Lorentz symmetry in a low-energy limit of a string theory is responsible for arising the LV theories which are thus treated as low-energy effective theories.

The paradigmatic mechanism for breaking the Lorentz symmetry in a spontaneous manner is based on the bumblebee model \index{bumblebee model} originally introduced within the gravity context \cite{KosGra}. However, this model can be treated as well in the plane space-time. By definition, this model describes the vector field $B_{\mu}$ called the bumblebee field, with a self-coupling potential:
\begin{eqnarray}
\label{bbee}
S=\int d^4x\Big(-\frac{1}{4}B_{\mu\nu}B^{\mu\nu}-V(B^{\mu}B_{\mu}\mp b^2)
\Big).
\end{eqnarray}
Here $B_{\mu\nu}=\pa_{\mu}B_{\nu}-\pa_{\nu}B_{\mu}$ is a stress tensor, and $b^2>0$. The potential $V(B^{\mu}B_{\mu}\mp b^2)$ is suggested, by renormalizability reasons, to be either of the form $V=\frac{f}{4!}(B^{\mu}B_{\mu}\mp b^2)^2$ with $f$ being a coupling constant, or of the form $V=\frac{1}{2}\sigma(B^{\mu}B_{\mu}\mp b^2)$, with $\sigma$ being a Lagrange multiplier field; for the $(+---)$ signature, the sign in the expression $B^{\mu}B_{\mu}\mp b^2$ is chosen to be $(-)$ for the time-like $B_{\mu}$, and $(+)$ for the space-like one. 

It is clear that for both these forms of the potential the minimum is achieved for $B^{\mu}B_{\mu}=\pm b^2$. Therefore one concludes that the value of $B^{\mu}$ equal to the constant vector $b^{\mu}$  (satisfying the relation $b^{\mu}b_{\mu}=\pm b^2$), which can be chosen to be, e.g., parallel to one of coordinate axes, say $x^0$ or $x^i$, introduces a privileged direction. Afterward, one can proceed in a manner similar to the Higgs mechanism, that is, we can expand the $B_{\mu}$ near the minimum as $B_{\mu}\to \tilde{B}_{\mu}+b_{\mu}$, and as a result, we get a mass term for the new field $\tilde{B}_{\mu}$ looking like $\frac{f}{4!}(4\tilde{B}^{\mu}\tilde{B}^{\nu}b_{\mu}b_{\nu}+\tilde{B}^2b^2)$. Thus, we arrive at the essentially Lorentz-breaking contribution to the quadratic action, looking like $\tilde{B}^{\mu}\tilde{B}^{\nu}b_{\mu}b_{\nu}$. Namely this mechanism is responsible for arising of LV terms in a low-energy limit of string theory \cite{KosSam}, so that it is natural to treat LV theories as effective theories for a low-energy limit of some more fundamental theory. If the bumblebee action, including, what is especially important, its potential, is induced through radiative corrections, the Lorentz symmetry is said to be broken dynamically. This is the simplest example of a situation where the Lorentz symmetry breaking can be treated as an emergent phenomenon. Also, in \cite{Seifert}, it was shown that if the action (\ref{bbee}) is extended to a curved space-time, one can introduce the bumblebee perturbation $\tilde{B}_{\mu}$ through the replacement $B_{\mu}\to b_{\mu}+\tilde{B}_{\mu}$, so that $b_{\mu}$ is the v.e.v. of $B_{\mu}$, the induced metric is $g^{\mu\nu}=\eta^{\mu\nu}+\beta b^{\mu}b^{\nu}$, with $\beta$ is some constant, $\tilde{B}_{\mu}$ is small, and both the potential $V$ and its derivative vanish at $B_{\mu}=b_{\mu}$ (i.e. at $\tilde{B}_{\mu}=0$). As a result, one gets the LV action for the $\tilde{B}_{\mu}$, composed by the sum of the usual Maxwell term, the aether-like term (\ref{keven}), where $\kappa^{\mu\nu\lambda\rho}$ is constructed on the base of the vector $b^{\mu}$, and the corresponding potential. Therefore, this mechanism allows to generate a CPT-even LV terms for the vector field. In \cite{Altschul:2009ae}, this approach was generalized for the case of the antisymmetric tensor field.

 Let us consider now examples of dynamical Lorentz symmetry breaking, for theories describing vectors and antisymmetric tensors.  Some first discussions of this approach can be found in \cite{spont}, and its further development, for the vector field, was presented in \cite{spont1}, and for the antisymmetric tensor one, in \cite{spont2}. Although these studies involve calculations of quantum corrections, it is very natural to discuss their results namely here, within a general context of spontaneous Lorentz symmetry breaking.
	
	We start with the following four-fermion model, chosen to be massless for the sake of the simplicity \cite{spont1}:
	\bea
	\label{mlff}
	{\cal L}_0=i\bar{\psi}\ds\psi-\frac{G}{2}(\bar{\psi}\gamma_{\mu}\gamma_5\psi)^2.
	\eea
	As usual, we introduce the vector field $B_{\mu}$ in order to avoid arising of the non-renormalizable four-fermion vertex:
	\bea
	{\cal L}=\frac{g^2}{2}B^{\mu}B_{\mu}+\bar{\psi}(i\ds-e\Bs\gamma_5)\psi.
	\eea
	It is immediate to see that, eliminating the $B_{\mu}$ from this action through purely algebraic equations of motion, we recover (\ref{mlff}) with $G=\frac{e^2}{g^2}$. We can write down the one-loop corrected effective potential for this theory as
	\bea
	\label{1lcep}
	V_{eff}=-\frac{g^2}{2}B^{\mu}B_{\mu}+i{\rm Tr}\ln(\ps-e\Bs\gamma_5).
	\eea
	To break the Lorentz symmetry, we must first find the minimum of the $V_{eff}$. It is obtained from the condition $\frac{dV_{eff}}{dB_{\mu}}|_{eB_{\mu}=b_{\mu}}=0$. The resulting $b_{\mu}$ vector will introduce the privileged space-time direction. As a result, we obtain
	\bea
	\label{gap} \frac{dV}{dB_{\mu}}|_{eB=b}=-\frac{g^2}{e}b_{\mu}-i\Pi_{\mu}=0,
	\eea
	where
	$\Pi_{\mu}={\rm tr}\int\frac{d^4k}{(2\pi)^4}\frac{i}{\ks-\bs\gamma_5}(-ie\gamma_{\mu}\gamma_5)$.
	Calculating the trace in this expression with use of the exact propagator of the fermionic field $(\ks-\bs\gamma_5)^{-1}$ instead of its expansion in power series in $b_{\mu}$ (we note that in the massless case this exact propagator has a very simple form; different expressions for the spinor propagator will be discussed in the next chapter, details of the calculation presented here can be found in \cite{spont1}), we find that 
	\begin{equation}\label{Pimu}
		\Pi^\mu = \frac{ieb^2}{3\pi^2}b^\mu.
	\end{equation}		
We note that the $\Pi^{\mu}$ is finite: during its calculation, instead of an expected UV divergence one finds a removable singularity, pole terms cancel out, and no renormalization is needed. In next chapters we will see that removable singularities sometimes arise within one-loop calculations in LV theories, providing finiteness of corresponding contributions.

	So, the gap equation (\ref{gap}) yields 
	\bea\label{DVef2}
	\frac{dV_{eff}}{dB_\mu}\Big|_{eB_{\mu}=b_{\mu}} = \left(-\frac{1}{G}+\frac{b^2}{3\pi^2}\right)eb_\mu= 0,
	\eea
	and its integration gives the effective potential:
	\bea
	V_{eff}=\frac{e^4}{12\pi^2}B^4-\frac{e^2b^2}{6\pi^2}B^2+\alpha,
	\eea
	where $\alpha$ is an arbitrary constant, and for $\alpha=\frac{b^4}{12\pi^2}$, with $b^2=\frac{3\pi^2}{G}$, one has $V_{eff}=\frac{1}{12\pi^2}(e^2B^2-b^2)^2$, that is, the positively defined potential perfectly reproducing the bumblebee form. If the finite temperature is introduced, one can find that this potential generates phase transitions \cite{spont1}.
	
	It is interesting now to consider the dynamics of $B_{\mu}$ expanded about the non-zero vacuum $<B_{\mu}>=\frac{b_{\mu}}{e}$. In this case, we make a shift $B_{\mu}=\frac{b_{\mu}}{e}+A_{\mu}$, with $A_{\mu}$ is the new dynamical field. In this case, the effective action of $A_{\mu}$ is
	\bea
	S_{eff}=\frac{g^2}{2}(\frac{b_{\mu}}{e}+A_{\mu})(\frac{b^{\mu}}{e}+A^{\mu})-i{\rm tr}\int \frac{d^4k}{(2\pi)^4}\ln(\ks-\bs\gamma_5-e\As\gamma_5).
	\eea
	After one-loop calculations (see \cite{spont1} for details), we find
		\begin{eqnarray}
		\label{Lvecsp}
		{\cal L} &=& -\frac14 F_{\mu\nu}F^{\mu\nu} + \frac{e^2}{24\pi^2}b^\mu\epsilon_{\mu\nu\lambda\rho}A^\nu F^{\lambda\rho} -\nonumber\\
		&-&\frac{e^2}{48\pi^2}(\partial_\mu A^\mu)^2 - \frac{e^4}{12\pi^2}\left(A_{\mu} A^\mu + \frac{2}{e} A\cdot b\right)^2.
	\end{eqnarray}
	i.e. we obtained an appropriate structure of the bumblebee Lagrangian, involving a Maxwell-like kinetic term (we note that such a form of the kinetic term is necessary to avoid ghosts, see \cite{Bled}), a positively defined potential, and also, Proca-like and CFJ terms.
	
	Now, let us consider the antisymmetric tensor field case. We start with another four-fermion model:
	\bea
	{\cal L}=\bar{\psi}(i\ds-m)\psi-\frac{G}{2}J_{\mu\nu}J^{\mu\nu},
	\eea
	where $J_{\mu\nu}=i\bar{\psi}\gamma_{[\mu}\partial_{\nu]}\gamma_5^q\psi$, with $q=1,2$, is a (pseudo)tensor current which can be coupled to an antisymmetric tensor field in the minimal manner, i.e. through the term $B_{\mu\nu}J^{\mu\nu}$. We note that, unlike the above case of the vector bumblebee model, in this theory the choice of zero mass will not simplify calculations essentially. Again, we introduce the bumblebee field, in this case, tensorial one, $B_{\mu\nu}$, in order to eliminate the four-fermion coupling:
	\begin{eqnarray}
		\label{rewrite}
		{\cal L} &=& {\cal L}_0 + \frac{g^2}{2} \left(B_{\mu\nu}-\frac{e}{g^2}J_{\mu\nu}\right)^2 \nonumber \\
		&=& \frac{g^2}{2}B_{\mu\nu} B^{\mu\nu} + \bar\psi(i\ds-ieB^{\mu\nu}\gamma_{[\mu}\partial_{\nu]}\gamma_5^q-m)\psi.
	\end{eqnarray}
Similarly to (\ref{1lcep}), we can write the one-loop corrected effective potential of our theory as
\begin{equation}\label{Vef}
	V_\mathrm{eff} = -\frac{g^2}{2}B_{\mu\nu} B^{\mu\nu} +i \,\mathrm{tr} \int\frac{d^4p}{(2\pi)^4}\, \ln(\ps-eB_{\mu\nu}\gamma^{[\mu}p^{\nu]}\gamma_5^q-m).
\end{equation}
In this case, we find that at $q=2$, the one-loop effective potential can be obtained exactly \cite{spont2}. However, for study of spontaneous Lorentz symmetry breaking, assuming that fields are small, it is sufficient to find only the terms up to the fourth order in dynamical fields. In \cite{spont2} our effective potential was shown to look like, after renormalization:
\begin{equation}
	V^{(q=2)}_\mathrm{eff} = \frac{m^4}{4} (e^2B_{\mu\nu}B^{\mu\nu}-b_{\mu\nu}b^{\mu\nu})^2+\ldots,
\end{equation} 
where the $b_{\mu\nu}$ is the non-zero v.e.v. of $B_{\mu\nu}$. 

For $q=1$, we can find the one-loop effective potential only order by order. The relevant contribution looks like
\bea
V^{(q=1)}_{eff} &=& \frac{7 m^4}{12} (e^2B_{\mu\nu}B^{\mu\nu}-b_{\mu\nu}b^{\mu\nu})^2  +\frac{7e^4m^4}{48} (B_{\mu\nu}\tilde{B}^{\mu\nu})^2+\ldots, 
\eea
where $\tilde{B}_{\mu\nu}$ is the dual of $B_{\mu\nu}$: $\tilde{B}_{\mu\nu}=\frac{1}{2}\epsilon_{\mu\nu\rho\sigma}B^{\rho\sigma}$.
So, in both cases we have continuous sets of minima allowing for spontaneous symmetry breaking. The derivative dependent contributions to the effective action, both for $q=1$ and $q=2$, can be obtained as well. Explicitly, the low-energy effective Lagrangians for these theories, after wave function renormalizations and adding some finite constants, look like
\begin{eqnarray}
	{\cal L}_{B,1} &=& -\frac{1}{12} H_{\mu\nu\lambda}H^{\mu\nu\lambda} -\frac{7 m^4}{12} (e^2B_{\mu\nu}B^{\mu\nu}-b_{\mu\nu}b^{\mu\nu})^2 -\frac{2}{5} (\partial_\alpha B^{\mu\alpha})^2 -\nonumber\\
	&-& \frac{7e^4m^4}{48} (B_{\mu\nu}\tilde{B}^{\mu\nu})^2+{\cal O}(B^3),
\end{eqnarray}
and
\begin{eqnarray}
	{\cal L}_{B,2} &=& -\frac{1}{12} H_{\mu\nu\lambda}H^{\mu\nu\lambda} -\frac{m^4}{4} (e^2B_{\mu\nu}B^{\mu\nu}-b_{\mu\nu}b^{\mu\nu})^2 -\frac{e^4m^4}{16} (B_{\mu\nu}\tilde{B}^{\mu\nu})^2+\nonumber\\ &+&
	{\cal O}(B^3).
\end{eqnarray}
Thus, we found that each of these Lagrangians displays the same structure as in the vector field case (\ref{Lvecsp}), i.e. it includes a gauge invariant kinetic term, a potential allowing to break the Lorentz symmetry spontaneously, and some other terms.

To close the section, we note, first, that the Hamiltonian formulation of theories with spontaneous Lorentz symmetry breaking displays some peculiarities \cite{Seif2019,EscPot} (for the discussion of degrees of freedom in such theories, see also \cite{Seif2018}), second, that the spontaneous Lorentz symmetry breaking plays a special role in a curved space-time where it appears to be the only consistent mechanism to break the Lorentz symmetry, third, that in some cases, dynamical Lorentz symmetry breaking can generate a nontrivial topology \cite{Seif2010}. The detailed discussion of Lorentz symmetry breaking in gravity will be presented in the Chapter 6.

\section{Conclusions}

In this chapter, we discussed the main classical impacts of the Lorentz symmetry breaking, related with general structure of classical actions of various field theory models, corresponding propagators and plane wave solutions. It is clear that the list of results obtained at the classical level is much wider. For example, one can mention many studies of exact solutions and geometric phases. One of the first important results in this direction is the explicit expression for static solutions of the modified Maxwell equations in the  three-dimensional LV extended QED, whose action involves the Julia-Toulouse term (\ref{JT}). It was shown in \cite{Ferr2003} that in this case the static equations are the fourth order ones, and while at small distance the scalar  potential displays the usual logarithmic behavior, at large distance, instead of the usual exponential-like decay occurring in the usual Chern-Simons modified QED, the potential is dominated by the LV term and logarithmically grows.  Further, solutions of classical equations of motion in different LV extensions of QED were studied also, e.g. in \cite{CasFerr}  (see also references therein). The paper \cite{FerrBel}, where a possibility of arising the Aharonov-Casher phase in the extended QED, with the LV term was introduced through the already mentioned magnetic coupling (\ref{magn}), was studied, started a line of investigation of geometrical phases in LV theories. Further, the impact of Lorentz symmetry breaking in solutions of the Dirac equation has been also considered in \cite{BelEO,SchReis,SchReis1} and many other papers; for the review of these studies, we recommend also \cite{BaBel}.

All these results clearly show that the LV effects should have nontrivial implications already at the tree level. At the same time, one of the most important problems is -- what possible mechanisms could be responsible for arising the LV modifications of known field theory models? One of these mechanisms uses the spontaneous  Lorentz symmetry breaking \cite{KosSam} discussed briefly in this chapter, with its key feature consists in arising of LV terms at the tree level as a consequence of choosing some minimum of the potential introducing the privileged vector or tensor, so that the Lagrangian of the theory considered near this minimum displays dependence on this vector (tensor), and the another one, based on the old idea of emergent dynamics \cite{Bjorken}, is a perturbative generation approach, intensively used to justify the possibility for arising the LV modifications of the field theory models, and applied for the first time already in \cite{JaKo}. Following this idea, there is some fundamental spinor matter field coupled in a LV manner with other fields (scalar, vector or gravity ones), so that different LV additive terms for these fields arise as one-loop quantum corrections. We note that, in a certain sense, dynamical symmetry breaking, involving both generating of a quantum correction and choosing of one minimum, combines these approaches. At the same time we note that the coupling of a gauge field to a scalar field instead of a spinor one is less convenient for perturbative generation of LV terms since the corresponding action will not include the Dirac matrices necessary to generate the Levi-Civita symbol which is present in CPT-Lorentz breaking terms discussed in this book. The perturbative generation methodology allows to obtain many of the terms proposed in \cite{KosGra}, and we will discuss it in the next chapter.

\newpage

\chapter[Perturbative generation]{Perturbative generation of LV terms}

In this chapter, we demonstrate the methodology allowing for generating various LV terms, involving scalar and gauge fields. Conceptually, we follow the already mentioned idea of the emergent dynamics \cite{Bjorken}, suggesting that the new contributions to Lagrangians of different field models can arise as quantum corrections in some fundamental theories. Namely this method has been used in \cite{JaKo}, where an appropriate LV coupling of the spinor field to a constant axial vector $b_{\mu}$  within a LV extension of QED was applied to generate the CFJ term. Just this approach became further the main tool allowing to obtain new LV terms. Besides of this, it is worth to mention also other methodologies allowing to do this, such as: dimensional reduction \cite{BelFerr}, spontaneous Lorentz symmetry breaking, that was briefly discussed in the previous chapter, and known, in some cases, like in \cite{TM3D}, to occur in the one-loop corrected effective action of a corresponding theory, so that one can speak about the dynamical Lorentz symmetry breaking, and the noncommutative fields method \cite{Gamboa}. However, the perturbative generation seems to be the most natural manner for obtaining LV terms. We note nevertheless that this way is especially appropriate just in the case when the corresponding LV quantum corrections are finite, since otherwise one should, in order to ensure the multiplicative renormalizability, introduce the corresponding LV term already at the tree level, which clearly contradicts to the interpretation of the corresponding new LV term as an emergent phenomenon.

\section{Problem of ambiguities}

First of all, we should discuss the reasons for the most controversial phenomenon characteristic for four-dimensional LV gauge contributions, that is, the arising of the \index{ambiguities} ambiguities. The characteristic situation \cite{JackAmb} is the following one: the one-loop integral over the internal momenta is an expression looking like a sum of two divergent terms, with the pole parts of these terms mutually cancel. However, any divergent term represents itself as a sum of a pole part and a finite part. In principle, this finite part can be fixed with use of an appropriate normalization condition \cite{Bonneau}. Nevertheless, in general, there is no reasons to prefer any of different normalization conditions in LV theories, especially taking into account that the corresponding terms arising after the calculations are actually finite and thus do not depend on any renormalization scale. Therefore, within different regularizations, or, as is the same, different normalization conditions, the finite parts turn out to depend on the regularization scheme, and hence the result for the corresponding integral over internal momenta is ambiguous. From the formal viewpoint, this ambiguity is a perfect example of the indeterminate form $\infty-\infty$, whose fixing requires special conditions.
Typically, two most known ambiguous integrals are considered. The first of them arises within study of the CFJ term, see e.g. \cite{ourYM}, looking like
\bea
\label{ambCFJ}
J_{\mu\nu}=\int\frac{d^4p}{(2\pi)^4}\frac{\eta_{\mu\nu}p^2-4p_{\mu}p_{\nu}+3m^2\eta_{\mu\nu}}{(p^2-m^2)^3}.
\eea
One can easily verify that by carrying out the replacement $p_{\mu}p_{\nu}\to\frac{1}{4}\eta_{\mu\nu}p^2$, with a subsequent integration in four dimensions, and the replacement $p_{\mu}p_{\nu}\to\frac{1}{4+\epsilon}\eta_{\mu\nu}p^2$, with a subsequent integration in $4+\epsilon$ dimensions, and,  afterward, taking the limit $\epsilon\to 0$, one arrives at different results for $J_{\mu\nu}$ \cite{ourYM}. This difference stays at the finite temperature as well, and, moreover, these two possibilities do not exhaust a list of other manners of calculating this integral, with more results can be obtained.

Another ambiguous integral was found in \cite{aether}. It reads as
\bea
\label{ambaether}
I_{\mu\nu}=\int\frac{d^4p}{(2\pi)^4}\frac{m^2\eta_{\mu\nu}+2p_{\mu}p_{\nu}-\eta_{\mu\nu}p^2}{(p^2-m^2)^2},
\eea
and also yields different results under the same replacements as above. Thus, we can conclude that in LV theories, in certain cases, formally divergent contributions can be actually finite, but display ambiguities as a consequence of arising the indeterminate form $\infty-\infty$. 

In principle, more ambiguous integrals can exist. For example, in a manner similar to the above discussions, one can show that logarithmically divergent integrals of the form 
\bea
J_{\mu\nu}^{(n)}=\int\frac{d^4p}{(2\pi)^4}\frac{p^{2n}(\eta_{\mu\nu}p^2-4p_{\mu}p_{\nu}+3m^2\eta_{\mu\nu})}{(p^2-m^2)^{n+3}},
\eea 
for all non-negative $n$, will be ambiguous by the same reasons, and one of the possibilities for an analogous generalization for $I_{\mu\nu}$ is described by the following superficially quadratically divergent integrals: 
\bea
I_{\mu\nu}^{(n)}=\int\frac{d^4p}{(2\pi)^4}\frac{p^{2n}(m^2\eta_{\mu\nu}+2p_{\mu}p_{\nu}-\eta_{\mu\nu}p^2)}{(p^2-m^2)^{n+2}}.
\eea
However, up to now, there is no known constructive examples of physically motivated scenarios where both these ambiguous integrals really arise, except of the usual case $n=0$ corresponding to eqs. (\ref{ambCFJ},\ref{ambaether}) and occurring in the above-mentioned LV extensions of QED. In principle, other, more involved, ambiguous loop integrals can exist as well, see f.e. \cite{d5arb}.

At the same time, it should be noted that the chiral QED, whose Lagrangian is given by \cite{nonamb}
\bea
\label{chiqed}
{\cal L}=\bar{\psi}(i\ds+e\As(1-\gamma_5)+\bs(1+\gamma_5)-m)\psi,
\eea
displays a highly nontrivial behavior. Indeed, since the spinor propagator in this theory is the common one, that is, $\frac{i}{\ks-m}$, the CFJ-like contribution to the effective Lagrangian will be given by the expression:
\bea
\label{s2p}
S_2(p)&=&-\frac{e^2}{2}{\rm tr}\int\frac{d^4k}{(2\pi)^4}\Big[\frac{1}{\ks+\ps-m}\As(-p)(1-\gamma_5)\frac{1}{\ks-m}\bs(1+\gamma_5)\times\nonumber\\&\times& \frac{1}{\ks-m}\As(p)(1-\gamma_5)+\\
&+&
\frac{1}{\ks+\ps-m}\bs(1+\gamma_5)\frac{1}{\ks+\ps-m}\As(-p)(1-\gamma_5)
\times\nonumber\\&\times&
\frac{1}{\ks-m}\As(p)(1-\gamma_5)
\Big].\nonumber
\eea
Let us now take into account only the first order in the external momentum $p$, use the expression $\frac{1}{\ks-m}=\frac{\ks+m}{k^2-m^2}$, and consider e.g. only the first contribution of $S_2$. Then, we get
\bea
S_{2,1}(p)&=&\frac{e^2}{2}{\rm tr}\int\frac{d^4k}{(2\pi)^4}\frac{\ks+m}{(k^2-m^2)^4}\ps(\ks+m)\As(-p)
\times\nonumber\\ &\times&
(1-\gamma_5)(\ks+m)\bs(1+\gamma_5)(\ks+m)\As(p)(1-\gamma_5).
\eea
This expression is superficially logarithmically divergent and involves the factor $(1-\gamma_5)(\ks+m)\bs(1+\gamma_5)$. However, $\gamma_5$ commutes with even degrees of $\gamma^{\mu}$ and anticommutes with odd ones, hence, the UV leading momentum dependent factor, which only contributes to the divergence, turns out to be proportional to $(1-\gamma_5)(1+\gamma_5)=0$. Thus, the potential divergence disappears. The similar situation occurs in another contribution to (\ref{s2p}), implying also vanishing of the leading (divergent) contribution. As a result, unlike the usual calculation of the CFJ term, after the calculation of the trace, the resulting expression is superficially finite and hence contains no ambiguity. Therefore, in the chiral QED (\ref{chiqed}) the CFJ term is ambiguity-free. The explicit result for it is \cite{nonamb}:
\bea
{\cal L}_{CFJ}=\frac{e^2}{3\pi^2}\epsilon^{\mu\nu\lambda\rho}b_{\mu}A_{\nu}\pa_{\lambda}A_{\rho}.
\eea
However, one can show that in this case, the absence of an ambiguity is paid by a price of existence of the gauge anomaly. Indeed, it is easy to see that in this case the gauge-breaking Proca-like divergent term is generated. Hence, the consistent elimination of ambiguities requires a strong extension of the model which has been performed  in \cite{Scarpnew}. 

\section{LV contributions in the scalar sector}

In this section, we discuss the simplest possible LV modifications for the scalar field models. There are two typical terms, namely, the aether-like term (\ref{scalaether}) and the essentially two-dimensional Chern-Simons-like term (\ref{scal2d}).

The simplest manner to generate the aether-like term is based on the use of the LV Yukawa-like action
\bea
S=\int d^Dx\Big[\bar{\psi}(i\ds-m)\psi+\frac{1}{2}(\pa^{\mu}\phi\pa_{\mu}\phi+m^2\phi^2)-g\bar{\psi}\as\psi\phi\Big],
\eea
where the LV vector $a_{\mu}$ is dimensionless, and the new vertex, given by (\ref{yulv}), is one of the ingredients of the LV SME \cite{ColKost2}.

The lower contribution to the aether-like term is given by the Feynman diagram depicted by the Fig.~3.1 \cite{aether}:

\begin{figure}[!ht]
\begin{center}
\includegraphics[angle=0,scale=1.0]{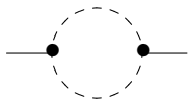}
\end{center}
\caption{Two-point function of the scalar field.}
\end{figure}

After calculating the trace and disregarding the terms proportional to $a^2$ (denoted here by dots), which do not break the Lorentz symmetry, we arrive at the following result:
\bea\label{i2}
S_2(p)&=& -\frac{d}{2}g^2\phi(p)\phi(-p)[\eta^{\mu\nu}\eta^{\rho\sigma}+
\eta^{\mu\sigma}
\eta^{\rho\nu}]a_{\mu}a_{\rho}\times\nonumber\\&\times&
\int\frac{d^Dk}{(2\pi)^D}\frac{k_{\nu}(k_{\sigma}+p_{\sigma})}{[k^2-
  m^2][(k+p)^2 -m^2]}+ \ldots,
\eea
where $d$ is a dimension of Dirac matrices in the corresponding space-time, and $D$ is a space-time dimension. It is clear that the aether-like contribution from this integral is finite in two- and three-dimensional space-times, and, within the framework of the dimensional regularization, also in all space-times with higher odd dimensions. Explicitly, one gets, after returning to the coordinate space:
\bea
S_2=-\frac{dg^2\Gamma(2-\frac{D}{2})}{6(4\pi)^{D/2}(m^2)^{2-D/2}}\phi(a\cdot\pa)^2\phi.
\eea
This term evidently replays the structure (\ref{scalaether}). We see that at $D=4$, the aether-like term should be present already at the tree level, in order to introduce the one-loop counterterm of the form $\frac{1}{2}Z_{aether}\phi(a\cdot\pa)^2\phi$, with $Z_{aether}=1-\frac{g^2}{12\pi^2\epsilon}$. We note that in this case the renormalizability of the theory is not spoiled since no couplings with a negative mass dimension are present. In principle, the aether term for the scalar field can be generated also through other couplings to a spinor matter, e.g. the CPT-even one
\bea
S=\int d^Dx\Big[\bar{\psi}(i\ds+ik^{\mu\nu}\gamma_{\mu}\partial_{\nu}-h\phi-m)\psi+\frac{1}{2}(\pa^{\mu}\phi\pa_{\mu}\phi+m^2\phi^2)\Big],
\eea
with $k^{\mu\nu}$ is a constant tensor, and again, the aether-like term will be finite in two, three and all higher odd dimensions, and divergent in four dimensions.

There is one more possible LV additive term for the scalar theory, that is, the essentially two-dimensional term (\ref{scal2d}) (see \cite{BM} for its applications for studies of topological defects). To generate it, we start with the following two-dimensional action, see \cite{PP}:
\bea
S=\int d^2x\bar{\psi}\Big[i\ds-m-g\as\phi-mg\gamma_5\chi\Big]\psi,
\eea
where $\gamma_5=\gamma_0\gamma_1$ now is a two-dimensional analogue of the chirality matrix, $\phi$ is a scalar field (denoted by a solid line), and $\chi$ is a pseudoscalar one (denoted by a dashed-and-dotted line).

\begin{figure}[!ht]
\begin{center}
\includegraphics[angle=0,scale=1.0]{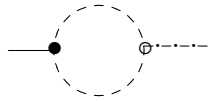}
\end{center}
\caption{Two-point function of $\phi$ and $\chi$ fields.}
\end{figure}

Here the dark circle is for $\as$ insertion, and the light one is for $\gamma_5$ one.
The contribution of the Fig.~3.2 looks like
\bea
\Gamma=\int\frac{d^2p}{(2\pi)^2}\phi(-p)\Pi(p)\chi(p),
\eea
with $\Pi(p)$ being a self-energy tensor given by
\bea
\Pi(p)=-img^2{\rm tr}\int\frac{d^2k}{(2\pi)^2}\Big[\as\frac{i}{\ks-m}\gamma_5\frac{i}{\ks+\ps-m}\Big].
\eea
After straightforward calculations, we find that, in the first order in the external $p_{\mu}$,
\bea
\Pi(p)=-i\frac{g^2}{2\pi}\epsilon^{\mu\nu}a_{\mu}p_{\nu}.
\eea
So, our final result for this contribution to the effective action is
\bea
\Gamma=\frac{g^2}{2\pi}\int d^2x a_{\mu}\epsilon^{\mu\nu}\phi \pa_{\nu}\chi.
\eea
This is just the term (\ref{scal2d}). We note that it is ambiguity-free as is should be, since the result is superficially finite. In principle, such a term, for the special case of a space-like $a_{\mu}$ (namely, the case $a_0=0$), can be also generated through the noncommutative field method proposed in \cite{Gamboa}.

In principle, other LV additive terms, involving scalar fields only, can be introduced, e.g., the Myers-Pospelov-like ones (cf. \cite{MP}):
\bea
{\cal L}_{MP,scal}=\frac{1}{\Lambda_N}\bar{\phi}(n\cdot\pa)^N\phi,
\eea
where $\Lambda_N$ is some scale factor of an appropriate mass dimension, $n_{\mu}$ is a dimensionless vector, and $N\geq 3$. The importance of these terms consists in the fact that for a space-like $n_{\mu}$, there will be no higher time derivatives and hence no ghosts, so, one has a consistent scheme to incorporate higher derivatives to a field theory model without use of a Horava-Lifshitz approach. These terms can be generated through schemes we presented above, at higher orders of a derivative expansion of the corresponding effective action.

\section{Mixed LV scalar-vector contribution}

In three space-time dimensions, there is an unique possibility to construct the quadratic LV additive term, involving fields of different natures, scalar one and gauge one, that is, the Julia-Toulouse term (\ref{JT}). In the section 2.3, we already noted that this term naturally emerges within the gauge embedding of the special Lorentz-CPT breaking extension of the self-dual model. However, this is not an unique way to generate the term (\ref{JT}).

Following \cite{JTours}, we start with the following three-dimensional action:
\bea
S=\int d^3x\bar{\psi}\Big[i\ds-m-e\As-g\as\phi\Big]\psi.
\eea
Here $a^{\mu}$ is a LV constant vector, and $\phi$ is a scalar.
The one-loop effective action is clearly given by the functional trace:
\bea
\label{futra}
\Gamma^{(1)}=-i{\rm Tr}\ln\Big[i\ds-m-e\As-g\as\phi\Big].
\eea
The desired two-point ``mixed'' function is described by the Feynman diagram given by the Fig.~3.3.

\begin{figure}[!ht]
\begin{center}
\includegraphics[angle=0,scale=1.0]{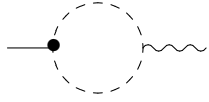}
\end{center}
\caption{Mixed two-point vector-scalar function.}
\end{figure}

Here the black dot is for the $\as$ insertion, the dashed line is for a spinor propagator, and solid and wavy lines denote external scalar and gauge fields, respectively.

The contribution of this diagram yields
\bea
I&=&-eg\,{\rm tr}\int\frac{d^3p}{(2\pi)^3}\int \frac{d^3k}{(2\pi)^3}\As(-p)(\ks+m)\phi(p)\as(\ks+\ps+m)
\times\nonumber\\&\times& \frac{1}{(k^2-m^2)[(k+p)^2-m^2]}.
\eea
For the $diag(+--)$ signature, and Dirac matrices $(\gamma^0)^{\alpha}_{\phantom{\alpha}\beta}=\sigma^2, (\gamma^1)^{\alpha}_{\phantom{\alpha}\beta}=i\sigma^1, (\gamma^0)^{\alpha}_{\phantom{\alpha}\beta}=i\sigma^3$, obeying relations: $\{\gamma^{\mu},\gamma^{\nu}\}=2\eta^{\mu\nu}$, ${\rm tr}(\gamma^{\mu}\gamma^{\nu}\gamma^{\lambda})=2i\epsilon^{\mu\nu\lambda}$, we arrive at
\bea
I&=&-2i\epsilon_{\alpha\mu\nu}egm\int\frac{d^3p}{(2\pi)^3}p^{\alpha}A^{\mu}(-p)\phi(p)a^{\nu}
\times\nonumber\\&\times&
\int \frac{d^3k}{(2\pi)^3}\frac{1}{(k^2-m^2)[(k+p)^2-m^2]}.
\eea
Then, after integrating over internal momenta, in the first order in $p_{\mu}$ we obtain
\bea
I=\epsilon_{\lambda\mu\nu}eg\frac{m}{4\pi|m|}\int\frac{d^3p}{(2\pi)^3}p^{\lambda}A^{\mu}(-p)\phi(p)a^{\nu}.
\eea
This expression can be represented as
\bea
\label{resultJT}
I=-\frac{eg}{8\pi}\sgn(m)\int d^3x \epsilon^{\lambda\mu\nu}F_{\lambda\mu}a_{\nu}\phi.
\eea
This is exactly the Julia-Toulouse term (\ref{JT}). We note that here the sign of the mass arises. It is rather typical for the three-dimensional theories, see e.g. \cite{Redlich}, where the similar situation was shown to take place for the Chern-Simons term. 

This calculation can be performed through evaluating (\ref{futra}) not only with the Feynman diagram approach as we did now, but also through the proper-time method. To proceed with it, we add to the (\ref{futra}) the field independent term $-i{\rm Tr}\ln(i\ds+m)$, in order to have a trace of an operator of second order in derivatives, as the proper-time method requires. So, our one-loop effective action becomes:
\bea
\Gamma^{(1)}=-i{\rm Tr}\ln(-\Box-m^2-e\As(i\ds+m)-g\phi \as(i\ds+m)).
\eea
Since we need only the first order in $a_{\mu}$, we expand this expression up to the first order in $a_{\mu}$ and then employ the Schwinger proper-time representation \index{proper-time representation} $A^{-1}=\frac{1}{i}\int_0^{\infty}ds e^{isA}$. We arrive at
\bea
\Gamma^{(1)}_1=-g{\rm Tr}\Big[\int_0^{\infty} ds e^{is(\Box+m^2+e\As(i\ds+m))}\phi \as(i\ds+m)\Big].
\eea
Then we expand this expression up to first derivatives and to the first order in $A_{\mu}$. We obtain
\bea
\Gamma^{(1)}_1&=&-eg{\rm Tr}\Big[\int_0^{\infty} ds e^{ism^2}\Big(is\As(i\ds+m)-s^2(\pa^{\mu}\As)(i\ds+m)\pa_{\mu}\Big)\times\nonumber\\&\times& \phi \as(i\ds+m)e^{is\Box}\Big].
\eea
After keeping only the terms with three Dirac matrices, with the subsequent Fourier transform and integrating over momenta, we arrive at the result (\ref{resultJT}). 

We note, that, in principle, our result for the Julia-Toulouse term could be obtained within usual calculations of the Chern-Simons term, where one could do the replacement $A_{\mu}\to A_{\mu}+a_{\mu}\phi$. Also, if one would like to consider the free scalar-vector theory whose action involves the usual kinetic term for the scalar field, the Maxwell term, and the Julia-Toulouse term, it is possible to show that the dispersion relations are just those ones given by  (\ref{direl2}).

There are some other manners to generate the mixed term (\ref{JT}). One of them consists in a dual immersion of the LV extension of the self-dual theory, namely this way has been employed in the section 2.3. Another one consists in a dimensional reduction of the electrodynamics with the CFJ term carried out in \cite{BelFerr} (see also \cite{SchRed}). We start with its four-dimensional Lagrangian looking like
\bea
{\cal L}=-\frac{1}{4}F_{\mu\nu}F^{\mu\nu}+\epsilon^{\mu\nu\lambda\rho}k_{\mu}A_{\nu}\pa_{\lambda}A_{\rho},
\eea
and ``freeze'' the dynamics along the third spatial dimension (perhaps, this method can be an useful tool to describe essentially two-dimensional quantum systems, for example, graphene), so, one can split the indices like $\nu=(a,3)$, with the index $a$ takes values $0,1,2$, and $A^3=\phi$, and the dependence on $x_3$ will be suppressed (thus, $\pa_3\phi=\pa_3A_a=0$), so, one arrives at the following dimensionally reduced Lagrangian (with $\epsilon^{abc}\equiv\epsilon^{3abc}$):
\bea
{\cal L}_{red}=-\frac{1}{4}F_{ab}F^{ab}-\frac{1}{2}\pa_a\phi\pa^a\phi+\epsilon^{abc}(\mu A_a\pa_bA_c-k_a F_{bc}\phi).
\eea
We see that within this reduction, the Chern-Simons term with a mass $\mu=k_3$ and the JT term are generated.

One more way to generate this term is based on the dynamical Lorentz symmetry breaking approach, presented in \cite{TM3D}. We proceed in a manner similar to that one described in the section 2.4 (see also \cite{spont}). The key idea is the following one: in the three-dimensional space-time, we consider several four-fermion interactions which, as it is usually done within the Gross-Neveu approach, afterwards are presented in terms of vertices involving corresponding different scalar or vector auxiliary  fields ${\cal A}_I$, where $I$ is a generalized summation index. As a result, we have the equivalent theory represented by the following Lagrangian: 
\bea
{\cal L}=\frac{g^2_I}{2}A_IA^I+\bar\psi(i\ds - m - e_I{\cal A}_I\Gamma^I)\psi,
\eea
where ${\cal A}_I=\{V_\mu,\Theta,A_\mu,\Phi,T_\mu\}$ is a set of the vector and scalar fields coupled to spinors with use of matrices $\Gamma_I=\{\gamma_\mu,\gamma_3,\gamma_\mu\gamma_5,\gamma_3\gamma_5,\gamma_\mu\gamma_3\gamma_5\}$ playing here the role of generalized Dirac matrices,  and $\psi$ is the four-component spinor. Here we use $4\times 4$ matrices, whose explicit form is given in \cite{TM3D}, so, our spinor representation of the Lorentz group is reducible. 

The one-loop corrected effective potential of the theory defined in a manner analogous to that one used in the section 2.4, i.e. as
\bea
V_{eff}=-\frac{g^2_I}{2}{\cal A}_I{\cal A}^I+i{\rm tr}\int\frac{d^3k}{(2\pi)^3}\ln(\ks-m-e_I{\cal A}_I\Gamma^I),
\eea
where only derivative independent terms are taken into account, evidently possesses nontrivial minima at some $\left\langle{\cal A}_I\right\rangle=b_I/e_I$, with $b_I=\{0,0,b_{\mu},\phi_0,0\}$.
Now, let us make the shift of the vacuum by the rule ${\cal A}_I\to {\cal A}_I+b_I/e_I$, so that now $\left\langle{\cal A}_I\right\rangle=0$.
As a result, the one-loop contribution to the effective action (we note that here the overall sign is changed as it must be!) takes the form
\begin{equation}
S_{eff}[{\cal A},b] = -i {\rm Tr} \ln(i\ds-m-b\cdot\Gamma-e_I{\cal A}_I\Gamma^I).
\end{equation}
In this effective action we take into account only the lower, quadratic order in fields ${\cal A}_I$, given by
\begin{eqnarray}
S_{eff}^{(2)}[{\cal A},b] &=& \frac i2 \int d^3x\, {\cal A}_I\Pi^{IJ}{\cal A}_J,
\end{eqnarray}
with the corresponding self-energy tensor $\Pi^{IJ}$ written within the derivative expansion methodology \cite{Panigrahi} (the detailed discussion of this approach will be given in the next section) looks like
\begin{eqnarray}\label{Pimunu}
\Pi^{IJ}=-e_Ie_J {\rm tr} \int \frac{d^3p}{(2\pi)^3} \frac {i}{i\ds-m-b\cdot\Gamma}\Gamma^{\it I} \frac {i}{\,\ps\,-\,i\ds-m-b\cdot\Gamma}\Gamma^{\it J}.
\end{eqnarray}
Expanding this expression up to the first order in derivatives and calculating corresponding traces and integrals, we obtain the following mixed and conventional Chern-Simons terms:
\bea
{\cal L}_{CS} &=& - \frac{e_V\,e_{V_3}}{2\pi m}\Theta\,\epsilon^{\mu\lambda\nu}b_\mu\partial_\lambda V_\nu - \frac{e_V\,e_T}{\pi}\epsilon^{\mu\lambda\nu}V_\mu\partial_\lambda T_\nu + \frac{e_{V_3}\,e_T}{\pi}\Theta\,g^{\mu\nu}b_\mu T_\nu \nonumber\\
&& + \frac{e_{\!A}\,e_{\!A_3}}{6\pi m}\Phi\,\epsilon^{\mu\lambda\nu}b_\mu \partial_\lambda A_\nu + \frac{e_{\!A}^2}{6\pi m} \phi_0\,\epsilon^{\mu\lambda\nu} A_\mu\partial_\lambda A_\nu.
\eea
We note here also the presence of Julia-Toulouse terms, i.e., the first and fourth terms.

By analyzing the above Chern-Simons-like Lagrangian, we can discuss two models that exhibit charge fractionalization without the breaking of $T$ symmetry. First, if we restrict ourselves to the case $T_{\mu}=A_{\mu}=\Phi=0$, i.e. we have only the usual vector $V_{\mu}$ and the scalar $\Theta$, we arrive at
\bea\label{CS1}
{\cal L}_{CS} &=& \frac{e_{V_3}\,e_V}{2\pi m}\,\vartheta\,\epsilon^{\mu\lambda\nu}\partial_\mu\Theta\,\partial_\lambda V_\nu,
\eea
where we have introduced $\vartheta(x)=x_\mu b^\mu+ C$, with $C$ being a constant. At the same way, if we choose $V_\mu=\Theta=T_\mu=0$, we have
\bea{\label{CS2}}
{\cal L}_{CS} &=& - \frac{e_{\!A_3}\,e_{\!A}}{6\pi m}\vartheta\,\epsilon^{\mu\lambda\nu}\partial_\mu\Phi\,\partial_\lambda A_\nu,
\eea
which, up to an adjustment of numerical constants, is the low-energy mixed Chern-Simons term for graphene discussed in \cite{Ser1,Ser2}. By considering the properties of the bilinear terms associated to the fields ${\cal A}_I$, with respect to $C$, $P$, and $T$ transformations, we note that both terms (\ref{CS1}) and (\ref{CS2}) violate the $C$ symmetry, whereas the $P$ and $T$ symmetries are preserved. This, obviously, leads to a CPT symmetry violation, with the $C$ symmetry violation is a direct consequence of the charge fractionalization \cite{TM3D}, whereas the $T$ symmetry is preserved.
This is another manner to generate the mixed scalar-vector term. In principle, it can be obtained also in an essentially non-perturbative way, that is, the Julia-Toulouse technique used for study of condensation of topological defects \cite{JTours}.
To conclude this section, we note that this term, first, can arise by different reasons, second, apparently can be useful to study condensed matter systems. 

\section{The Carroll-Field-Jackiw term}

\subsection{General situation and zero temperature case}

The CFJ term (\ref{CFJ}) is, without doubts, a paradigmatic example of the Lorentz-breaking term. 
As we noted in the beginning of this chapter, it can be generated as a quantum correction from the following in the appropriate LV extension of the spinor QED. Explicitly, this theory is described by the Lagrangian:
\bea
\label{lvqed0}
{\cal L}_{QED}=-\frac{1}{4}F_{\mu\nu}F^{\mu\nu}+\bar{\psi}(i\ds-e\As-m+\bs\gamma_5)\psi.
\eea
This possibility was for the first time described in \cite{JaKo} and, afterward, discussed in numerous papers. 
The one-loop effective action is clearly given by the fermionic determinant:
\bea
\label{fundet}
\Gamma^{(1)}=-i{\rm Tr}\ln (i\ds-e\As-m+\bs\gamma_5).
\eea
The CFJ term is evidently generated by the expansion of this determinant up to the second order in $A_{\mu}$, and to the first order in derivatives.
Its final form is
\bea
\label{cfjfinal}
{\cal L}_{CFJ}=Ce^2\epsilon^{\mu\nu\lambda\rho}b_{\mu}A_{\nu}\pa_{\lambda}A_{\rho},
\eea
with $C$ being some numerical constant. To obtain the CFJ term, and consequently the constant $C$, one considers the contributions to the two-point function described by the following Feynman diagrams:


\begin{figure}[!ht]
\begin{center}
\includegraphics[angle=0,scale=1.0]{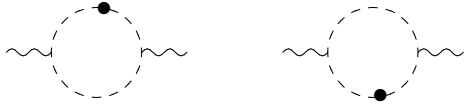}
\end{center}
\caption{Contributions to the CFJ term.}
\end{figure}


The external lines are the gauge fields, and the $\bullet$ symbol here is for the $\bs\gamma_5$ insertion. So, let us describe some ways to calculate the contributions of these two graphs.
 
Within the first approach we use the usual spinor propagators. In this case, the sum of these graphs yields
\bea
\Gamma_2[A]&=&-\frac{e^2}{2}\int \frac{d^4p}{(2\pi)^4}A^{\mu}(-p)\Pi_{\mu\nu}(p)A^{\nu}(p),
\eea
where
\bea
\label{self}
\Pi_{\mu\nu}(p)&=&{\rm tr}\int\frac{d^4k}{(2\pi)^4}\Big[
\gamma_{\mu}\frac{1}{\ks-m}\bs\gamma_5\frac{1}{\ks-m}\gamma_{\nu}\frac{1}{\ks+\ps-m}+
\nonumber\\
&+&
\gamma_{\mu}\frac{1}{\ks-m}\gamma_{\nu}\frac{1}{\ks+\ps-m}\bs\gamma_5\frac{1}{\ks+\ps-m}
\Big]
\eea
is the self-energy tensor for this theory up to the first order in $b^{\mu}$.
Following one manner, to obtain the CFJ term, one should expand the integrand up to the first order in the external momentum $p$. Namely this way was employed in the original paper \cite{JaKo}. However, already in that paper, it was argued that the CFJ term, being superficially logarithmically divergent, is actually finite but ambiguous, Indeed, this procedure yields the result
\bea
\Pi^{\mu\nu}=\frac{1}{2\pi^2}\epsilon^{\mu\nu\alpha\beta}b_{\alpha}p_{\beta}\left(\frac{\theta}{\sin\theta}-\frac{1}{4}
\right),
\eea
where $\theta=2\arcsin(\sqrt{p^2}/2m)$. Therefore, in the low-energy limit $p\to 0$, one has $\Pi^{\mu\nu}=\frac{3}{8\pi^2}\epsilon^{\mu\nu\alpha\beta}b_{\alpha}p_{\beta}$, and, then, $C=\frac{3}{16\pi^2}$. One can develop as well another procedure, within which the $b^{\mu}$ becomes an external non-constant field depending on one more external momentum, so, the only gauge invariant result for the CFJ contribution to the effective Lagrangian is zero. This suggestion was originally made in \cite{CG} and discussed in \cite{JaKo}. Further, in \cite{AltKarki,AltKarki2} it has been argued that, in a certain sense, the zero result for the CFJ term is preferable since, if one would treat $b_{\mu}$ as a dynamical field, even the contribution of the CFJ term to the effective action will break the gauge invariance, so, the zero value for the coefficient $C$ is necessary to maintain the gauge symmetry.  The same argumentation can be applied to the four-dimensional gravitational CS term as well \cite{AltKarki}.

Instead of expanding the effective action in series in $b_{\mu}$, one can perform the sum over all possible $\bs\gamma_5$ insertions and obtain the exact propagator of the spinor field \index{exact propagator of the spinor field} \cite{Andr}:
\bea
\label{exactprop}
S(p)&=&i(\ps-m+\bs\gamma_5)^{-1}=i\frac{p^2+m^2-b^2-2(b\cdot p+m\bs)\gamma_5}{(p^2+b^2-m^2)^2-4[(b\cdot p)^2-m^2b^2]}\times\nonumber\\&\times&
(\ps+m-\bs\gamma_5),
\eea
together with the dimensional regularization, one arrives at the result \cite{Andr}:
\bea
\Pi^{\mu\nu}=-\frac{1}{2\pi^2}\epsilon^{\mu\nu\alpha\beta}b_{\alpha}p_{\beta}\left(1-\theta(-b^2-m^2)\sqrt{1+\frac{m^2}{b^2}}
\right).
\eea 
Thus, for $b^2>0$ (that is, time-like $b^{\mu}$), one has $\Pi^{\mu\nu}=-\frac{1}{2\pi^2}\epsilon^{\mu\nu\alpha\beta}b_{\alpha}p_{\beta}$, and $C=\frac{1}{4\pi^2}$, while for the zero mass and the space-like $b^{\mu}$ one finds a zero result, so, $C=0$ (see details in \cite{Andr}). We note that the exact propagator (\ref{exactprop}) is especially convenient for the calculations in the massless fermions case since it displays no infrared singularities at $m=0$.
As we already mentioned in the section 3.1, the ambiguity is a price we must pay for a finiteness of the superficially (logarithmically) divergent result. It was argued in \cite{JackAmb} that this ambiguity is strongly related with the \index{Adler-Bell-Jackiw (ABJ) anomaly} Adler-Bell-Jackiw anomaly.

Let us also discuss some other ways to calculate the CFJ term. One of the main manners to do it is based on the derivative expansion method \index{derivative expansion} \cite{Panigrahi}. The essence of this method is as follows. We carry out a formal Fourier transformation of the determinant (\ref{fundet}) which we then rewrite as
\bea
\label{fundet10}
\Gamma^{(1)}&=&-i{\rm Tr}\ln (\ps-e\As-m+\bs\gamma_5)=\nonumber\\
&=&-i{\rm Tr}(1-e\As\frac{1}{\ps-m+\bs\gamma_5})+\cdots,
\eea
where dots are for the field-independent contribution. However, now, to extract possible contributions depending on derivatives of $A_{\mu}$, we use the following identity which is straightforward in the coordinate representation, it is the key identity of this approach reflecting the fact that the fields and derivatives do not commute, with $p_{\mu}$ being an integration momentum in the corresponding loop:
\bea
A_{\alpha}p_{\mu}=p_{\mu}A_{\alpha}-i(\pa_{\mu}A_{\alpha}).
\eea
This identity can be used to put all internal momenta in positions allowing to integrate over them (with the "correct" position of internal momenta is at the left-hand side from all fields). It can be straightforwardly generalized for higher orders in derivatives. Equivalently, we can start with the expression (\ref{fundet10}) with the subsequent replacements: $i\pa_{\mu}\to p_{\mu}$, $A_{\mu}(x)\to A_{\mu}(x-i\frac{\pa}{\pa p})$, and further
we expand $A_{\mu}(x-i\frac{\pa}{\pa p})=A_{\mu}(x)-i(\pa_{\nu}A_{\mu})(x)\frac{\pa}{\pa p_{\nu}}+\ldots$, and the derivatives with respect to $p_{\mu}$ must act on propagators $\frac{1}{\ps-m+\bs\gamma_5}$. Keeping only the terms of
  the second order in fields $A_{\alpha}$ and of the first order in their derivatives and in the constant $b_{\mu}$ vector, we arrive at the following expression for the one-loop effective Lagrangian \cite{TM0509008}:
\bea
\label{TM0509008}
{\cal L}^{(1)}&=&2ie^2\int\frac{d^4p}{(2\pi)^4}\frac{1}{(p^2-m^2)^3}[\epsilon^{\nu\lambda\mu\rho}(3m^2+p^2)-4\epsilon^{\nu\lambda\mu\alpha}p_{\alpha}p^{\rho}]b_{\rho}A_{\nu}\pa_{\lambda}A_{\mu}\nonumber\\ 
&=& e^2\epsilon^{\nu\lambda\mu\alpha}k_{\alpha}A_{\nu}\pa_{\lambda}A_{\mu},
\eea
where $k_{\mu}=Cb_{\mu}$, and $C$ is defined by the Eq. (\ref{cfjfinal}). 
The constant $C$ is easily shown to be finite and ambiguous, just as we discussed in the Section 3.1: if we replace $p^{\mu}p^{\nu}\to\frac{p^2}{4}\eta^{\mu\nu}$ and integrate in 4 dimensions, we get $C=\frac{3}{16\pi^2}$, and if we replace $p_{\mu}p_{\nu}\to\frac{1}{4+\epsilon}\eta_{\mu\nu}p^2$ and integrate in $4+\epsilon$ dimensions, we arrive at $C=\frac{1}{4\pi^2}$ (the same values of $C$ are presented in \cite{ourYM}, arising not only within calculating the usual CFJ term but also within computing its non-Abelian extension). Besides of these results, the coefficient $C$ was calculated with use of the exact propagator (\ref{exactprop}) in the massless case \cite{EPCW} and found to be equal to $\frac{1}{8\pi^2}$. Within the Pauli-Villars regularization, the coefficient $C$ vanishes \cite{ColKost2}.

One more methodology to treat the CFJ term in the massless case was proposed in \cite{ScarpReg}. To do it, we express the exact propagator (\ref{exactprop}) for $m=0$ as
\bea
\label{exactm0}
S_{exact}(p)=\frac{i}{\ks+\bs\gamma_5}=\frac{i}{\ps+\bs}P_L+\frac{i}{\ps-\bs}P_R,
\eea
where $P_R=\frac{1+\gamma_5}{2}$ and $P_L=\frac{1-\gamma_5}{2}$ are the left and right chiral projectors \index{chiral projectors} Actually, namely these projectors were used in the section 3.1. The self-energy tensor is expressed in terms of exact propagators in the simplest form,
\bea
\label{self0}
\Pi_{\mu\nu}(p)&=&\frac{1}{2}{\rm tr}\int\frac{d^4q}{(2\pi)^4}\gamma_{\mu}S_{exact}(q)\gamma_{\nu}S_{exact}(q+p),
\eea
so, it remains to substitute here directly the exact massless propagator (\ref{exactm0}) and extract only the CFJ-like terms, that is, those ones involving only one $\gamma_5$, which clearly can yield the Levi-Civita symbol.
The corresponding result is \cite{ScarpReg}:
\bea
\Pi^{\mu\nu}_{CFJ}(p)&=&4ib_{\beta}\epsilon^{\mu\nu\alpha\beta}\int\frac{d^4q}{(2\pi)^4}\frac{(p_{\alpha}+q_{\alpha})}{(q+b)^2(q+p+b)^2},
\eea
with the integral over $k$ is supposed to be regularized in some way. Afterwards, we apply the methodology of the implicit regularization \index{implicit regularization} (see e.g. \cite{implicit}), which allows us to find that
\bea
\Pi^{\mu\nu}_{CFJ}(p)&=&-4i\alpha_2b_{\beta}p_{\alpha}\epsilon^{\mu\nu\alpha\beta},
\eea
where $\alpha_2$ is some completely undetermined constant, finite but strongly depending on the regularization prescription. It is clear that this is the only possible one-derivative contribution to the self-energy tensor proportional to $\epsilon^{\mu\nu\alpha\beta}$. In \cite{ScarpReg} the $\alpha_2$ is treated as a surface term arising within the integration, explicitly defined through the relation
$$
\alpha_2 g_{\mu\nu}=\int\frac{d^4q}{(2\pi)^4}\frac{\pa}{\pa q^{\mu}}\left[\frac{q_{\nu}}{(q^2-\lambda^2)^2}
\right],
$$
with $\lambda$ is a mass scale.
From a formal viewpoint the arising of $\alpha_2$ is related with the fact that integrals $I_{0\alpha}=p_{\alpha}\int\frac{d^4q}{(2\pi)^4}\frac{1}{(q+b)^2(q+p+b)^2}$ and $I_{\alpha}=\int\frac{d^4q}{(2\pi)^4}\frac{q_{\alpha}}{(q+b)^2(q+p+b)^2}$ diverge, and the $\Pi^{\mu\nu}_{CFJ}$ will be proportional to their linear combination looking like a momentum $p_{\alpha}$ multiplied by some undetermined expression behaving as $\infty-\infty$, with $\alpha_2$ will be just the value of this expression, which is not fixed by any physical reasons.

Once more manner to argue that the CFJ term is strongly ambiguous being proportional to a completely undetermined constant is based on the \index{ambiguities!functional integral approach} functional integral argumentation \cite{Chung}. To do it, we start with the functional integral corresponding to the one-loop effective action (\ref{fundet}):
\bea
Z[A]&=&\int D\psi D\bar{\psi}\exp(iS[A])=\nonumber\\
&=&\int D\psi D\bar{\psi}\exp(i\int d^4x\bar{\psi}(i\Ds-m-\bs\gamma_5)\psi),
\eea
where $D_{\mu}=\pa_{\mu}-ieA_{\mu}$ is a usual gauge covariant derivative. Then, after the changes of variables $\psi(x)\to e^{i\alpha(x)\gamma_5}\psi(x)$, $\bar{\psi}(x)\to\bar{\psi}(x)e^{i\alpha(x)\gamma_5}$, the integral measure is modified as $D\psi D\bar{\psi}\to D\psi D\bar{\psi} \exp\left(-\frac{i}{8\pi^2}\int d^4x\alpha(x)\tilde{F}^{\alpha\beta}F_{\alpha\beta}\right)$, see details in \cite{Chung,Fuji}. The action $S[A]$ under this changes of variables transforms to $S^{\prime}[A]=\int d^4x[\bar{\psi}(i\Ds-me^{2i\alpha(x)\gamma_5}-\bs\gamma_5)\psi-\pa_{\beta}\alpha(x)J_5^{\beta}]$, with $J_5^{\beta}$ being a conserved axial current which, however, can be determined in an ambiguous manner like $J_5^{\beta}=\bar{\psi}\gamma^{\beta}\gamma_5\psi+c\tilde{F}^{\beta\gamma}A_{\gamma}$, with an arbitrary constant $c$, and the $\tilde{F}^{\beta\gamma}$ is a tensor dual to $F_{\mu\nu}$, since the only restriction for this axial current is that it should satisfy the conservation law $\pa_{\beta}J^{\beta}_5=0$. Choosing $\alpha(x)=-x^{\alpha}b_{\alpha}$, one rests with
\bea
\label{gfu}
Z[A]&=&e^{i(c-\frac{1}{4\pi^2})\int d^4x b_{\alpha}\tilde{F}^{\alpha\beta}A_{\beta}}\times\nonumber\\&\times& 
\int D\psi D\bar{\psi}\exp(i\int d^4x\bar{\psi}(i\Ds-me^{2i\gamma_5x^{\mu}b_{\mu}})\psi).
\eea
We calculate the functional determinant of $i\Ds-me^{2i\gamma_5x^{\mu}b_{\mu}}$ up to the first order in $b_{\mu}$ and the second order in the field $A_{\mu}$. In the momentum space, multiplication by the coordinate $x^{\mu}$ is equivalent to differentiation with respect to the corresponding momenta, so we can perform the loop integration as above in this section, with the result for the exponential of the fermionic determinant will be \cite{Chung}:
\bea
& &\int D\psi D\bar{\psi}\exp(i\int d^4x\bar{\psi}(i\Ds-me^{2i\gamma_5x^{\mu}b_{\mu}})\psi)= \nonumber\\ &=& \exp(\frac{i}{4\pi^2}\int d^4x b_{\alpha}\tilde{F}^{\alpha\beta}A_{\beta}).
\eea
Substituting this result to the generating functional $Z[A]$ (\ref{gfu}), we find that the only surviving term is that one proportional to the constant $c$. Thus, the corresponding contribution to the effective action $\Gamma[A]=-i\ln Z[A]$ is completely undetermined being equal to
\bea
\Gamma[A]=c\int d^4x b_{\alpha}\tilde{F}^{\alpha\beta}A_{\beta},
\eea
that is, the CFJ coefficient $C$ from (\ref{cfjfinal}) within this prescription is equal to the factor $c$.
We note that this fact completely matches the conclusion achieved in \cite{ScarpReg} on the base of the implicit regularization. Regarding the argumentation presented in \cite{Bonneau}, we note that it is based on choosing appropriate normalization conditions, whereas there is no unique prescription to  define these conditions once and forever, so, the ambiguity can be eliminated only within a separate calculation, but there is no profound reasons to say that one value of the constant $C$ is correct but others are not. Actually, the papers \cite{Chung,ScarpReg} argued that the ambiguity of the CFJ term is intrinsic. We note that the result $C=\frac{1}{4\pi^2}$ has been unambiguously obtained also within the Schwinger proper time approach \cite{ptime}. However, using of the proper time formulation itself involves some specific regularization prescription.

\subsection{Finite temperature case}

The ambiguity we found for the CFJ term can be naturally generalized for the \index{finite temperature} finite temperature case. To illustrate the situation, we study the contribution (\ref{TM0509008}) at the finite temperature within two different regularization schemes analogous to those ones used in a previous subsection, that is, we suggest the purely spatial dimension $d$ to be either 3 or $3+\epsilon$ with $\epsilon\to 0$, so, actually, we again face the result which is finite but undetermined at $\epsilon=0$, i.e. a removable singularity.

As usual within the finite temperature approach, we follow the Matsubara formalism, that is, we assume that the system stays in a thermal equilibrium at the temperature $T=\frac{1}{\beta}$. Then, we carry out the Wick rotation and suggest that the time-like component of the moment is discrete, $p_4=2\pi T(n+\frac{1}{2})$, with $n$ is integer. In this case we can replace the integral over $p_0$ by the sum over $n$. It is instructive to consider (\ref{TM0509008}) separately for the time-like and space-like $b_{\mu}$.

Within one manner we start with the expression (\ref{TM0509008}), and, after applying the Matsubara formalism, promote spatial integral to $d$ dimensions, and choose $b_{\mu}$ to be time-like. We arrive at
\bea
k_4=b_4T\sum_{n=-\infty}^{\infty}\int\frac{d^d\vec{p}}{(2\pi)^d}\frac{3\omega^2_n-\vec{p}^2}{(\vec{p}^2+\omega^2_n)^3}=\frac{b_4}{4\pi^2},
\eea
that is, in this case we have $C=\frac{1}{4\pi^2}$ which matches one of zero-temperature results obtained in \cite{Andr}.
If within (\ref{TM0509008}) we suggest the vector $b_{\mu}$ to be space-like, we replace $p_ip_j\to\delta_{ij}\frac{\vec{p}^2}{d}$, with $d=3+\epsilon$ is a spatial dimension, and arrive at
\bea
k_i=b_iT\sum_{n=-\infty}^{\infty}\int\frac{d^d\vec{p}}{(2\pi)^d}\frac{4m^2-\omega^2_n+(\frac{4}{d}-1)\vec{p}^2}{(\vec{p}^2+\omega^2_n)^3},
\eea
which is not ambiguous within this calculation, and the final result is
\bea
k_i=\frac{b_i}{4\pi^2}(1+F(\xi)),
\eea
where $\xi=\frac{m}{2\pi T}$, and
\bea
\label{fxi}
F(\xi)=2\pi^2\int_{|\xi|}^{\infty}dz(z^2-\xi^2)^{1/2}\frac{\tanh(\pi z)}{\cosh^2(\pi z)}
\eea
is a temperature depending function vanishing at $T\to 0$, i.e. $\xi\to\infty$, and tending to 1 at $T\to\infty$.

Alternatively, we could first replace $p^{\mu}p^{\nu}\to\frac{1}{4}\eta^{\mu\nu}p^2$ in (\ref{TM0509008}), and only afterward apply the Matsubara formalism. In this case we would arrive at
\bea
k_{\mu}=\frac{3b_{\mu}}{16\pi^2}(1+F(\xi)),
\eea
where $F(\xi)$ is given by (\ref{fxi}).
 We conclude that the CFJ term continues to be ambiguous at non-zero temperature as well, so, the finite temperature is not sufficient to eliminate the ambiguity of the result.

It is clear that this list of possible finite-temperature values for $k_{\mu}$ we gave here is not complete since other regularizations exist as well. Moreover, it is interesting to note that only within some specific schemes one finds that all components of the $k_{\mu}$ vector characterizing the CFJ term are proportional to the original LV vector $b_{\mu}$ with the same temperature dependent coefficient $\alpha$, that is, $k_{\mu}=\alpha(T) b_{\mu}$, as occurs, for example, in the second scheme we used here, whereas in general case it is not so. The fact that in general $k_{\mu}$ is not proportional to $b_{\mu}$, clearly reflects the intrinsic Lorentz symmetry breaking occurring in a finite temperature case due to the existence of the privileged reference frame of the thermal bath, which could break the proportionality between $k_{\mu}$ and $b_{\mu}$. In \cite{schemes} the strong asymmetry between spatial and temporal components of the $k_{\mu}$ stemming from this breaking of the proportionality has been obtained with use of non-covariant splitting the integration momentum, applied by the rule $p_{\mu}\to\hat{p}_{\mu}+\delta_{\mu 0}p_0$ so that $\hat{p}_0=0$, with subsequent arising the "covariant" and "non-covariant" CFJ-like contributions (this asymmetry has been further confirmed in \cite{ourJHEP}). Actually, a lot of manners to calculate the CFJ term, starting from the Lagrangian (\ref{lvqed0}), on the base of different regularization schemes, including the finite temperature case, is known (see e.g. \cite{ourJHEP} and references therein). 

\subsection{Non-Abelian generalization}

The natural generalization of the CFJ term consists in its non-Abelian generalization which can be introduced by analogy with the well-known non-Abelian CS term. Originally, the non-Abelian extension of the CFJ term (\ref{nab}) has been proposed in \cite{ColMac}. To show that it could emerge as a perturbative correction, we start with the expression for the one-loop effective action of the gauge field (\ref{fundet}), where, however, we suggest that the gauge field is Lie-algebra valued, $A^{\mu}=A^{\mu a}T^a$, where $T^a$ are generators of some Lie group satisfying the relations $[T^a,T^b]=if^{abc}T^c$ and ${\rm tr}(T^aT^b)=\delta^{ab}$.
To find the non-Abelian CFJ term we should, besides of the contribution involving two external gauge fields and one derivative acting on one of these fields, as in the subsection 3.4.1, consider also the contribution with three external fields and no derivatives.

This scheme of calculations has been pursued in \cite{ourYM}. In this case, the corresponding generalization of the fermionic determinant (\ref{fundet}) is
	\bea
	\label{fundet1}
	\Gamma^{(1)}=-i{\rm Tr}\ln \left[(i\ds-m+\bs\gamma_5)\delta^{ij}-g\gamma^{\mu}A^a_{\mu}(T^a)^{ij}\right].
	\eea
After expansion of this determinant up to relevant orders and calculating traces and integrals, the non-Abelian CFJ term was found to be
\bea
\label{resnab}
S_{CFJ,non-Abelian}=\int d^4x\epsilon^{\nu\lambda\rho\mu}k_{\mu}(A^a_{\nu}\pa_{\lambda}A^a_{\rho}-\frac{2}{3}igf^{abc}A_{\nu}^aA_{\lambda}^bA_{\rho}^c),
\eea
with the constant vector $k_{\mu}$ is just the same one that can be read off from the expression (\ref{TM0509008}). Repeating the discussion of the previous subsections, we conclude that this vector is finite and ambiguous. Moreover, in the finite temperature case it behaves just in the same way as in the Abelian theory (see the subsection 3.4.2).

Another way to obtain this term is based on the Schwinger proper time approach \cite{ptime}: we consider the same expression for the one-loop effective action (\ref{fundet}), but with the Lie-algebra valued vector field $A^{\mu}=A^{\mu a}T^a$. After adding the field-independent factor $-i{\rm Tr}\ln(i\ds-m-\bs\gamma_5)$ in order to deal with the second-order operator of the standard form $\Box+\ldots$ which is the most convenient one within the proper time method, and expanding the effective action up to the first order in the LV vector $b_{\mu}$, we obtain the desired contribution to the effective action in the form
\bea
\Gamma=-i{\rm Tr}(\Box+ig\As\ds+mg\As+m^2)^{-1}[(g\As+2m)\bs\gamma_5-2i(b\cdot\pa)\gamma_5].
\eea
Now, we employ the Schwinger representation by the rule
\bea
(\Box+ig\As\ds+mg\As+m^2)^{-1}=\int_0^{\infty}dse^{-s(\Box+ig\As\ds+mg\As+m^2)},
\eea
and, after manipulations described in \cite{ptime}, we see that the divergent part of this effective action completely vanishes,  and the result identically reproduces the expression (\ref{resnab}), with $k_{\mu}=\frac{g^2}{4\pi^2}b_{\mu}$, this is one of the results obtained in \cite{ourYM}. Formally this result is ambiguity-free, however, we note that the Schwinger representation itself plays the role of the specific regularization eliminating the ambiguity. We note nevertheless that the number of methods to generate the CFJ term, both in Abelian and non-Abelian versions, is not exhausted by the approaches described here, since it can be generated from much more sophisticated theories, including non-minimal ones, see f.e. \cite{d5arb}.

Another interesting question is the study of topological properties of the non-Abelian CFJ term. In principle, one could consider the simplest topological structure, that is, to suggest the vector $k^{\mu}$ to be directed along the $z$ axis, therefore, for any value of $z$ coordinate one has a topological structure of the three-dimensional non-Abelian Chern-Simons theory, and the topological structure of the whole four-dimensional theory is some foliation obtained as a direct sum of the three-dimensional topological structures. However, this issue certainly requires further studies. 

\section{CPT-even term for the gauge field}

The CPT-even extension for the free gauge field has been originally suggested \cite{ColKost2} to look like
\bea
\label{leven}
{\cal L}_{even}=-\frac14\kappa^{\mu\nu\rho\sigma}F_{\mu\nu}F_{\rho\sigma},
\eea
where $\kappa^{\mu\nu\rho\sigma}$ is a constant tensor possessing the symmetries of the Riemann curvature tensor. One of the most interesting particular cases is that one when this tensor is expressed in terms of only one constant vector (or pseudovector) $u_{\mu}$ through the expression (\ref{kappavect}).
In this case the term (\ref{leven}) takes the form:
\bea
\label{leven1}
{\cal L}_{even}=u^{\mu}u_{\rho}F_{\mu\nu}F^{\rho\nu}.
\eea
Namely this expression was discussed in \cite{Carroll} within the context of higher dimensions.
The natural question evidently consists in the possibility for its perturbative generation. Certainly it could be done in different ways.

The most straightforward way to obtain this term is based on use of the magnetic (non-minimal) coupling, namely this way was followed in \cite{aether}. The magnetic coupling has been introduced in \cite{FerrBel} in the four-dimensional space in the form (\ref{magn}).
We have showed in Section 2.4 that just this coupling arises within the dual embedding procedure of the LV self-dual model (see (\ref{new})), where the current is the usual spinor one, that is, $j^{\mu}=\bar{\psi}\gamma^{\mu}\psi$. However, it is clear that, first, it would be interesting to consider the theory involving not only the non-minimal coupling only but also the usual interaction vertex $e\bar{\psi}\gamma^{\mu}\psi A_{\mu}$, second,  it is possible to obtain not only a purely non-minimal contribution to the aether term but other ones as well.  Although the non-minimal coupling is non-renormalizable, we note that if we restrict our study by the one-loop order, or, as is the same, by the fermionic determinant, the non-renormalizability does not give any problems.

To demonstrate how the aether term can arise, let us consider the extended QED with the magnetic coupling whose action, in the spinor sector, is \cite{QEDNM}:
\bea
\label{qednm}
S=\int d^4x\bar{\psi}(i\ds-m-e\As-g\epsilon_{\mu\nu\rho\sigma}\gamma^{\mu}b^{\nu}F^{\rho\sigma}-\bs\gamma_5)\psi,
\eea
i.e., the role of $u^{\mu}$ used in (\ref{leven1}) is played by the axial vector $b^{\mu}$.
The corresponding fermionic determinant would be
\bea
\label{fundet11}
\Gamma^{(1)}=-i{\rm Tr}\ln (i\ds-e\As-g\epsilon_{\mu\nu\rho\sigma}\gamma^{\mu}b^{\nu}F^{\rho\sigma}-m-\bs\gamma_5).
\eea

There will be three different one-loop aether-like contributions in this theory: the first of them is a purely non-minimal one, the second one is mixed, involving both minimal and non-minimal vertices, and the third one involves two minimal vertices. Let us study all these cases. 

First, the purely non-minimal contribution is given by the Feynman diagram depicted at Fig. 3.5.

\begin{figure}[!ht]
\begin{center}
\includegraphics[angle=0,scale=1.0]{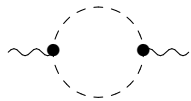}
\end{center}
\caption{Contributions to the aether term for the electromagnetic field with non-minimal vertices.}
\end{figure}

The corresponding analytic expression for this diagram looks like \cite{aether}:
\bea
\label{aethercontr}
S_2(p)&=&-\frac{g^2}{2}\epsilon^{\mu\nu\rho\sigma}\epsilon^{\mu'\nu'\rho'\sigma'}b_{\mu}F_{\nu\rho}b_{\mu'}
F_{\nu'\rho'}
\int\frac{d^4k}{(2\pi)^4}\frac{1}{[k^2-m^2]^2}\times\nonumber\\&\times& 
{\rm tr}\big[m^2\gamma_{\sigma}\gamma_{\sigma'}+k^{\alpha}k^{\beta}\gamma_{\alpha}\gamma_{\sigma}\gamma_{\beta}
\gamma_{\sigma'}\big].
\eea
There are several manners to calculate this expression. Here we discuss two of them, those ones developed in \cite{QEDNM}. To start, we evaluate the trace in the four-dimensional space-time and obtain
\bea
\label{s2int}
S_2(p)&=&-2g^2\epsilon^{\mu\nu\rho\sigma}
\epsilon^{\mu'\nu'\rho'\sigma'}b_{\mu}F_{\nu\rho}b_{\mu'}
F_{\nu'\rho'}
\int\frac{d^4k}{(2\pi)^4}\frac{1}{[k^2-m^2]^2}
\big[m^2\eta_{\sigma\sigma'}+\nonumber\\&+&k^{\alpha}k^{\beta}(\eta_{\alpha\sigma}\eta_{\beta\sigma'}-\eta_{\alpha\beta}\eta_{\sigma\sigma'}
+\eta_{\alpha\sigma'}\eta_{\beta\sigma})\big].
\eea
Actually, the integral in this expression has just the structure given by (\ref{ambaether}), up to the overall factor. Now, let us perform its calculation.

Within the first manner, we replace $k^{\alpha}k^{\beta}\to\frac{1}{4}\eta^{\alpha\beta}k^2$, as it should be done in the four-dimensional space-time, and only after that, we introduce the dimensional regularization, promoting the integral to $d=4-\epsilon$ dimensions. 
Carrying out Wick rotation, we arrive at the Euclidean result
\bea
S_2(p)=-ig^2\epsilon^{\mu\nu\rho\sigma}\epsilon^{\mu'\nu'\rho'\sigma'}b_{\mu}F_{\nu\rho}b_{\mu'}
F_{\nu'\rho'}\eta_{\sigma\sigma'}\int\frac{d^{4-\epsilon}k_E}{(2\pi)^{4-\epsilon}}
\frac{k^2_E+2m^2}{[k^2_E+m^2]^2}.
\eea
In this case we have the mutual cancellation of divergences, reproducing the scenario discussed in the Section 3.1, so, our integral turns out to be finite:
\bea
& &\int\frac{d^{4-\epsilon}k_E}{(2\pi)^{4-\epsilon}}
\frac{k^2_E+2m^2}{[k^2_E+m^2]^2}=\frac{m^{2(1-\epsilon/2)}}{(4\pi)^{2-\epsilon/2}}
\left[\Gamma(\frac{\epsilon}{2})+\Gamma(-1+\frac{\epsilon}{2})\right]=\nonumber\\ &=&
-\frac{m^2}{16\pi^2}+O(\epsilon).
\eea
Collecting all together, multiplying two
Levi-Civita symbols, disregarding the Lorentz-invariant term proportional to $b^2$, in order to keep track of the aether term only, and returning to the Minkowski space, we arrive at
\bea
\label{s2d4}
S_2(p)&=g^2\frac{m^2}{4\pi^2}(b^{\mu}F_{\mu\nu})^2.
\eea
Within another manner, we first promote our integral to $d=4-\epsilon$ dimensions, then we replace $k^{\alpha}k^{\beta}\to\frac{1}{d}\eta^{\alpha\beta}k^2$. In this case, in the Euclidean space we have
\bea
S_2(p)=-ig^2\epsilon^{\mu\nu\rho\sigma}\epsilon^{\mu'\nu'\rho'\sigma'}b_{\mu}F_{\nu\rho}b_{\mu'}
F_{\nu'\rho'}\eta_{\sigma\sigma'}\int\frac{d^dk_E}{(2\pi)^d}
\frac{2(\frac{d-2}{d})k^2_E+2m^2}{[k^2_E+m^2]^2}.
\eea
If  one substitutes $d=4-\epsilon$, with $\epsilon\neq 0$ in this expression, the integral identically vanishes. So, the purely non-minimal contribution displays a removable singularity at $\epsilon=0$. In principle, other results can be obtained for the expression (\ref{aethercontr}), see the discussion in \cite{aetherT}, where also the finite temperature behavior of the generated aether term is considered. So, for these two examples one can write this result as
\bea
\label{s2d4ff}
S_{mixed}&=C_0g^2m^2(b^{\mu}F_{\mu\nu})^2,
\eea
with $C_0$ is a finite ambiguous constant. 

We note that in other space-time dimensions one can introduce the non-minimal vertex as well, it will look like $S_{magn}=g\bar{\psi}\epsilon_{\nu\rho\sigma}b^{\nu}F^{\rho\sigma}\psi$ in $3D$ and $S_{magn}=g\bar{\psi}\epsilon_{\lambda\mu\nu\rho\sigma}\sigma^{\lambda\mu}b^{\nu}F^{\rho\sigma}\psi$ in $5D$. In both cases, the aether term will be finite within the dimensional regularization framework \cite{aether}. 

Second, we study the contributions involving one non-minimal vertex and one minimal one, and one insertion $\bs\gamma_5$ in one of the propagators \cite{QEDNM}. In this case one can consider just the same Feynman diagrams given by Fig. 3.4, used in the Section 3.4.1 for the calculation of the CFJ term, with the only difference that one of the external $A_{\mu}$ fields in this case is replaced by     $\epsilon_{\mu\nu\rho\sigma}b^{\nu}F^{\rho\sigma}$, and the coupling $e$ accompanying just this field -- by $g$. The structure of the integral over the internal momenta continues to be the same as in the case of the CFJ term, that is, (\ref{TM0509008}). As a result, we repeat all calculations from \cite{TM0509008} and arrive at
\bea
\label{s2d4af}
S_{mixed}&=-2Ceg(b^{\mu}F_{\mu\nu})^2,
\eea
with the finite constant $C$ is just that one characterizing the value of the CFJ term, it is defined in (\ref{cfjfinal}), and its value is discussed in details in the Sec. 3.4.1. We note that this contribution arises only in the four-dimensional space-time, where the vector $\epsilon_{\mu\nu\rho\sigma}b^{\nu}F^{\rho\sigma}$ can be defined, and has no analogues in other space-time dimensions. 

Third, there is a purely minimal aether-like contribution. In this case we have two $\bs\gamma_5$ insertions (some preliminary studies of this contribution were also carried out in \cite{Bonneau} within the renormalization group context), and  the explicit form of this contribution is
\begin{eqnarray}
\label{1}
S_{AA}(p)&=&\frac{ie^2}{2}\int\frac{d^4l}{(2\pi)^4}(\gamma^{\mu}\frac{1}{\ls-m}\gamma^{\nu}\frac{1}{\ls+\ps-m}
\gamma_5\bs\frac{1}{\ls+\ps-m}\gamma_5\bs\frac{1}{\ls+\ps-m}+\nonumber\\
&+&
\gamma^{\mu}\frac{1}{\ls-m}\gamma_5\bs\frac{1}{\ls-m}
\gamma^{\nu}\frac{1}{\ls+\ps-m}\gamma_5\bs\frac{1}{\ls+\ps-m}+\nonumber\\
&+&
\gamma^{\mu}\frac{1}{\ls-m}\gamma_5\bs\frac{1}{\ls-m}
\gamma_5\bs\frac{1}{\ls-m}\gamma^{\nu}\frac{1}{\ls+\ps-m})
A_{\mu}(-p)A_{\nu}(p),
\end{eqnarray}
where $p$ is an external momentum. 

This contribution is a sum of Feynman diagrams given by Fig. 3.6.

\begin{figure}[!ht]
\begin{center}
\includegraphics[angle=0,scale=1.0]{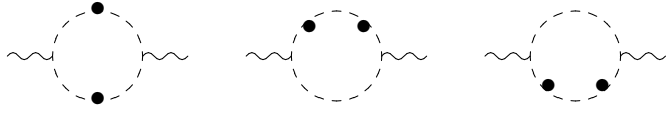}
\end{center}
\caption{Contributions to the aether term for the electromagnetic field with minimal vertices.}
\end{figure}

Taking into account only the terms of the second order in $p$, we get for the sum of these graphs:
\begin{eqnarray}
\label{sd24aa}
S_{AA}=-\frac{e^2}{6m^2\pi^2}b_{\mu}F^{\mu\nu}b^{\lambda}F_{\lambda\nu}.
\end{eqnarray}
We note that this expression is superficially finite and hence ambiguity-free. 

So, the complete aether-like contribution is a sum of (\ref{s2d4ff},\ref{s2d4af},\ref{sd24aa}), and it is finite and involves two ambiguous constants $C_0$ and $C$. While the constant $C$ is related with the CFJ anomaly, the question of relation of $C_0$ with any possible anomaly is still open. 

This calculation can be straightforwardly generalized for a non-Abelian case, see \cite{aethernab}. For the presence of a minimal coupling only, the result for the aether term is obtained through expansion of the functional determinant (\ref{fundet1}) up to relevant orders, and the result is a straightforward generalization of (\ref{sd24aa}):
	\begin{eqnarray}
		\label{sd24aa1}
		S_{AA}=-\frac{e^2}{6m^2\pi^2}b_{\mu}F^{\mu\nu a}b^{\lambda}F^a_{\lambda\nu},
	\end{eqnarray}
	where $F^a_{\lambda\nu}=\partial_{\mu}A_{\nu}^a-\partial_{\nu}A_{\mu}^a-igf^{abc}A^b_{\mu}A^c_{\nu}$ is the non-Abelian field strength.
It is clear that including of the non-Abelian analogue of magnetic coupling will yield non-Abelian analogues of (\ref{s2d4ff}) and (\ref{s2d4af}).

Besides of the study performed in \cite{aether,QEDNM}, a very interesting discussion of ambiguity of the aether term for the gauge field is presented in \cite{MCN}. In this paper, it was argued that the same ambiguous integral over momenta as in (\ref{s2int}), that is,
\bea
I_{\sigma\sigma'}&=&\int\frac{d^4k}{(2\pi)^4}\frac{1}{[k^2-m^2]^2}\times\nonumber\\ &\times&
\big[m^2\eta_{\sigma\sigma'}+k^{\alpha}k^{\beta}(\eta_{\alpha\sigma}\eta_{\beta\sigma'}-
\eta_{\alpha\beta}\eta_{\sigma\sigma'}
+\eta_{\alpha\sigma'}\eta_{\beta\sigma})\big].
\eea
will accompany not only the aether term, but also the CFJ term and the Proca term, so that the whole contribution will look like
\bea
\Gamma_2=-2(eA^{\sigma}+g\epsilon^{\sigma\lambda\mu\nu}b_{\lambda}F_{\mu\nu})I_{\sigma\sigma'}(eA^{\sigma'}+g\epsilon^{\sigma'\lambda'\mu'\nu'}b_{\lambda'}F_{\mu'\nu'}),
\eea
and it is natural to choose the value of the $I_{\sigma\sigma^{\prime}}$ to be zero in order to rule out the $A^{\mu}A_{\mu}$ term, thus avoiding the breaking of the gauge symmetry. Nevertheless, since this integral is ambiguous, nobody forbids to choose its values to be different for all three contributions, that is, to be zero when it accompanies the Proca term, and non-zero when it accompanies the gauge invariant aether and CFJ terms.

To close the discussion of generating the aether term from the extended QED Lagrangian given by (\ref{qednm}), we note its advantage consists in the fact that only in this theory one can arrive at finite CPT-even contributions. All another couplings will yield divergent aether-like results in the four-dimensional space-time. There are two most important examples of obtaining these results. In the paper \cite{Maluf1}, the starting point is the following extension for the spinor sector of QED:
\bea
S=\int d^4x\bar{\psi}(i\ds-e\As+\frac{\lambda}{2}\kappa^{\mu\nu\rho\sigma}\sigma_{\mu\nu}F_{\rho\sigma}-m)\psi.
\eea
We see that the LV coupling is CPT-even. The corresponding one-loop effective action of the gauge field, that is, the fermionic determinant, is given by the expression:
\bea
\Gamma^{(1)}[A]=-i{\rm Tr}\ln(i\ds-e\As+\frac{\lambda}{2}\kappa^{\mu\nu\rho\sigma}\sigma_{\mu\nu}F_{\rho\sigma}-m).
\eea
In this case, it is easy to calculate the contributions of first and second orders in $\kappa_{\mu\nu\rho\sigma}$ to the two-point function of the gauge field. They diverge, explicitly looking like \cite{Maluf1}:
\bea
I_1&=&\frac{me\lambda}{8\pi^2\epsilon}\kappa_{\mu\nu\alpha\beta}F^{\mu\nu}F^{\alpha\beta};\nonumber\\
I_2&=&-\frac{\lambda^2}{16\pi^2\epsilon}\kappa_{\mu\nu\rho\sigma}\kappa^{\rho\sigma}_{\phantom{\rho\sigma}\alpha\beta}F^{\mu\nu}F^{\alpha\beta}.
\eea
It is clear that if the constant LV tensor $\kappa_{\mu\nu\alpha\beta}$ is characterized by only one vector, looking like (\ref{kappavect}), both these results reproduce the aether term (\ref{aether}). We note that, first, to subtract the arising divergence in a consistent manner, one should introduce the aether term in the tree-level action from the very beginning, second, for the dimensionless $\kappa$, the coupling $\lambda$ has a negative mass dimension, so, this theory is actually non-renormalizable and can be treated only as a kind of some low-energy effective model.

Another interesting LV extension of the QED is characterized by the following action of fermions:
\bea
S=\int d^4x\bar{\psi}\left[i(\ds+ie\As)+ic^{\mu\nu}\gamma_{\mu}(\pa_{\nu}+ieA_{\nu})-m\right]\psi.
\eea
Here, the $c^{\mu\nu}$ is a constant tensor which is usually assumed to be symmetric (actually, if $c^{\mu\nu}$ is antisymmetric, it can be ruled out through an appropriate field redefinition \cite{ColMac2}). Since it is dimensionless, the corresponding theory is all-loop renormalizable.
The corresponding fermionic determinant is
\bea
\Gamma^{(1)}[A]=-i{\rm Tr}\ln[i(\ds+ie\As)+ic^{\mu\nu}\gamma_{\mu}(\pa_{\nu}+ieA_{\nu})-m].
\eea
Expanding it up to the second order in $A_{\mu}$, for the simplest choice $c_{\mu\nu}=u_{\mu}u_{\nu}$ we arrive at the following result \cite{Maluf2}:
\begin{equation}
S_{eff}^{(2)} = -\frac{e^2}{6 \pi ^2 \epsilon}\int d^4x\, \frac{1}{4}\left(1+u^2\right)^{-1}\tilde F_{\mu\nu} \tilde F^{\mu\nu},
\end{equation}
where
\begin{equation}
\tilde F_{\mu\nu} = (g_{\mu\alpha}+u_\mu u_\alpha)(g_{\nu\beta}+u_\nu u_\beta) F^{\alpha\beta},
\end{equation}
so, we arrive at the divergent aether-like contribution, and again the aether term must be present in the tree-level action from the very beginning for the sake of the multiplicative renormalizability of the theory. 
The analogues of this result for other choices of $c_{\mu\nu}$ can be obtained as well \cite{Maluf2}. In the five-dimensional case, the corresponding result is finite within the dimensional regularization.

\section{Higher-derivative LV terms}

The next step of our study consists in generating the higher-derivative terms, attracting great attention due to highly nontrivial dispersion relations allowing for birefringence of electromagnetic waves, rotation of polarization plane in a vacuum (see section 2.2) and many other interesting effects, a general review on wave propagation in higher-derivative LV extensions of QED can be found in \cite{KosMewED} for the gauge sector, and in \cite{KosMewSpin} for the spinor sector, more results on higher-derivative LV extensions of QED and their relation to massive photons \index{massive photons} are discussed in \cite{Lehn2012}. As we already mentioned, the first known example of such terms is the Myers-Pospelov (MP) term (\ref{MP}) which is very convenient since it allows to conciliate higher derivatives with unitarity, for certain choices of the LV vector. Another important example is the higher-derivative CFJ-like term
\bea
\label{hdcfj}
{\cal L}^{hd}_{CFJ}=\frac{\alpha}{M^2}\epsilon^{\mu\nu\rho\sigma}b_{\mu}A_{\nu}\Box\pa_{\nu}A_{\rho}.
\eea
Here $\alpha$ is some dimensionless constant, and $M$ is a mass scale. Both the MP term and the higher-derivative CFJ-like term can be generated as perturbative corrections within the same LV extended QED whose action is given by (\ref{qednm}). We start again with the fermionic determinant (\ref{fundet11}). Expanding it in power series in $b_{\mu}$ we find that the three-derivative term is described by three contributions, that is, those ones with two non-minimal vertices, with one minimal vertex and one non-minimal one, and, finally, with two minimal ones \cite{MNP}.

For the first contribution, we have only one insertion in one of propagators and therefore consider again the Feynman diagrams given by the Fig. 3.4, so, we reproduce identically the situation described in the Section 3.4.1, with the only difference is that in our case we should replace both external $eA_{\mu}$  legs by external $g\epsilon_{\mu\nu\rho\sigma}b^{\nu}F^{\rho\sigma}$ legs, while the integral over momenta is given again by the (\ref{self}). After extracting only the first order in external momenta which is the only relevant one for this fragment, we arrive at the same integral over momenta as in (\ref{TM0509008}), whose result, in a pure analogy with the Section 3.4.1, is
\bea
\Gamma^{(1)}_1=2g^2C\epsilon^{\nu\lambda\mu\alpha}b_{\alpha}\tilde{A}_{\nu}\pa_{\lambda}\tilde{A}_{\mu},
\eea
where $\tilde{A}_{\mu}=\epsilon_{\mu\nu\rho\sigma}b^{\nu}F^{\rho\sigma}$, and $C$ is the same finite ambiguous constant arising within calculations of the CFJ term (see Section 3.4) and defined by (\ref{cfjfinal}). Carrying out contractions of the Levi-Civita symbols, we arrive at
\bea
\label{term1}
\Gamma^{(1)}_1=2g^2C \left[b^\alpha F_{\alpha\mu}(b\cdot\partial)b_\beta\epsilon^{\beta\mu\nu\lambda}F_{\nu\lambda}+b^2b_\beta\epsilon^{\beta\mu\nu\lambda}A_\mu\Box F_{\nu\lambda}\right].
\eea
We see that this contribution is, first, finite and ambiguous (which reflects the fact that the Adler-Bell-Jackiw (ABJ) anomaly can be extended to involve higher-derivative terms \cite{Scarpanom}), second, involves both the MP term (the first one in this expression) and the higher-derivative CFJ-like term (the second one in this expression). We note that here the last term is of third order in the vector $b_{\mu}$ while in principle one can obtain the analogous term without the constant $b^2$ multiplier, that is, the one given by (\ref{hdcfj}), using the one-loop effective action of the minimal LV QED (\ref{fundet}) as a starting point \cite{MWS}. In that case the higher-derivative CFJ-like contribution is finite and ambiguity-free.

For the second contribution, we consider Feynman diagrams with two insertions in propagators. The corresponding contributions are again depicted at Fig. 3.6, and in this case we repeat the calculation from the previous section and replace only one $eA_{\mu}$  leg by the external $g\epsilon_{\mu\nu\rho\sigma}b^{\nu}F^{\rho\sigma}$ leg in the expressions (\ref{1}) and (\ref{sd24aa}). As a consequence, we arrive at the expression analogous to (\ref{sd24aa}), where this replacement of one external leg is performed.  Hence, the  corresponding result can be written as 
\bea
\label{term2}
S_{b2}=\frac{eg}{6\pi^2m^2}\int d^4x\, \left[b^\alpha F_{\alpha\mu}(b\cdot\partial)b_\beta\epsilon^{\beta\mu\nu\lambda}F_{\nu\lambda}+b^2b_\beta\epsilon^{\beta\mu\nu\lambda}A_\mu\Box F_{\nu\lambda}\right].
\eea
Again, we obtain both the MP and the higher-derivative CFJ-like terms. We note that this result is non-ambiguous.

For the third contribution, we have three insertions in the propagators. 
The corresponding Feynman diagrams are depicted at Fig. 3.7.

\begin{figure}[!ht]
\begin{center}
\includegraphics[angle=0,scale=1.0]{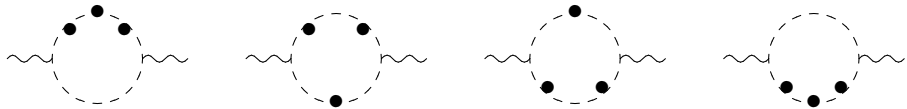}
\end{center}
\caption{Higher-derivative contributions with three insertions.}
\end{figure}

The corresponding result looks like
\bea
\label{term3}
S_{b3}=\frac{4e^2}{45\pi^2m^4} \left[b^\alpha F_{\alpha\mu}(b\cdot\partial)b_\beta\epsilon^{\beta\mu\nu\lambda}F_{\nu\lambda}+\frac54b^2b_\beta\epsilon^{\beta\mu\nu\lambda}A_\mu\Box F_{\nu\lambda}\right].
\eea
The sum of these three contributions (\ref{term1},\ref{term2},\ref{term3}) gives our final result for the higher-derivative contribution to the effective Lagrangian. It reads as
\bea
\label{term}
S_{hd}&=&\left(2g^2C+\frac{eg}{6\pi^2m^2}+\frac{4e^2}{45\pi^2m^4}\right) b^\alpha F_{\alpha\mu}(b\cdot\partial)b_\beta\epsilon^{\beta\mu\nu\lambda}F_{\nu\lambda}\nonumber\\
&+&\left(2g^2C+\frac{eg}{6\pi^2m^2}+\frac{e^2}{9\pi^2m^4}\right)\,b^2b_\beta\epsilon^{\beta\mu\nu\lambda}A_\mu\Box F_{\nu\lambda}.
\eea
We see that this result is finite and gauge invariant. Also, one can observe that for the light-like $b_{\mu}$, the CFJ-like term vanishes, and only the MP term survives. The finite temperature behavior of this term is studied in \cite{Celeste}.

There are other possibilities to generate the higher-derivative LV terms. First of all, it is worth to mention the study based on using the third-rank tensor $g^{\mu\nu\rho}$ (it is antisymmetric with respect to two first indices), so that the following extension of the QED is introduced \cite{TMHD}:
\bea
S=\int d^4x\bar{\psi}[i(\gamma^{\mu}+\frac{1}{2}g^{\kappa\lambda\mu}\sigma_{\kappa\lambda})(\pa_{\mu}+ieA_{\mu})-m]\psi.
\eea
We note that this theory is renormalizable since $g^{\kappa\lambda\mu}$ is dimensionless.
In this case, the first-order LV contribution is given by the sum of Feynman diagrams given by Fig. 3.8, where the dark ball is for $g^{\mu\nu\lambda}$ insertion.

\begin{figure}[!ht]
\begin{center}
\includegraphics[angle=0,scale=1.0]{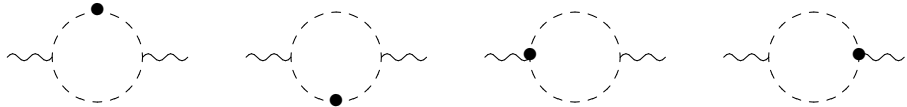}
\end{center}
\caption{Higher-derivative contributions with $g^{\mu\nu\lambda}$ insertion.}
\end{figure}

The propagator in this theory is usual, so, one can straightforwardly sum the contribution of these diagrams. As a result, we arrive at the following higher-derivative contribution \cite{TMHD}:
\bea
\label{tmhd}
\Gamma_{g}=\frac{e^2}{12m\pi^2}A_{\mu}(g^{\mu\nu\alpha}\Box\pa_{\alpha}-g^{\mu\rho\alpha}\pa_{\rho}\pa_{\alpha}\pa^{\nu}-g^{\rho\nu\alpha}\pa_{\rho}\pa_{\alpha}\pa^{\mu})A_{\nu}.
\eea
We note that {\it a priori} the only restriction for the $g^{\mu\nu\lambda}$ is its antisymmetry with respect to the first two indices. Therefore we can present the tensor $g^{\mu\nu\alpha}$ as a sum of irreducible (totally antisymmetric and partially symmetric) parts as $g^{\mu\nu\alpha}=\epsilon^{\mu\nu\alpha\beta}g_{\beta}+\bar{g}^{\mu\nu\alpha}$, where $g_{\gamma}=-\frac{1}{6}g^{\mu\nu\alpha}\epsilon_{\mu\nu\alpha\gamma}$. In this case the first term in (\ref{tmhd}) will yield the higher-derivative CFJ-like result (\ref{hdcfj}); evidently, if $g^{\mu\nu\lambda}$ is totally antisymmetric, only the first term of (\ref{tmhd}) is non-trivial. It was claimed in \cite{TMHD} that other components of (\ref{tmhd}) can describe some MP-like corrections.

Also, we have noted above that the higher-derivative CFJ-like term (\ref{hdcfj}) can arise already in the first order in $b_{\mu}$. The corresponding prescription has been discussed in \cite{MWS}. Our starting point is the usual fermionic determinant (\ref{fundet}), which, as we already noted, corresponds to the simplest LV extension of the QED. Then we apply the derivative expansion approach just as in the section 3.4, but in this case we expand our fermionic determinant up to the third derivative of the background field $A_{\mu}$ keeping at the same time only the first order in $b_{\mu}$. Effectively it means that we should expand the self-energy tensor given by (\ref{self}) up to the third order in the external $p_{\mu}$. The result will be finite and ambiguity-free. Its explicit form is given by (\ref{hdcfj}), with $\frac{\alpha}{M^2}=\frac{e^2}{24m^2\pi^2}$. In principle, since the constant vector $b_{\mu}$ is small, this term will dominate in comparison with the CFJ-like part of (\ref{term}) which we calculated above. The finite temperature behavior of this term is also described in \cite{MWS}. 

\section{Divergent contributions in gauge and spinor sectors of LV QED}

Up to now, we, being motivated by the concept of emergent dynamics, paid our main attention to finite contributions to the one-loop effective action. The key idea of our approach consists in suggesting that various LV terms in the gauge sector are generated as a consequence of integrating over a spinor field in some fundamental LV extension of QED, which can be done consistently only if the results are finite, otherwise one must introduce these terms from the very beginning, jeopardizing the concept. At the same time, possible divergent contributions to the effective action for different LV extensions of QED also must be considered, in order to obtain a more complete perturbative description of a corresponding theory. We note that study of divergent corrections is especially important for non-minimal LV theories because of their non-renormalizability.
	 
	The first step in study of divergent LV contributions has been performed already in \cite{Kost2001a}, where the one-loop renormalization of the minimal LV extension of QED (\ref{genrenmod}) in the first order in LV parameters was carried out. Further discussions of these first-order results, including the methodology allowing to adapt the known Lehmann-Symanzik-Zimmermann (LSZ) formalism \index{Lehmann-Symanzik-Zimmermann (LSZ) formalism}
	to LV theories, are presented in \cite{Lehn2014}. Clearly, the natural continuations of this study are, first, calculating divergent corrections of second and higher orders in LV constant vectors (tensors), second, obtaining quantum corrections in the spinor sector of corresponding theories, third, investigation of higher loop LV corrections. Let us briefly review the main results achieved along these directions.

The divergent corrections in effective action, involving second and higher orders in LV parameters, were studied in a number of papers. In \cite{aether}, the aether-like contribution to the effective action in the spinor sector of the non-minimal LV QED with the action
$$
S=\int d^4x\bar{\psi}(i\ds-m-e\As-g\epsilon_{\mu\nu\rho\sigma}\gamma^{\mu}b^{\nu}F^{\rho\sigma})\psi,
$$
similar to (\ref{qednm}), but without $\bar{\psi}\bs\gamma_5\psi$ term, was found to look like
\bea
\label{spinaet}
\Gamma_{sp}^{(1)}\propto \frac{ig^2m^2}{\pi^2\epsilon}\bar{\psi}\bs(b\cdot\partial)\psi,
\eea
which matches the form of the CPT-even aether term for the spinor field $iu^{\mu}u^{\nu}\bar{\psi}\gamma_{\mu}\partial_{\nu}\psi$ proposed in \cite{Carroll}. It worth mentioning that in three- and five-dimensional space-times the analogous aether-like corrections arise as well, being finite in these cases within the framework of the dimensional regularization. It is clear that the aether term for the spinor field will arise as well in the case of presence of the $\bar{\psi}\bs\gamma_5\psi$ additive term. The divergent aether-like results were also shown to arise in the spinor sector in \cite{Scarp3} where the contributions of the second order in LV parameters of the minimal LV extension of QED \cite{Kost2001a}, given by (\ref{genrenmod}), were obtained.  Also, divergent contributions in the spinor sector, up to the first order in derivatives, and proportional to the first order of non-minimal LV parameters listed in \cite{KosLi}, were obtained in the paper \cite{ourrev} and in some other papers.

In the gauge sector, besides of divergent results for the CPT-even LV term discussed in the section 3.5, the most interesting divergent contributions of the second order in LV parameters were obtained in the paper \cite{2order}, where the minimal LV extension of QED \cite{Kost2001a}, given by (\ref{genrenmod}), was studied, and the one-loop divergent corrections were shown to have the aether-like form (\ref{leven}). It is worth to mention also higher-derivative divergent LV contributions obtained in \cite{Maluf1,AntAly} for the CPT-even case (of fourth order in derivatives), and in \cite{MarReyes}, for the CPT-odd case (of third order in derivatives).

As for higher-loop calculations, it should be noted that, up to now, there are a few higher-loop results for LV theories. The examples are, first, two-loop renormalization in the LV scalar field theory involving the aether-like LV term (\ref{scalaether}), performed in \cite{Altscal}, second, discussion of higher-loop contributions in the simplest LV extension of QED (\ref{lvqed0}) presented in \cite{nohighloop}, where it was argued that no CFJ term can be generated in higher loops. 

There are other interesting discussions of perturbative aspects of LV theories and related papers, we discussed only part of results obtained up to now. Nevertheless, still there are many perturbative calculations to be performed, especially, those ones treating impacts of various non-minimal vertices listed in \cite{KosLi}.

\section{Discussion and conclusions}

We discussed the possibilities of perturbative generation of different LV terms. Explicitly, we saw that these terms arise within the context of the emergent dynamics in a theory involving coupling of scalar, gauge and, as it will be discussed in the Chapter 6, gravitational fields to some primordial spinor field which afterwards is being integrated out. Here we were mostly interested in the calculation schemes allowing to obtain essentially finite results, to avoid problems with the renormalization. Indeed, in an opposite case the corresponding terms in gauge/scalar sector should be introduced already at the tree level, to get the multiplicatively renormalizable theory, hence these terms cannot be more considered as an emergent phenomenon.

We showed that it is possible to obtain both the new LV terms defined in many space-time dimensions, as e.g. the aether terms for gauge and scalar fields, and the new terms well defined only in specific dimensions, such as the CFJ-like term for scalar fields defined only in two dimensions, the mixed scalar-vector term defined only in three dimensions, and the CFJ term defined only in four dimensions. We demonstrated that there are manners to generate various terms suggested in \cite{KosLi}, which apparently shows that the perturbative mechanism is a consistent way to get new LV terms. It is interesting to note that in certain cases the LV couplings nevertheless can yield non-zero Lorentz-invariant quantum corrections.  Besides of simple arising the terms proportional to $b_{\mu}b^{\mu}$, as occurs e.g. within studies of the aether terms for scalar and gauge fields \cite{aether}, a very interesting  example is presented in \cite{axion}, where the Lorentz-invariant axion term has been generated in the LV Abelian gauge theory from the triangle Feynman diagram similar to that one used in this chapter to obtain the CFJ term.

An interesting feature of some of our results is their ambiguity. Actually, we showed that at the one-loop order there are two typical ambiguities, the first of them, being a paradigmatic example, arises within the calculation of the CFJ term, and of the terms which can be obtained from the CFJ term through replacing external legs, while the second one arises in the purely non-minimal sector of the LV extended QED. We indicated that, in principle, other ambiguous integrals can exist, and the ambiguity persists in the finite temperature case as well. It is interesting to note that while the first ambiguity is closely related with the Adler-Bell-Jackiw (ABJ) anomaly (for details, see \cite{JackAmb}), the possible anomaly related to the second ambiguity is not known yet, and, moreover, the problem of its existence is still open, see also the discussion in \cite{Scarpanom}.

Besides of finite terms, which are of special interest, providing a possible mechanism for generating LV terms in the gauge sector, there are many divergent LV corrections which arise in various LV extensions of QED. We reviewed briefly some results related to obtaining these terms presented in a number of papers. Still, many studies of divergent LV terms, as well as of finite ones, are to be done. It is natural to expect that one of main lines in these studies will consist in explicit calculations of perturbative impacts of various LV terms listed in the review \cite{KosLi}. In this context, we can mention the paper \cite{d5arb} where perturbative generation of the CFJ term in the extended spinor QED involving all dimension-5 LV operators was studied, with the result was shown to be finite and ambiguous.

We close the discussion claiming that the perturbative generation shows itself to be an appropriate mechanism for obtaining at least some of the LV extensions of known theories, especially, of the spinor electrodynamics. In the Chapter 6, we demonstrate that this methodology can be applied in the case of gravity as well.

\chapter[Lorentz symmetry breaking and noncommutativity]{Lorentz symmetry breaking and space-time noncommutativity}

As it was claimed in \cite{KostCarr}, one of important motivations for the Lorentz symmetry breaking stems from the space-time noncommutativity. Indeed, the initial statement for the noncommutative field theory is the suggestion that the commutator of two space-time coordinates $\hat{x}^{\mu},\hat{x}^{\nu}$ is {the constant antisymmetric matrix $\Theta^{\mu\nu}$ (see \cite{SW}):
\bea
[\hat{x}^{\mu},\hat{x}^{\nu}]=i\Theta^{\mu\nu}.
\eea
This matrix is clearly not Lorentz-invariant, so, this form of the space-time noncommutativity breaks the Lorentz symmetry. At the same time, some of various additive LV terms proposed in \cite{KosGra,KosLi} are based on second-rank constant tensors, which can be, in many cases, antisymmetric as well. Hence, an action of a classical noncommutative field theory expanded up to a first order in $\Theta^{\mu\nu}$, can be naturally treated as a particular case of certain models discussed in \cite{KosGra}. Following \cite{SW}, this relation emerges from the low-energy limit of the superstring action in the presence of the constant antisymmetric tensor field $B_{\mu\nu}$, so that  the $\Theta^{\mu\nu}$ is related with $B_{\mu\nu}$ through the algebraic relation:  
\bea
\Theta^{\mu\nu}=2\pi\alpha^{\prime}\left(\frac{1}{g+2\pi\alpha^{\prime}B}\right)^{\mu\nu}_A,
\eea
where the subscript $A$ is for the antisymmetric part, and $\alpha^{\prime}$ is a constant describing a coupling of the string to the $B_{\mu\nu}$. Here, the space-time metric $g_{\mu\nu}$ is assumed to be the Minkowski one.
We note that  this manner of introducing the Lorentz symmetry breaking within the string context essentially differs from that one used in \cite{KosSam}, based on the spontaneous Lorentz symmetry breaking.  Therefore, the problem of systematic description of noncommutative field theories is based on the methodology distinct from that one described in previous chapters, but, nevertheless, there are some strong analogies between noncommutative field theories and "usual" LV theories, which will be  discussed in this chapter.

\section{Formulations for noncommutativity}

As it was argued in \cite{SW}, a systematic study of field theories defined in a noncommutative space-time can be done through an appropriate mapping of these theories to the usual space-time. 
There are two known manners to map theories formulated in a noncommutative space-time to theories formulated in the usual space-time, that is, the Seiberg-Witten (SW) map \cite{SW} and the Moyal product \cite{SW,Filk}.

The Moyal product \index{Moyal product} is a universal formulation for noncommutative field theories allowing to define noncommutative extensions for any field theory except of gravity, since in that case one should generalize this product to the curved space-time as well, and the consistent manner of such a generalization is still unknown. To introduce the Moyal product, one assumes \cite{SW}, that any field $\phi(\hat{x})$ in a noncommutative $D$-dimensional space-time can be expressed in the form of the Fourier integral:
\bea
\phi(\hat{x})=\int\frac{d^Dk}{(2\pi)^D}\tilde{\phi}(k)e^{ik\hat{x}},
\eea
with $k$ are usual momenta, and $\tilde{\phi}(k)$ is a Fourier transform of this field,  so, using the Hausdorff formula, for the product of two noncommutative fields one has
\bea
&&\phi_1(\hat{x})\phi_2(\hat{x})=\int\frac{d^Dk_1d^Dk_2}{(2\pi)^{2D}}\phi_1(k_1)\phi_2(k_2)e^{ik_1\hat{x}}e^{ik_2\hat{x}}=\nonumber\\
&=&\int\frac{d^Dk_1d^Dk_2}{(2\pi)^{2D}}\phi_1(k_1)\phi_2(k_2)e^{i(k_1+k_2)\hat{x}}e^{-\frac{i}{2}\Theta^{\mu\nu}k_{1\mu}k_{2\nu}}.
\eea
It is clear that this product is associative and can be straightforwardly generalized to an arbitrary number of fields, with the only difference with the usual commutative theories consists in the presence of an additional factor like $e^{-i\Theta^{\mu\nu}k_{1\mu}k_{2\nu}}$. This allows us to introduce a following rule: the product of fields on the noncommutative space-time must be mapped into their Moyal product on the usual space-time defined as
\bea
&&\phi_1(x)*\phi_2(x)*\ldots*\phi_n(x)=\int\frac{d^Dk_1d^Dk_2\ldots d^Dk_n}{(2\pi)^{nD}}\phi_1(k_1)\times\nonumber\\&\times&\phi_2(k_2)\ldots \phi_n(k_n) e^{i(k_1+k_2+\ldots+k_n)x}\exp[-\frac{i}{2}\Theta^{\mu\nu}\sum\limits_{i<j\leq n}k_{i\mu}k_{j\nu}].
\eea
We see that the impact of the space-time noncommutativity is concentrated now in the $\Theta$-dependent phase factor. This formula evidently can be applied to a product of any fields.

Another manner to treat the noncommutativity is based on the SW map which is an especially powerful tool for the gauge theories. In this case we suggest that the noncommutative gauge theory is described in terms of the noncommutative gauge field $\hat{A}_{\mu}$ which can be represented as a power series in $\Theta^{\mu\nu}$. Let our starting point be a noncommutative (NC) Yang-Mills theory whose action is
\bea
\label{NCYM}
S=-\frac{1}{4}{\rm tr}\int d^Dx F_{\mu\nu}*F^{\mu\nu},
\eea
with $F_{\mu\nu}=\pa_{\mu}\hat{A}_{\nu}-\pa_{\nu}\hat{A}_{\mu}-ig[\hat{A}_{\mu},\hat{A}_{\nu}]_*$, so, the commutator is both algebraic and Moyal one. We see that it is nontrivial even for the $U(1)$ case. The $\hat{A}_{\mu}$ is a noncommutative gauge field.

Then we, following \cite{SW}, suggest that the NC gauge field is projected into a usual gauge field in such a way that if the noncommutative field $\hat{A}_{\mu}$ is mapped in some manner to the usual commutative field $A_{\mu}$, the infinitesimal gauge transformation of $\hat{A}_{\mu}$ is related through this mapping to the infinitesimal gauge transformation of $A_{\mu}$, so that
\bea
\hat{A}(A)+\delta_{\hat{\lambda}}\hat{A}(A)=\hat{A}(A+\delta_{\lambda}A),
\eea
with the parameter $\hat{\lambda}$ of the noncommutative gauge transformation is linked with the usual parameter of the gauge transformation $\lambda$ in some way. It was shown in \cite{SW} that the explicit expressions for $\hat{A}$, $\hat{\lambda}$ and $\hat{F}$ can be obtained as series in $\Theta$ order by order, with the lower terms of these series look like:
\bea
\hat{A}_{\mu}&=&A_{\mu}-\frac{1}{4}\Theta^{\nu\lambda}\{A_{\nu},\pa_{\lambda}A_{\mu}+F_{\lambda\mu}\}+{\cal O}(\Theta^2);\nonumber\\
\hat{\lambda}&=&\lambda+\frac{1}{4}\Theta^{\nu\lambda}\{\pa_{\nu}\lambda,A_{\lambda}\}+{\cal O}(\Theta^2);\\
\hat{F}_{\mu\nu}&=&F_{\mu\nu}+\frac{1}{4}\Theta^{\kappa\rho}(2\{F_{\mu\kappa},F_{\nu\rho}\}
-\{A_{\kappa},D_{\rho}F_{\mu\nu}+\pa_{\rho}F_{\mu\nu}
\})
+{\cal O}(\Theta^2).\nonumber
\eea
This mapping can be used to construct the SW analogue of the noncommutative Yang-Mills theory which is known to display many interesting effects at the tree level.

In the next sections we discuss some impacts of the noncommutativity within the LV context, both within Moyal product and SW map approaches.

\section{LV impacts of the Moyal product}

The Moyal product approach is the most used formulation of the noncommutative space-time being especially convenient for calculating of the loop corrections. It is easy to show that the quadratic part of the action for any noncommutative model is not affected by the noncommutativity, which nontrivially contributes to interaction vertices only. Then, in many cases, due to presence of (anti)symmetrized products of fields in vertices, the Moyal phase factors are reduced to sine or cosine forms. As a result, we can find new forms of contributions from different Feynman diagrams.

As the simplest example, we consider the noncommutative $\phi^4$ theory \cite{Arefeva}:
\begin{eqnarray}
S=\int d^Dx\left[-\frac{1}{2}\phi(\Box+m^2)\phi+\frac{\lambda}{4!}\phi*\phi*\phi*\phi
\right].
\end{eqnarray}
It is instructive to give here the explicit form of the noncommutative quartic vertex in momentum space reflecting the symmetry between all four $\phi$ legs:
\begin{eqnarray}
V_4&=&\frac{\lambda}{72}\int\frac{d^Dp_1d^Dp_2d^Dp_3d^dp_4}{(2\pi)^{4D}}\phi(p_1)\phi(p_2)\phi(p_3)\phi(p_4)\times\\&\times&
[\cos(p_1\wedge p_2)\cos(p_3\wedge p_4)+\cos(p_1\wedge p_3)\cos(p_2\wedge p_4)+\nonumber\\
&+&\cos(p_1\wedge p_4)\cos(p_2\wedge p_3)].\nonumber
\end{eqnarray}
Here we define $p\wedge q=\frac{1}{2}\Theta^{\mu\nu}p_{\mu}q_{\nu}$. 

The lower quantum correction is presented by an usual seagull graph given by Fig. 4.1.


\begin{figure}[!ht]
\begin{center}
\includegraphics[angle=0,scale=1.0]{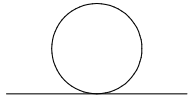}
\end{center}
\caption{Higher-derivative contributions with three insertions.}
\end{figure}


After carrying out all contractions, the contribution of this graph to the self-energy tensor, in the Euclidean space, can be cast as \cite{Arefeva}
\begin{eqnarray}
\Sigma(p)=-\frac{\lambda}{6}\int\frac{d^Dk}{(2\pi)^D}\frac{2+\cos(2k\wedge p)}{k^2+m^2}.
\end{eqnarray}
We see that this is a sum of two terms called planar and nonplanar contributions. The reasons for the names "planar" and "nonplanar" can be found in \cite{Arefeva} and in papers cited there. The  planar contribution is the one not affected by the noncommutativity.  It coincides with the contribution of the whole graph for the commutative analogue of this theory, up to the constant factor -- indeed, for $\Theta^{\mu\nu}=0$, the numerator reduces to 3 -- and, in this case, the planar contribution  displays the same divergence, up to a constant factor, as in the commutative counterpart of this theory. In principle, in some noncommutative field theories, contributions from certain Feynman diagrams can be completely planar due to cancellation of the exponential factors stemming from vertices, for example, such a situation occurs for a two-point function of the gauge superfield in the supersymmetric $CP^{N-1}$ model in the fundamental representation \cite{FerTeix}. 

It is clear that since the planar part does not depend on $\Theta^{\mu\nu}$, it is perfectly Lorentz invariant, for $D=4+\epsilon$ being equal to
\begin{eqnarray}
\Sigma_{pl}(p)=\frac{\lambda m^2}{48\pi^2\epsilon}+{\rm fin}.
\end{eqnarray}
The second, nonplanar term essentially depends on the noncommutativity matrix $\Theta^{\mu\nu}$. To proceed with it, we can use the formula \cite{Alv},
with $p\circ p\equiv \frac{1}{4}p_{\mu}\Theta^{\mu\nu}\Theta_{\nu\lambda}p^{\lambda}$:
\begin{eqnarray}
&&\int\frac{d^Dk}{(2\pi)^D}\frac{e^{ik\wedge p}}{(k^2+M^2)^N}=(-\frac{1}{2})^N\frac{1}{(2\pi)^{D/2}}\frac{1}{(N-1)!}(M^2)^{D/2-N}\times\nonumber\\&\times&
\frac{K_{D/2-N}(\sqrt{M^2p\circ p})}{[\sqrt{M^2p\circ p}]^{D/2-N}}.
\end{eqnarray}
It is well known that for small arguments $x\to 0$, the asymptotics of the modified Bessel function is  $K_n(x)|_{x\to 0}\sim x^{-n}$. Therefore we see that if $D/2\geq N$, the r.h.s. of the expression above is singular at $p\to 0$ being proportional to $\frac{1}{(p\circ p)^{D/2-N}}$. In particular, for $N=1$ and $D=4$, as in our case, the nonplanar contribution is proportional to $\frac{1}{p\circ p}$, thus displaying a quadratic infrared singularity and being at the same time of the order $\Theta^{-2}$, that is, the singularity of the same order as the UV divergence in a planar counterpart of the contribution. This mechanism for arising infrared singularities of such a form in noncommutative theories with ultraviolet divergences is called the ultraviolet/infrared (UV/IR) mixing \index{ultraviolet/infrared (UV/IR) mixing} \cite{Minw}, and these singularities are called the UV/IR infrared ones. 

Besides of this, we note that in general, 
$\Theta^{\mu\nu}\Theta_{\nu\lambda}$ is not required to be proportional to $\delta^{\mu}_{\lambda}$, hence we find that the nonplanar contributions, first, essentially break the Lorentz symmetry, second, are large if the UV/IR infrared singularity is present, since the $\Theta^{\mu\nu}$ is naturally treated to be small. Hence we face the problem of \index{large quantum corrections} large LV quantum corrections, similarly to the situation occurring in higher-derivative LV theories \cite{Reyes2}. From a formal viewpoint, the large quantum corrections reflect the fact that the noncommutative deformation of the theory plays the role of some regularization used to eliminate the ultraviolet divergences, similarly to higher-derivative additive terms, therefore, at $\Theta^{\mu\nu}\to0$, as well as if higher derivative additive terms in \cite{Reyes2} are switched off, the regularization is removed, and the singularities are recovered. Therefore we see that together with breaking of the perturbative expansion, we violate the restriction for LV terms to be small. Moreover, using the mechanism of the UV/IR mixing \cite{Minw}, one can obtain terms with an arbitrary large negative degree of $\Theta^{\mu\nu}\Theta_{\nu\lambda}$, so, the resulting effective theory cannot be well-defined if we suggest LV parameters to be small. This is the main problem of the noncommutative field theories, which, as is well known, is typically solved through an appropriate supersymmetric extension of a noncommutative theory which typically allows to eliminate dangerous quadratic and linear divergences,  both ultraviolet and the related UV/IR infrared ones. The paradigmatic example is the noncommutative Wess-Zumino model \cite{NCWZ}. However, to close this section, we emphasize again that the form of noncommutative theories, either supersymmetric or non-supersymmetric ones, is essentially distinct from conventional LV theories considered throughout this book, whose action is a sum of a usual Lorentz-invariant action and a small additive LV term involving only some lower degrees of LV constant parameters.

\section{Seiberg-Witten map}

Another description of noncommutative gauge field theories is based on the SW map. As we already noted in Section 4.1, the SW map is based on the gauge-preserving mapping of the noncommutative gauge field to the commutative one. It is easy to see \cite{SW} that while the action of the usual NC gauge theory is polynomial with respect to the fields $\hat{A}_{\mu}$, see (\ref{NCYM}), its commutative image will look like an infinite series in $A_{\mu}$, and there is no closed form for it. Nevertheless, the SW map is known to obtain many interesting tree-level results.

The SW mapping for the noncommutative Maxwell action up to the lower nontrivial order in $\Theta^{\alpha\beta}$ yields the following Lagrangian:
\begin{eqnarray}
\label{swmap}
{\cal L}_{SW}=-\frac{1}{4}F_{\mu\nu}F^{\mu\nu}+\frac{1}{8}\Theta^{\alpha\beta}F_{\alpha\beta}F_{\mu\nu}F^{\mu\nu}-\frac{1}{2}\Theta^{\alpha\beta}F_{\mu\alpha}F_{\nu\beta}F^{\mu\nu}+\ldots.
\end{eqnarray}

First of all, one can show that the wave propagation in the theory (\ref{swmap}) displays some interesting features. Indeed, the corresponding Euler-Lagrange equations will be highly nonlinear, therefore, whereas the first pair of the Maxwell equations, that one stemming from the antisymmetry of $F_{\mu\nu}$, will be the same as in the usual electrodynamics, the second one, corresponding to Euler-Lagrange equations, will be strongly modified \cite{Wulkmap}. So, let us briefly discuss the Maxwell equations in the new theory.

If we define as usual that $F^{0i}=E^i$, and $F_{ij}=-\epsilon_{ijk}B^k$, and choose the noncommutativity to be purely spatial, that is, $\Theta^{0i}=0$ and $\Theta^{ij}=\epsilon^{ijk}\Theta^k$, while the first pair of Maxwell equations has the standard form
\bea
\vec{\nabla}\cdot\vec{B}&=&0;\nonumber\\
\nabla\times\vec{E}&=&-\frac{1}{c}\frac{\pa\vec{B}}{\pa t},
\eea
the second pair can be formally written as
\bea
\vec{\nabla}\cdot\vec{D}&=&0;\nonumber\\
\nabla\times\vec{H}&=&\frac{1}{c}\frac{\pa\vec{D}}{\pa t},
\eea
but in this case the vectors $\vec{H}$ and $\vec{D}$ are strongly nonlinear in $\vec{E}$ and $\vec{B}$, so that up to the first order in $\vec{\Theta}$, one finds \cite{Guralnik}:
\bea
\vec{D}&=&(1-\vec{\Theta}\cdot\vec{B})\vec{E}+(\vec{\Theta}\cdot\vec{E})\vec{B}+(\vec{E}\cdot\vec{B})\vec{\Theta};\nonumber\\
\vec{H}&=&(1-\vec{\Theta}\cdot\vec{B})\vec{B}+\frac{1}{2}(E^2-B^2)\vec{\Theta}-(\vec{\Theta}\cdot\vec{E})\vec{E}.
\eea
The solutions of these equations display essentially new effects because of their nonlinear nature. For example, even for the simplest  electric field described by the plane wave $\vec{E}=\vec{E}(\omega t-\vec{k}\cdot\vec{r})$, one has  possibilities to obtain very nontrivial effects. First, the known transversality relations for electric and magnetic fields, and similarly for $\vec{H}$ and $\vec{D}$, look like \cite{Guralnik}:
\bea
\label{transv}
\vec{B}&=&\vec{\kappa}\times\vec{E}+\vec{b};\nonumber\\
\vec{D}&=&-\vec{\kappa}\times\vec{H}+\vec{d}.
\eea
Here $\vec{\kappa}=\frac{c\vec{k}}{\omega}$ is a dimensionless wave vector, and the vectors $\vec{b}$, $\vec{d}$ are time-independent ones, they serve to describe some primordial constant electric and magnetic fields.

One of the conclusions is that now the electric field is no more a transversal one, with its longitudinal component $E_L$ is related with transversal ones $E_T^i$ as
\bea
E_L=-\hat{\kappa}^i\beta^{ij}E^j_T,
\eea
with $\hat{\kappa}$ is a normalized wave vector, and $\beta^{ij}=\Theta^ib^j+\Theta^jb^i$.

The dispersion relation in this theory is  $\frac{\omega}{ck}=1-\vec{\Theta}_T\cdot\vec{b}_T$, where $\vec{b}_T$ is a  transversal part of the background magnetic field $\vec{b}$ (\ref{transv}), and $\vec{\Theta}_T$ -- of the vector $\vec{\Theta}$ (we note that, following \cite{Gamboa}, the space-time noncommutativity itself can give an origin for the background magnetic field). So, superluminal propagation and causality violation are ruled out for  certain choices of $\vec{\Theta}_T$ and $\vec{b}_T$. 

A more detailed discussion of dispersion relations in the SW formulation of the noncommutative theories was developed in \cite{MNR}. Following this paper, we start with covariant equations of motion which can be obtained from the Lagrangian (\ref{swmap}):
\bea\label{GAM}
\pa_{\mu}F^{\mu\nu}+\Theta^{\alpha\beta}F_{\alpha}^{\phantom{\alpha}\mu}(\pa_{\beta}F_{\mu}^{\phantom{\mu}\nu}+\pa_{\mu}F_{\beta}^{\phantom{\mu}\nu})=0.
\eea
Then, it is easy to show that for the plane wave solution $F_{\mu\nu}(x)=\tilde{F}_{\mu\nu}(kx)$, with $k^{\mu}$ being a constant vector, Eq.~(\ref{GAM}) trivially becomes 
\bea\label{SWeqmov}
k_\mu \tilde F'^{\mu\nu} + \Theta^{\alpha\beta} \tilde{F_\alpha}^\mu (
k_\beta \tilde{F}_{\mu}^{\prime\,\nu} + k_\mu \tilde{F}_{\beta}^{\prime\,\nu} ) = 0,
\eea
where the prime in $\tilde{F}^{\prime}_{\mu\nu}$ is for differentiation with respect to $kx$. In this context, remaining Maxwell equations are given by the standard Bianchi identity,
\bea
\label{BianchiSW}
\partial_\rho F_{\mu\nu} + \partial_\nu F_{\rho\mu} + \partial_\mu F_{\nu\rho} = 0,
\eea
and contracting it with $\partial^\rho$, we obtain
\bea\label{BianchiSW2}
k^2 \tilde{F}^\prime_{\mu\nu} + k^\rho k_\nu
\tilde{F}^\prime_{\rho\mu} + k^\rho k_\mu
\tilde{F}^\prime_{\nu\rho} = 0,
\eea
where we have used also the fact that our field is the plane wave solution. Now, by using Eq.~(\ref{SWeqmov}) in the above expression, we finally get
\bea 
k^2\tilde{F}^{\prime}_{\mu\nu}+2\Theta^{\alpha\beta}\tilde{F}_{\alpha}^{\phantom{\alpha}\rho}k_{\beta}k_{\rho}\tilde{F}^{\prime}_{\mu\nu}=0.
\eea
We note that since $k_\rho\tilde{F}_{\alpha}^{\phantom{\alpha}\rho}$ is already of the first order in $\Theta$ because of (\ref{SWeqmov}), the second term in above equation can be disregarded. Then, we conclude that the dispersion relation in this theory has the usual form $k^2=0$, and, by considering again Eq.~(\ref{BianchiSW}), the plane wave continues to be transversal as in the commutative case, i.e., $k_\mu \tilde{F}'^{\mu\nu} = 0$.

However, if we make the simplest modification by considering the case of a superposition of a constant background $B_{\mu\nu}$ and a plane wave $\tilde F_{\mu\nu}(kx)$, the field equation (\ref{GAM}) becomes
\bea
\label{fieldeqn}
\tilde{k}_\mu \tilde{F}^{\prime\mu\nu} = 0,
\eea
where $\tilde{k}_\mu = k_\mu + \Theta^{\alpha\beta} B_{\alpha\mu} k_\beta - {\Theta_\mu}^\alpha{B_\alpha}^\beta k_\beta$. Thus, by substituting the above expression into Eq.~(\ref{BianchiSW2}), we obtain the modified dispersion relation $k^2=-2\Theta^{\alpha\beta}B_{\alpha}^{\phantom{\alpha}\rho}k_{\beta}k_{\rho}$, which in certain cases reduces to the results of \cite{Guralnik}. Moreover, one can show as well that there exists a mapping between the solution of the equations of motion in the Moyal picture, where the Moyal product is expanded up to the first order, and the superposition of a constant background and a plane wave in the SW picture. This justifies that at the tree level, these two approaches are equivalent for a certain class of solutions, that is, the solutions described by a superposition of a constant field and a plane wave. 

Nevertheless, it should be noted that at the quantum level these two approaches are essentially distinct, since, unlike the Moyal product approach, first, the SW map yields an essentially non-renormalizable theory, second, it can be made to be consistent with the supersymmetry only with use of a nonlinear representation of the SUSY algebra \cite{MarTam}. 

We note that the behavior rather similar to that one described above is displayed by the plane wave solutions in another LV deformation of the electrodynamics based on the Heisenberg-Euler action \index{Heisenberg-Euler action} \cite{HeisEul}, so, it is quite typical for nonlinear extensions of the electrodynamics. As a by-product, we note that the noncommutativity and/or Lorentz symmetry breaking can be treated as possible ingredients for analogue models aimed to describe the propagation of electromagnetic field in a condensed matter as it was also suggested in \cite{MNR}. 

\section{Noncommutative fields}

Besides of the Moyal product and Seiberg-Witten map, there is one more manner to implement a noncommutativity within the field theory models, that is, the so-called noncommutative field approach proposed initially in \cite{Gamboa1}. Its starting point is the deformation of canonical commutation relations between two fields or two momenta. Following \cite{Gamboa}, if one considers a theory of two free real scalar fields $\phi_1$ and $\phi_2$ with equal masses described by the usual action corresponding to the Hamiltonian density
\bea
H=\frac{1}{2}\sum\limits_{i=1}^2\left(\pi_i\pi_i+\vec{\nabla}\phi_i\cdot\vec{\nabla}\phi_i+m^2\phi_i\phi_i\right).
\eea  
Then we deform the canonical commutation relation between two momenta as
\bea
\label{defren}
[\pi_i(\vec{x}),\pi_j(\vec{y})]=i\Theta_{ij}\delta(\vec{x}-\vec{y}),
\eea
whereas other canonical commutation relations stay untouched (in \cite{PP}, the similar analysis was performed for an equivalent theory involving a complex scalar field $\phi$, and the conjugated one $\phi^*$). In the case of free fields a deformation of the relation between two fields implies in analogous results, however, in a general theory such a deformation could yield much more complicated, in general nonpolynomial, expressions. In principle, such a deformation can be treated as a kind of Bopp shift. Also, it is clear that non-trivial noncommutative relations between space-time coordinates $x^{\mu}$ themselves will imply as well nontrivial noncommutative relations between their functions, in particular, fields and corresponding canonical momenta. Therefore, this approach can be naturally interpreted as another, alternative description for noncommutativity, although there is no straightforward mapping between Moyal product and noncommutative fields frameworks.

After the above mentioned deformation of commutation relations, we should find new canonical variables $\Pi_i$  whose commutator is the usual one, $[\Pi_i(\vec{x}),\Pi_j(\vec{y})]=0$. They are given by the definition
\bea
\Pi_i=\pi_i-\frac{1}{2}\Theta_{ij}\phi_j,
\eea
which allows us to write immediately the deformed Lagrangian, which, after expressing of momenta in terms of velocities, looks like
\bea
{\cal L}_{def}=\Pi_i\dot{A}_i-H=-\sum\limits_{i=1}^2\phi_i(\Box+m^2)\phi_i-\frac{1}{2}\sum\limits_{i, j=1}^2\Theta_{ij}\dot{\phi}_i\phi_j.
\eea
It is clear that the additive term proportional to $\Theta_{ij}$ breaks the Lorentz symmetry involving only the time derivative, which clearly will modify the dispersion relations. 

If we apply this methodology to the free electrodynamics \cite{Gamboa}, the situation becomes more interesting. We follow the usual Dirac prescription for quantization of constrained theories, so, we get the primary constraint $\Phi^{(1)}=\pi_0\simeq 0$, where $\pi_{\mu}=F_{0\mu}$ is a momentum canonically conjugated to $A^{\mu}$. The Hamiltonian density is
\bea
\label{h}
H=\frac{1}{2}\pi_i\pi_i+\frac{1}{4}F_{ij}F_{ij}+A_0\pa_i\pi_i.
\eea
The requirement of conservation of the primary constraint yields the secondary one, $\Phi^{(2)}=\pa_i\pi_i\simeq 0$, after which the Dirac algorithm ends, there is no new constraints more.

Now, we again deform the canonical commutation relations to the form (\ref{defren}). The key point of the noncommutative approach now that we want the new model, obtained through the deformation of the gauge theory, to be gauge invariant as well. To see the way to do it, we note that the gauge transformation of $A_i$ field with a parameter $\alpha(\vec{x})$ can be presented as 
\bea
\delta A_i(\vec{x})=[A_i(\vec{x}),\int d^3\vec{y}\alpha(\vec{y})\Phi^{(2)}(\vec{y})]=-\pa_i\alpha(x).
\eea
In the usual, non-deformed case, one finds the following gauge transformation of the canonical momentum:
$$
\delta \pi_i(\vec{x})=[\pi_i(\vec{x}),\int d^3\vec{y}\alpha(\vec{y})\Phi^{(2)}(\vec{y})]=0,
$$
as it should be. We want to have the same gauge transformations of the field and the conjugated momentum in the  deformed case as well which is very natural, while it is clear that in this case $[\pi_i(\vec{x}),\int d^3\vec{y}\alpha(\vec{y})\Phi^{(2)}(\vec{y})]\neq 0$ since the $\pi_i$'s do not commute more. To achieve, for fields and momenta, the same gauge transformations as in the undeformed case, we extend the constraint $\Phi^{(2)}$ by an additive term so that the new constraint $\tilde{\Phi}^{(2)}$ reads as
\bea
\tilde{\Phi}^{(2)}=\pa_i\pi_i-\Theta_{jk}\pa_jA_k.
\eea
 It is easy to check that in this case one has $[A_i(\vec{x}),\int d^3\vec{y}\alpha(\vec{y})\tilde{\Phi}^{(2)}(\vec{y})]=-\pa_i\alpha(x)$ and $\pi_i(\vec{x}),\int d^3\vec{y}\alpha(\vec{y})\tilde{\Phi}^{(2)}(\vec{y})]=0$ as it should be.
For the sake of the convenience, we introduce the vector $\Theta_i$ satisfying the relation $\Theta_i=\frac{1}{2}\epsilon_{ijk}\Theta_{jk}$, so, $\Theta_{jk}=\epsilon_{ijk}\Theta_i$.

Again, one needs to get a canonical momentum $\Pi_i$ satisfying the relation $[\Pi_i,\Pi_j]=0$, while other commutation relations stay unchanged, as above. The $\Pi_i$ evidently looks like
\bea
\Pi_i=\pi_i-\frac{1}{2}\Theta_{ij}A_j,
\eea
and the resulting Lagrangian is
\bea
{\cal L}=\pi_i\dot{A}_i-H,
\eea
where the Hamiltonian $H$ is given by (\ref{h}), and the momenta are afterward expressed in terms of velocities. As a result, we arrive at the new Lagrangian
\bea
{\cal L}_{new}=-\frac{1}{4}F_{\mu\nu}F^{\mu\nu}+\frac{1}{4}\epsilon^{\mu\nu\lambda\rho}\Theta_{\mu}A_{\nu}F_{\lambda\rho},
\eea
where the constant 4-vector $\Theta_{\mu}$ is defined as $\Theta_{\mu}=(0,\Theta_i)$. Therefore, we see that the electrodynamics with the noncommutative deformation of the canonical algebra is equivalent to the electrodynamics with usual commutation relations, but with the CFJ term. It is interesting to note that performing the analogous noncommutative deformation of the canonical algebra in the electrodynamics with the CFJ term will imply in modification of the CFJ coefficient only, but without arising other new terms.  Also, if we apply this methodology in the three-dimensional case, instead of a LV term, the usual Lorentz-invariant Chern-Simons term will arise, with the mass is proportional to $\Theta=\frac{1}{2}\epsilon^{ij}\Theta_{ij}$ \cite{NPR}.

This methodology can be applied as well to non-Abelian theories \cite{Gamboa2} where the non-Abelian CFJ term (\ref{nab}) was generated. Therefore, the noncommutative deformation of the canonical algebra can be treated as one more manner to generate the LV terms, which differs from the perturbative generation approach based on using additional LV couplings. Moreover, this methodology can be applied as well even in the linearized gravity case \cite{ncqugra}, where, however, the study is much more sophisticated -- first, the components of a metric fluctuation tensor have two indices, which makes calculations to be more involved, second, to achieve the transversality, one should implement the transverse projectors into deformed commutation relations playing the role of analogues of (\ref{defren}). The extended linearized gravity theory with a resulting additive term, under imposing some gauge fixing conditions, yields the same dispersion relations as the extended linearized gravity theory with the additive term given by (\ref{lingrav}), which justifies the equivalence of the term obtained by noncommutative fields approach and the term (\ref{lingrav}).

\section{Conclusions}

We considered the LV impacts of noncommutative extensions for field theory models. It should be noted that we treated the simplest form of the space-time noncommutativity based on the Moyal product together with its mapping to a nonpolynomial commutative theory through the SW map. In principle, one can introduce as well other forms of noncommutativity which are not based on constant noncommutativity parameters, e.g., the dynamical noncommutativity approach \cite{Morita}. However, this approach is Lorentz invariant since all $\Theta^{\mu\nu}$, being new coordinates in this approach, are integrated out at final steps of all calculations. As a result of using the Moyal product or the SW map, one arrives at a theory essentially depending on the LV constant antisymmetric tensor $\Theta^{\mu\nu}$. We see, however, that while the SW map description is rather similar to the approach developed for constructing the usual LV SME \cite{ColKost1,ColKost2}, so that the action is a sum of a known Lorentz-invariant classical action and small LV corrections, the approach based on the Moyal product is essentially different, because, first, instead of calculating order by order, it allows to obtain the exact dependence of quantum corrections on the noncommutativity parameters $\Theta^{\mu\nu}$, second, in many cases, as a consequence of the UV/IR mixing, the results display singularities in a commutative limit. Nevertheless, both these situations have certain analogies in "usual" LV theories: the exact dependence on the LV parameters is treated, e.g., within the exact propagator of the spinor field (\ref{exactprop}) allowing to take into account all orders of its expansion in the LV vector $b_{\mu}$ \cite{Andr}, and the singularities in the Lorentz-invariant limit, which in NC theories corresponds to the commutative limit, can occur in some higher-derivative LV extensions of the field theory models \cite{Reyes2}. This similarity is rather natural since the noncommutative field theories by their essence are nothing more as infinite-derivative LV extensions of usual field theories. We argued that, at the classical level, in lower orders in expansion in the $\Theta^{\mu\nu}$ parameter, for some solutions of equations of motion, there is some mapping between SW map and Moyal product which justifies their classical equivalence in certain cases. However, the situation is more involved when quantum corrections are studied, since, unlike the Moyal product approach, within the SW map framework, the QED is clearly non-renormalizable.

Also, we introduced a methodology of noncommutative fields allowing for mapping of a theory formulated in terms of noncommuting momenta into a theory formulated in terms of canonical variables, whose action differs from the original action by additive terms of different orders in noncommutativity parameters $\Theta_{ij}$, and their presence explicitly breaks the Lorentz symmetry in four- and higher-dimensional space-times. The relevance of the noncommutative fields approach within the LV context is confirmed by the fact that, as we have noted in previous chapters, it allows to generate some of known LV additive terms, such as the CFJ term, its non-Abelian generalization, and the two-dimensional CS-like term for scalar fields.

We noted already that to solve the main problem of noncommutative theories, that is, the problem of nonintegrable infrared singularities caused by the UV/IR mixing mechanism, one should carry out the supersymmetric extension of these theories. This is, clearly, one of motivations to the interest in LV extensions of supersymmetric, especially superfield, theories. We discuss this problem in the next chapter.

\chapter{Lorentz symmetry breaking and supersymmetry}

The supersymmetry began to be treated as a fundamental symmetry of Nature already in 1970s by many reasons: first, it essentially improves the renormalization behavior of field theory models, second, it allows to unite Poincare symmetry with internal symmetries in a nontrivial manner, third, it found broad applications within the string theory. However, for a long time there was a common belief among field theorists that since the supersymmetry represents itself as a nontrivial extension of the Poincare symmetry, in LV theories the supersymmetry should be broken as well. Nevertheless,  a first manner to construct the LV extension of the supersymmetry algebra was proposed already in 2001 \cite{BK}. In 2003, another manner to conciliate the Lorentz symmetry breaking with an unbroken supersymmetry, based on introduction of some extra superfield(s) \index{superfield} with some of whose components depending on the LV parameters \cite{HelSUSY}, was proposed,  we note that namely within this approach the supersymmetric extension of the CFJ term was successfully constructed. In \cite{Bolokhov}, an idea of a straightforward LV extension of supersymmetric field theories where additive LV terms are introduced already at the level of a superfield action, was proposed. In this chapter we review all these approaches, including their advantages and disadvantages, and present corresponding main results obtained with their use, especially, examples of quantum calculations. We essentially use the superfield formalism for the description of LV supersymmetric field theories, since it, first, allows for very compact formulation of superfield theories, second, is highly convenient for quantum calculations (for a review of superfield methodology in supersymmetric field theories see e.g. \cite{BK0}). As we claimed in the previous chapter, the noncommutative (super)field theories also display the Lorentz symmetry breaking, the methodology of their studies is well developed, however, in this chapter we will restrict ourselves by theories where the Lorentz symmetry breaking is implemented in a conventional manner, that is, via small additive terms. 

\section{Deformation of the supersymmetry algebra}

The Kostelecky-Berger method of LV extension for superfield theories based on the deformation of the supersymmetry algebra is perhaps the most elegant one. Its main idea looks like follows \cite{BK}. Let us start with the usual supersymmetry algebra whose anticommutation relations, in the four-dimensional case, are (see e.g. \cite{BK0}):
\bea
\label{comrel}
\{Q_{\alpha},\bar{Q}_{\dot{\beta}}\}=2i\sigma^{\mu}_{\alpha\dot{\alpha}}\partial_{\mu},
\eea
so, the supersymmetry generators can be chosen, e.g., in the form \index{supersymmetry generators}
\bea
\label{sg1}
Q_{\alpha}&=&-\pa_{\alpha}+i\bar{\theta}^{\dot{\beta}}\bar{\sigma}^{\mu}_{\dot{\beta}\alpha}\partial_{\mu};\nonumber\\
\bar{Q}_{\dot{\alpha}}&=&\pa_{\dot{\alpha}}-i\theta^{\beta}\sigma^{\mu}_{\dot{\alpha}\beta}\partial_{\mu}
\eea
Other representations of the supersymmetry algebra in a four-dimensional space-time are possible as well, the only requirement is that in any case, generators must  satisfy the relation (\ref{comrel}). The $\theta^{\alpha},\bar{\theta}^{\dot{\alpha}}$ are the Grassmannian (anticommuting) coordinates \index{Grassmannian coordinates} of the superspace, and $\pa_{\alpha}=\frac{\pa}{\pa\theta^{\alpha}}$, with an analogous definition for $\pa_{\dot{\alpha}}$. All other anticommutators are: $\{Q_{\alpha},Q_{\beta}\}=0$, $\{\bar{Q}_{\dot{\alpha}},\bar{Q}_{\dot{\beta}}\}=0$, and both $Q_{\alpha}$ and $\bar{Q}_{\dot{\alpha}}$ commute with usual space-time derivatives $\pa_{\mu}$. The supersymmetry transformation of an arbitrary superfield $\Sigma$ can be represented as superspace translations: $\delta\Sigma=(\epsilon^{\alpha}Q_{\alpha}+\bar{\epsilon}^{\dot{\alpha}}\bar{Q}_{\dot{\alpha}})\Sigma$, where $\epsilon_{\alpha}$, $\bar{\epsilon}_{\dot{\alpha}}$ are the infinitesimal constant Grassmannian parameters. 

Then, let us deform these generators through a simple replacement: $\pa_{\mu}\to\pa_{\mu}+k_{\mu\nu}\pa^{\nu}=\nabla_{\mu}$, with $\nabla_{\mu}$ can be called a ``twisted'' derivative, i.e. instead of generators (\ref{sg1}), one must use new, "twisted" generators:
	\bea
	\label{sg2}
	 Q_{\alpha}&=&-\pa_{\alpha}+i\bar{\theta}^{\dot{\beta}}\bar{\sigma}^{\mu}_{\dot{\beta}\alpha}\nabla_{\mu};\nonumber\\ \bar{Q}_{\dot{\alpha}}&=&\pa_{\dot{\alpha}}-i\theta^{\beta}\sigma^{\mu}_{\dot{\alpha}\beta}\nabla_{\mu}.
	\eea 
The $k_{\mu\nu}$ is a constant tensor breaking the Lorentz symmetry. Without loss of generality, we can require all its components to be much less than 1, enforcing the Lorentz symmetry breaking to be small. Such replacement is linear with respect to derivatives, in order to be consistent with the Leibnitz rule. In principle, in some simple situations one can also construct a consistent deformation of the supersymmetric field theory abandoning the Leibnitz rule \cite{Quei}, however, in many cases the linearity of supersymmetry generators in derivatives is essential. 

To construct the superfield theory, one needs to introduce supercovariant derivatives in such a manner that, for any superfield $\Sigma$, the supercovariant derivatives of this superfield would be superfields, i.e. these derivatives should anticommute with supersymmetry generators: $\{Q_{\alpha},D_{\beta}\}=0$, $\{\bar{Q}_{\dot{\alpha}},D_{\beta}\}=\{Q_{\alpha},\bar{D}_{\dot{\beta}}\}=\{\bar{Q}_{\dot{\alpha}},\bar{D}_{\dot{\beta}}\}=0$.
 Explicitly, the "twisted" supercovariant derivatives consistent with the deformed supersymmetry look like
\bea
D_{\alpha}&=&\pa_{\alpha}+i\bar{\theta}^{\dot{\beta}}\bar{\sigma}^{\mu}_{\dot{\beta}\alpha}\nabla_{\mu};\nonumber\\
\bar{D}_{\dot{\alpha}}&=&-(\pa_{\dot{\alpha}}+i\theta^{\beta}\sigma^{\mu}_{\dot{\alpha}\beta}\nabla_{\mu}),
\eea
and satisfy the following anticommutation relations:
\bea
\{D_{\alpha},\bar{D}_{\dot{\beta}}\}=-2i\sigma^{\mu}_{\alpha\dot{\alpha}}\nabla_{\mu}.
\eea
It is clear that the Leibnitz rule is maintained for these derivatives. From this relation, one can derive the following useful identities:
\bea
D^2\bar{D}^2D^2=16\Box_DD^2;\quad\; D_{\alpha}D_{\beta}D_{\gamma}=0,
\eea
with the analogous identities can be obtained after replacements of $D_{\alpha}$ by $\bar{D}_{\dot{\alpha}}$. Here $\Box_D=\nabla^{\mu}\nabla_{\mu}=\Box+2k_{\mu\nu}\pa^{\mu}\pa^{\nu}+k_{\mu\nu}k^{\rho\nu}\pa^{\mu}\pa_{\rho}$ is the deformed d'Alembertian operator.

As a next step, one can introduce a chiral superfield \index{chiral superfield} $\Phi$ satisfying the relation $\bar{D}_{\dot{\alpha}}\Phi=0$, and similarly the antichiral one $\bar{\Phi}$ so that $D_{\alpha}\bar{\Phi}=0$. Their components can be defined through projections:
\bea
\phi=\Phi|, \quad\, \psi_{\alpha}=\frac{1}{2}D_{\alpha}\Phi|, \quad\, F=\frac{D^2}{4}\Phi|,
\eea
where the $|$ symbol means that after the differentiation one puts $\theta=\bar{\theta}=0$, and the expressions for components of $\bar{\Phi}$ are the analogous ones. This allows us to define the LV Wess-Zumino model whose action being written in terms of superfields formally reproduces the usual expression
\bea
\label{LBWZ}
S=\int d^8z\Phi\bar{\Phi}+\left[\int d^6z \left(\frac{m}{2}\Phi^2+\frac{\lambda}{3!}\Phi^3\right)+h.c.\right],
\eea
but the component form of this action involves extra, LV terms and looks like
\bea
S&=&\int d^4x \left(\phi\Box_D\bar{\phi}+F\bar{F}+i\bar{\psi}^{\dot{\alpha}}\sigma^{\mu\alpha}_{\dot{\alpha}}\nabla_{\mu}\psi_{\alpha}+\right. \nonumber\\
&+&\left.\Big(m(\psi^2 +\phi F)+\lambda(\phi \psi^2 +\frac{1}{2}F\phi^2)+h.c.\Big)\right).
\eea
One can immediately see that this action involves the aether-like term for the spinor field $i\bar{\psi}^{\dot{\alpha}}\sigma^{\mu\alpha}_{\dot{\alpha}}k_{\mu\nu}\pa^{\nu}\psi_{\alpha}$ which for the simplest, aether-like choice $k_{\mu\nu}=\alpha u_{\mu}u_{\nu}$, with $\alpha$ is some number, and $u^{\mu}$ is a constant vector (it is natural to require $|\alpha|\ll 1$, with $u^{\mu}u_{\mu}$ is either $-1$, 0 or 1), exactly reproduced the aether term introduced in \cite{Carroll}, and, under the same choice, the aether-like term 
$\alpha\phi(u\cdot\pa)^2\bar{\phi}$ and the new term of fourth order in the vector $u^{\mu}$ for the scalar field. Actually, the last term also displays the aether form looking like $\alpha^2u^2\phi(u\cdot\pa)^2\bar{\phi}$. So, we see that this theory involves aether terms for scalar and spinor fields as ingredients.

The superfield propagators in the theory (\ref{LBWZ}) are exact analogues of those ones in the usual Wess-Zumino model, with only difference in replacement of a simple derivative $\pa_{\mu}$ by a "twisted" one $\nabla_{\mu}$:
\bea
&&<\Phi(z_1)\bar{\Phi}(z_2)>=-\frac{i}{\Box_D-m^2}\delta^8(z_1-z_2);\\
&&<\Phi(z_1)\Phi(z_2)>=<\bar{\Phi}(z_1)\bar{\Phi}(z_2)>^*=-
\frac{im}{\Box_D-m^2}(\frac{D^2}{4\Box_D})\delta^8(z_1-z_2),\nonumber
\eea
with $D^2$, $\bar{D}^2$ factors are associated with the vertices by the same rules as in the usual Wess-Zumino model \cite{BK0}, that is, a vertex with an external chiral field carries the factor $-\frac{\bar{D}^2}{4}$, and with an external antichiral one -- the factor $-\frac{D^2}{4}$.

Here, some words about dispersion relations are in order. Disregarding the factor $\frac{1}{\Box_D}$ which is really needed only to rewrite integrals over a chiral (antichiral) subspace as those ones over a whole superspace, and does not correspond to new degrees of freedom, one finds the physical spectra are completely described by the denominator $\Box_D-m^2$, or, after the Fourier transform, $\tilde{p}^2+m^2$, where $\tilde{p}^2=(p_{\mu}+\alpha u_{\mu}u_{\alpha}p^{\alpha})(p^{\mu}+\alpha u^{\mu}u^{\beta}p_{\beta})$ is a twisted scalar square of the momentum. We find that for different choices of $u^{\mu}$, we have different dispersion relations:

(i) For the time-like $u^{\mu}=(1,0,0,0)$, one has $E^2(1-\alpha)^2=\vec{p}^2+m^2$.

(ii) For the space-like $u_{\mu}=(0,1,0,0)$, one has $E^2=\vec{p}^2+m^2+(2\alpha+\alpha^2)(\vec{u}\cdot \vec{p})^2$.

(iii) For the light-like $u_{\mu}=(1,1,0,0)$, and $\vec{p}$ along the $x$ axis, one has $E=\frac{1}{1-2\alpha}(-2\alpha p\pm\sqrt{p^2(1+2\alpha+4\alpha^2)+m^2}$.

It is clear that the dynamics in all these cases is consistent if $\alpha$ is enough small.

As an example of quantum calculation in this theory we present here the computation of the one-loop low-energy effective action which is completely described by the \index{K\"{a}hlerian effective potential} K\"{a}hlerian effective potential, by the definition depending only on the superfields themselves but not on their derivatives. We employ the methodology of summation over an infinite number of Feynman supergraphs \index{Feynman supergraphs} discussed in details in \cite{PW}.  

\begin{figure}[!ht]
\begin{center}
\includegraphics[angle=0,scale=1.0]{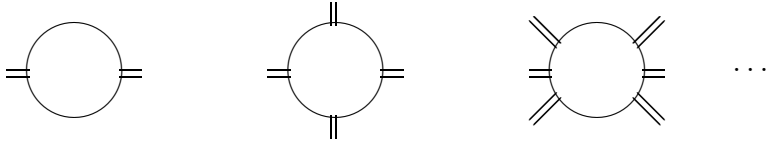}
\end{center}
\caption{Contributions to the K\"{a}hlerian effective potential.}
\end{figure}


In supergraphs given by the Fig. 5.1, the Feynman rules are modified: we incorporated the mass into background fields $\Psi=m+\lambda\Phi$, $\bar{\Psi}=m+\lambda\bar{\Phi}$ denoted here by double lines, so, we have
\bea
&&<\Phi(z_1)\bar{\Phi}(z_2)>=-\frac{i}{\Box_D}\delta^8(z_1-z_2);
\nonumber\\
&&<\Phi(z_1)\Phi(z_2)>=<\bar{\Phi}(z_1)\bar{\Phi}(z_2)>^*=0.
\eea

These Feynman supergraphs can be summed, just as it is done for usual superfield theories, yielding the following expression for the one-loop K\"{a}hlerian effective potential
\bea
K^{(1)}=-\frac{i}{2}\sum\limits_{n=1}^{\infty}\frac{1}{n}\int d^4\theta\left[\Psi\bar{\Psi}\frac{D^2\bar{D^2}}{16\Box_D^2}\right]^n\delta(z-z')|_{z=z'},
\eea
We simplify this expression taking into account that the $\frac{D^2\bar{D^2}}{16\Box_D}$ is a projecting operator even in our LV case, so, $[\frac{D^2\bar{D^2}}{16\Box_D}]^n=\frac{D^2\bar{D^2}}{16\Box_D}$. Then, we shrink the loop into a point through the known identity $\frac{D^2\bar{D^2}}{16}\delta(\theta-\theta^{\prime})|_{\theta=\theta^{\prime}}=1$ \cite{BK0}, and, after Fourier transform, remembering the structure of the deformed d'Alembertian operator, we arrive at
\bea
K^{(1)}=\frac{i}{2}\int d^4\theta\int\frac{d^4q}{(2\pi)^4}\frac{1}{(q_{\mu}+k_{\mu\nu}q^{\nu})^2}\ln\left(1-\frac{\bar{\Psi}\Psi}{(q_{\rho}+k_{\rho\sigma}q^{\sigma})^2}\right).
\eea
To integrate, we make a change of variables by the rule $q_{\mu}+k_{\mu\nu}q^{\nu}\to\tilde{q}_{\mu}$, which implies arising the Jacobian of this replacement through the rule $d^4q=\Xi d^4\tilde{q}$, with $\Xi=\det\frac{\partial q_{\nu}}{\partial\tilde{q}_{\mu}}=\det^{-1}(\delta^{\mu}_{\nu}+k^{\mu}_{\nu})$. This Jacobian is a constant, it does not depend on momenta. We note that if one suggests $k_{\mu\nu}$ to be antisymmetric and small, which does not imply any inconsistencies, this Jacobian reduces to 1. In the case of the effective potential, the only impact of the Lorentz symmetry breaking consists just in the presence of this $\Xi$ multiplier. After this change of variables and the Wick rotation, we have
\bea
K^{(1)}=-\frac{1}{2}\Xi\int d^4\theta\int\frac{d^4\tilde{q}}{(2\pi)^4}\frac{1}{\tilde{q}^2}\ln(1+\frac{\bar{\Psi}\Psi}{\tilde{q}^2}).
\eea 
Namely this expression (of course, except of the $\Xi$ factor) arises within the analogous calculations in Lorentz-invariant theories (see e.g. \cite{PW}), so, we immediately can write down the result:
\bea
K^{(1)}=-\frac{1}{32\pi^2}\Xi\int d^4\theta\Psi\bar{\Psi}\ln\frac{\Psi\bar{\Psi}}{\mu^2}.
\eea 
We see that the result differs from that one in the Lorentz-invariant theory only by the Jacobian factor. This allows us to formulate the following rule: for any supersymmetric theory whose superfield action does not involve explicit space-time derivatives, within the Kostelecky-Berger deformation of the supersymmetry algebra, the results of quantum calculations can be obtained from the usual ones through replacement of all derivatives (both spinor supercovariant and usual ones) by their "twisted" analogues, and multiplying of the $L$-loop expression by $\Xi^L$, with $\Xi$ is the above-mentioned Jacobian. This rule has been also explicitly verified for the coupling of the chiral matter with a gauge superfield \cite{SUSY2013}.
Moreover, the same rule is valid in the three-dimensional case as well, where there is only one type of Grassmannian variables $\theta^{\alpha}$, and the supercovariant derivatives are given by
\bea
D_{\alpha}&=&\pa_{\alpha}+i\theta^{\beta}\gamma^{\mu}_{\beta\alpha}\nabla_{\mu},
\eea
and the Dirac matrices for the three-dimensional space are the following $2\times 2$ matrices: $(\gamma^0)^{\a}_{\phantom{a}\b}=-i\sigma^2,
(\gamma^1)^{\a}_{\phantom{a}\b}=\sigma^1,(\gamma^2)^{\a}_{\phantom{a}\b}=\sigma^3$. One can see that since in the three-dimensional space the same deformed d'Alembertian $\Box_D$ arises, the dispersion relations in three- and four-dimensional superfield theories within this methodology are the same.

In this context, some interesting observations for supersymmetric Lorentz-breaking field theories within this approach can be done. First of all, this formulation possesses a nontrivial geometrical interpretation. Indeed, if we consider the three-dimensional theory of the scalar superfield $\Phi$, with the action
\bea
S=\int d^5z\left(\frac{1}{2}\Phi(D^2+m)\Phi+\frac{\lambda}{3!}\Phi^3\right)
\eea
and calculate the lower "fish" contribution to the two-point function, in the low-energy limit we arrive at \cite{SUSY2013}
\bea
\Gamma_2=\frac{\lambda^2\Xi}{48\pi|m|}\int d^5z \Phi(D^2-2m)\Phi.
\eea
Projecting this expression to components, we have:
\bea
\Gamma_2=-\frac{\lambda^2}{48\pi|m|}\int d^3x \Xi(\eta^{\mu\nu}\nabla_{\mu}\phi\nabla_{\nu}\phi+\ldots),
\eea
where $\phi$ is a lower component of the scalar superfield, and dots are for other terms. If we remind that $\nabla_{\mu}=\pa_{\mu}+k_{\mu\nu}\pa^{\nu}$, we can introduce a new "inverse" metric $g^{\rho\sigma}=\eta^{\mu\nu}(\delta_{\mu}^{\rho}+k_{\mu}^{\rho})(\delta_{\nu}^{\sigma}+k_{\nu}^{\sigma})$, and it is clear that $\Xi=|\det(g_{\rho\sigma})|^{1/2}\equiv \sqrt{|g|}$. So, this quantum correction can be represented as
\bea
\Gamma_2=-\frac{\lambda^2}{48\pi|m|}\int d^3x\sqrt{|g|}g^{\mu\nu}\pa_{\mu}\phi\pa_{\nu}\phi+\ldots,
\eea
so, this result formally replays the action of the scalar field in a curved space-time, thus, actually the Lorentz symmetry breaking generates a new geometry, which, however, also corresponds to a flat space since the new metric $g^{\mu\nu}$ is composed by constant components, so, our new geometry is affine. The question about possibility of a further generalization of this concept in order to get a space with a nontrivial curvature is still open since if the tensor $k_{\mu\nu}$ will not be constant, the possible extension of the supersymmetry algebra evidently would be much more involved. For a four-dimensional theory of a chiral scalar superfield, the same situation occurs as well with the only difference that the lower scalar component of this superfield is complex. However, a Lorentz-breaking gauge superfield theory apparently does not admit a geometrical interpretation, since there is no way to obtain the modified metric $g^{\mu\nu}$ contracted with vector fields.

Another observation is that if we consider the generalization of the free supersymmetric QED with an action of a gauge superfield $V$:
\bea
S=-\frac{1}{16}\int d^8z V D^{\alpha}\bar{D}^2D_{\alpha} V,
\eea
its component form will be
\bea
S=-\frac{1}{4}d^4x \bar{F}_{\mu\nu}\bar{F}^{\mu\nu},
\eea
where $\bar{F}_{\mu\nu}=\nabla_{\mu}A_{\nu}-\nabla_{\nu}A_{\mu}$ is a deformed field strength of the electromagnetic field. It is invariant under the Abelian transformations $\delta A_{\mu}=\nabla_{\mu}\xi$, so, the gauge transformation itself in this case depends on the LV parameter. We note that, unlike the cases of scalar and spinor fields, this action strongly differs from the QED with an additive aether term (\ref{aether}) which is invariant under usual gauge transformations $\delta A_{\mu}=\pa_{\mu}\xi$. The one-loop K\"{a}hlerian effective potential for a super-QED deformed in this way was calculated also in \cite{SUSY2013}. The generalization of these results for a non-Abelian supergauge theory can be easily developed.

\section{Lorentz symmetry breaking through introducing extra superfields}

One more approach to implement Lorentz symmetry breaking in superfield theories is based on introducing Lorentz breaking parameters through some (extra) superfields. Initially this idea was proposed in \cite{HelSUSY}, and it got further development in \cite{HelSUSY2,HelSUSY3}.

The key idea of this approach is the following one. Let us consider the gauge superfield \index{gauge superfield} whose component form, in the gauge sector, is (see e.g. \cite{BK0}):
\bea
V(x,\theta,\bar{\theta})=\bar{\theta}^{\dot{\alpha}}\sigma^{\mu}_{\dot{\alpha}\alpha}\theta^{\alpha}A_{\mu}(x)+\ldots,
\eea
where $A_{\mu}$ is the usual vector gauge field, $\theta_{\alpha},\bar{\theta}_{\dot{\alpha}}$ are the Grassmannian coordinates, and dots are for terms irrelevant for our purposes.
The corresponding Abelian superfield strength $W_{\alpha}=-\frac{1}{4}\bar{D}^2D_{\alpha}V$, in a purely vector sector, has the following component structure:
\bea
W_{\alpha}(x,\theta,\bar{\theta})=(\sigma^{\mu\nu})_{\alpha}^{\phantom{\alpha}\beta}\theta_{\beta}F_{\mu\nu}-\frac{i}{2}(\sigma^{\mu\nu})_{\alpha}^{\phantom{\alpha}\beta}(\sigma^{\lambda})_{\beta\dot{\alpha}}\theta^2\bar{\theta}^{\dot{\alpha}}\pa_{\lambda}F_{\mu\nu}+\ldots.
\eea
Then, let us consider the chiral scalar superfield $S$ \index{chiral superfield} (i.e. $\bar{D}_{\dot{\alpha}}S=0$):
\bea
S(x,\theta,\bar{\theta})=s(x)+\ldots,
\eea
with $s(x)$ is a scalar, and the lower component of the corresponding antichiral field is $s^*(x)$. We suggest the superfield $S$ to be purely external, displaying no dynamics. 

Let us extend the super-QED described by the usual supersymmetric Maxwell action (cf. \cite{BK0})
\bea
S_{Maxw}=\frac{1}{4}\int d^6z W^{\alpha}W_{\alpha},
\eea
through adding the following term:
\bea
S_{odd}=\int d^8z (SW^{\alpha}D_{\alpha}V+\bar{S}\bar{W}_{\dot{\alpha}}\bar{D}^{\alpha}V).
\eea
In components, one has (cf. \cite{HelSUSY}):
\bea
S_{odd}=\int d^4x\Big(-\frac{1}{2}(s+s^*)F_{\mu\nu}F^{\mu\nu}+\frac{i}{2}\pa_{\mu}(s-s^*)\epsilon^{\mu\nu\lambda\rho}F_{\nu\lambda}A_{\rho}\Big)+\ldots.
\eea
Here the dots are for other terms which do not contribute to the purely gauge sector of this theory.
We can choose $s(x)=-ik_{\mu}x^{\mu}$, and, consequently, $s^*(x)=ik_{\mu}x^{\mu}$, with $k_{\mu}$ is a constant vector. Therefore, $s+s^*=0$, and we rest with 
\bea
S_{odd}=\int d^4x\epsilon^{\mu\nu\lambda\rho}k_{\mu}F_{\nu\lambda}A_{\rho}+\ldots.
\eea
As a result, we see that the CFJ term is the only purely gauge contribution to the $S_{odd}$ (actually, it turns out to be that the gauge invariance of $S_{odd}$ in the component form is much more transparent than in the superfield form). Further, in \cite{HelSUSY2}, the dispersion relations for the theory with such an additive term were studied in different sectors. It is interesting to notice that, up to now, namely this approach is the only one allowing to generate the supersymmetric extension of the CFJ term, since the Kostelecky-Berger approach discussed in the previous section is essentially CPT even, and the approach based on applying extra derivatives to superfields, which we discuss in the next section, can only generate the higher-derivative terms for the ``main'' component of the corresponding superfield, that is, for the scalar $s$ within the chiral superfield and the vector $A_{\mu}$ within the real one.

Further, the CPT-even term can also be introduced within this method. It was shown in \cite{HelSUSY3} that, if we for the same superfields $W_{\alpha}$ and $S$ as above (and the conjugated ones) consider the additive term
\bea
S_{even}=\int d^8z (D^{\alpha}S) W_{\alpha}(\bar{D}_{\dot{\alpha}}\bar{S})\bar{W}^{\dot{\alpha}},
\eea
after projecting this action to components we arrive at \cite{HelSUSY3}
\bea 
S_{even}=\int d^4x\Big[
-4F_{\mu\nu}F^{\mu\nu}\pa^{\lambda}s\pa_{\lambda}s^*-8F^{\mu\lambda}F_{\nu}^{\phantom{\nu}\lambda}(\pa_{\mu}s\pa_{\nu}s^*+\pa_{\mu}s^*\pa_{\nu}s)
\Big].
\eea
For $s(x)=ik_{\mu}x^{\mu}$ and $s^*(x)=-ik_{\mu}x^{\mu}$, we arrive at
\bea 
S_{even}=\int d^4x\Big[
-4k^2F_{\mu\nu}F^{\mu\nu}-16k^{\mu}k^{\nu}F_{\mu\lambda}F_{\nu}^{\phantom{\nu}\lambda}
\Big],
\eea
where the first term replays the Maxwell term with the extra multiplier depending on the square of the LV vector, and the second one -- the CPT-even contribution introduced in \cite{aether}.
We see that the aether-like structure arises  naturally within this method. In \cite{HelSUSY3}, some applications of this term were studied.
It is natural to expect that within this approach some other LV terms can arise as well. Also, a natural problem consists in searching for the possibility of emerging of such a term as a quantum correction from an appropriate coupling. Up to now this scenario never was demonstrated.

\section[Straightforward breaking]{Straightforward Lorentz symmetry breaking in a superfield action}

The third manner to break the Lorentz symmetry in superfield theories is based on straightforward adding terms proportional to constant LV vectors (tensors) to the superfield action, it was firstly proposed in \cite{Bolokhov}.  The simplest way to do it consists in consideration of additive terms like
\bea
\label{deltahor}
\delta S=\int d^8z k^{\mu\nu}\pa_{\mu}\Phi\pa_{\nu}\bar{\Phi},
\eea
with the analogous extensions are possible as well for chiral Lagrangians (some examples will be given further) and for terms depending on gauge superfields. 
Nevertheless, the examples for LV extensions of superfield theories we consider here differ from those ones proposed in \cite{Bolokhov}, moreover, unlike that paper, here we concentrate on calculating the one-loop effective potential. It is clear that within this approach the supersymmetric LV terms essentially involve higher derivatives, e.g. for the expression above, one gets terms like $\int d^4x k^{\mu\nu}\pa_{\mu}\phi\Box\pa_{\nu}\phi$. However, for an appropriate choice of the LV parameters, in this case it is $k^{00}=k^{0i}=k^{i0}=0$, one can avoid the presence of the higher time derivatives, and, hence,  eliminate the ghosts, so, in a certain sense in this case we can speak about an attempt of supersymmetric extension of Horava-Lifshitz-like theories introduced in \cite{Horava:2009uw}. 

Our first example will be the theory proposed in \cite{HorSUSY}:
\bea
\label{sfigeral}
S=\int d^8z\Phi(1+\rho\Delta^{z-1})\bar{\Phi}+(\int d^6z W(\Phi) +h.c.),
\eea
with $\rho$ is some constant characterizing the energy scale at which higher derivatives become relevant, and the number $z\geq 2$ (typically, integer), in analogy with \cite{Horava:2009uw}, is called the critical exponent. It is clear that for $z=2$, the additive LV term reproduces the expression (\ref{deltahor}), with $k^{00}=k^{0i}=k^{i0}=0$ and $k^{ij}=-\rho\delta^{ij}$. We note that for any $z$, this LV term is CPT-even. One can consider $W(\Phi)=\frac{m}{2}\Phi^2+\frac{\lambda}{3!}\Phi^3$ as in the Wess-Zumino model, but it is not mandatory.

One can obtain the component structure of this action:
\bea
\label{compwz}
S&=&\int d^4x \left[\bar{\phi}\, \Box (1+\rho\Delta^{z-1}) \phi-i\bar{\psi}^{\dot{\alpha}}\sigma^m_{\dot{\alpha}\alpha}\partial_m(1+\rho\Delta^{z-1})\psi^{\alpha}+\right.\\&+&\left.\bar{F}(1+\rho\Delta^{z-1})F-
\left(\frac{m}{2}(\phi F+\frac{1}{2}\psi^{\alpha}\psi_{\alpha})+\frac{\lambda}{2}(\phi^2F+\frac{1}{2}\phi\psi^{\alpha}\psi_{\alpha}+h.c.\right)
\right].\nonumber
\eea
We see that in the scalar sector, one has two time derivatives and, maximally, $2z$ spatial ones, so, the critical exponent for the scalar component is equal to $z$. At the same time, the critical exponent for the spinor field is $2z-1$, i.e. the critical exponents for different components of the superfield do not coincide. 

The propagators in this theory look like
\bea
&&<\Phi(z_1)\bar{\Phi}(z_2)>=i\frac{1+\rho\Delta^{z-1}}{\Box(1+\rho\Delta^{z-1})^2-m^2}\delta^8(z_1-z_2);\\
&&<\Phi(z_1)\Phi(z_2)>=-
\frac{im}{\Box(1+\rho\Delta^{z-1})^2-m^2}(-\frac{D^2}{4\Box})\delta^8(z_1-z_2).\nonumber
\eea
As usual, $\Phi(z_1)\Phi(z_2)>=<\bar{\Phi}(z_1)\bar{\Phi}(z_2)>^*$.
It is not difficult to show that for $z\geq 2$ the theory is finite. 

To find the K\"{a}hlerian effective potential, we proceed just as in the Sec. 5.1: consider the same sequence of the supergraphs given by Fig. 5.1, again we incorporate the mass into the background field $\Psi=-W^{\prime\prime}$, and modify the propagators to be
\bea
&&<\Phi(z_1)\bar{\Phi}(z_2)>=\frac{i}{\Box(1+\rho\Delta^{z-1})}\delta^8(z_1-z_2);
\nonumber\\
&&<\Phi(z_1)\Phi(z_2)>=0.
\eea
Doing the D-algebra transformations, Wick rotation and summation as in the Sec. 5.1 together with the replacement $\rho\to \rho(-1)^{z-1}$, we arrive at
\bea
\label{k1min}
K^{(1)}=-\frac{1}{2}\int d^4\theta \int\frac{dk_{0E}d^3\vec{k}}{(2\pi)^4}\frac{1}{k^2}\ln\left[1+\frac{\Psi\bar{\Psi}}{k^2(1+\rho(\vec{k}^2)^{z-1})^2}
\right].
\eea
Unfortunately, this integral cannot be evaluated exactly. Using the scheme of approximated computations based on disregarding of subleading orders in $\vec{k}^2$, we find
\cite{HorSUSY}:
\bea
\label{epmin}
K^{(1)}=\frac{1}{12\pi}\csc(\frac{\pi}{z})
(\frac{4\rho}{3})^{-1/z} \int d^4\theta (\Psi\bar{\Psi})^{1/z}.
\eea
This expression displays singularity at $z\to 1$ as it should be, since the Lorentz-invariant case, where the theory displays a divergence, corresponds namely to this value of the critical exponent. 

If we introduce the similar deformation into the chiral effective potential, we have the action
\bea
\label{sfields}
S=\int d^8z\Phi\bar{\Phi}+\left[\int d^6z (\frac{1}{2}\Phi(m+a(-\Delta)^z)\Phi+\frac{\lambda}{3!}\Phi^3)+h.c.\right].
\eea
Proceeding in a similar way as above, after the Wick rotation we have the one-loop K\"{a}hlerian effective potential in the form
\bea
K^{(1)}=-\frac{1}{2}\int d^4\theta\int\frac{dk_{0E}d^3\vec{k}}{(2\pi)^4}\frac{1}{k^2}
\ln\left(
1+\frac{(\Psi+a\vec{k}^{2z})(\bar{\Psi}+a\vec{k}^{2z})}{k^2_{0E}+\vec{k}^2}
\right),
\eea
and, under the same approximate manner, we get
\bea
K^{(1)}=-\frac{1}{8\pi a^{1/z}}\csc(\frac{\pi}{2z})\int d^4\theta (\Psi\bar{\Psi})^{1/2z}.
\eea 
In this case the singularity occurs at $z=1/2$. However, this is not unusual since namely for $z=1/2$ in this theory spatial and time derivatives in UV leading terms enter the theory in the same order making the behaviour of loop integrals to be similar to the Lorentz-invariant case. Coupling of the chiral matter to gauge fields yields the analogous results.

A slightly different approach within this line is based on use of Myers-Pospelov-like terms (see Section 2.1) where some derivatives are contracted to constant LV vectors (tensors) whose specific choice allows to avoid arising the higher time derivatives. In this case we start with the following action \cite{CMP}:
\bea
\label{sfigeral1}
S&=&\int d^8z[\Phi(1-\frac{1}{\Lambda^2}(n\cdot\pa)^2)\bar{\Phi}+\phi\bar{\phi}]+
\nonumber\\ &+&
\left[\int d^6z (\frac{M}{2}\Phi^2+\frac{1}{2}\lambda\Phi\phi^2+\frac{1}{2}f\phi\Phi^2) +h.c.\right].
\eea
Here the $\phi$ is a light (massless) chiral superfield which we treat as a purely background one, and $\Phi$ is a heavy superfield which we assume to be a purely quantum one. Here $M$ is a large mass, and $\Lambda$ is an energy scale at which the higher derivatives become important (from phenomenological estimations \cite{CMP}, it is reasonable to choose $\Lambda$ to be the Planck mass, and $M$ to be a characteristic string mass, so, $M\simeq 10^{-2}\Lambda$). The $n^{\mu}$ is a dimensionless LV vector, in the Euclidean space $n^{\mu}n_{\mu}=1$.
After summation of the same graphs depicted at Fig. 5.1, we find the one-loop K\"{a}hlerian effective potential to be
\bea
K^{(1)}=-\frac{1}{2}\int d^4\theta\int\frac{d^4k}{(2\pi)^4}\frac{1}{k^2}\ln\left[k^2\left(1+\frac{(n\cdot k)^2}{\Lambda^2}\right)^2+|\Psi|^2
\right].
\eea
This integral can be calculated approximately. Since, as we already said, $M\simeq 10^{-2}\Lambda$, we can expand this expression in series in $M/\Lambda$ and $1/M$. As a result, we find
\bea
K^{(1)}=\frac{\lambda^2}{32\pi^2}\int d^4\theta\phi\bar{\phi}[3+\ln\frac{M^2}{4\Lambda^2}]+\ldots,
\eea 
where dots are for suppressed terms. We see that in this case the contributions arising due to the Lorentz symmetry breaking are significant, so, we observe the effect of large quantum corrections. However, this is rather natural since the LV term here effectively plays the role of the higher-derivative regularization. 
The same situation occurs within the first approach described in this section. Indeed, if we compare this result with (\ref{epmin}) with rescaling $\rho=\frac{\alpha}{\Lambda^{2z-2}}$, in order to have a dimensionless $\alpha$ and a scale $\Lambda$, we see that the result (\ref{epmin}) is proportional to $(\frac{M}{\Lambda})^{\frac{2}{z}-2}$, thus, for $z\geq 2$ the quantum correction in this case is also not suppressed, being large instead of this. To close this section, we  conclude that within this ``straightforward'' approach, both the Horava-Lifshitz-like and, for a space-like $n_{\mu}$, the Myers-Pospelov-like theories display no ghosts being therefore perturbatively consistent.

\section{Conclusions}

Now, let us compare the results obtained within different approaches aimed to introduce Lorentz symmetry breaking in superfield theories. First of all, it is interesting to note that all these ways allow for introducing CPT-even terms into classical actions, and, further, into quantum corrections, whereas only the way based on introducing of the extra superfield \cite{HelSUSY} can yield CPT-odd expressions, that is, first of all, the CFJ term. Up to now, no other manner to construct a supersymmetric extension of the CFJ term is known.

Second, a nontrivial fact consists in a possibility to generate a nontrivial geometry from quantum corrections. A discussion on this fact is presented also in \cite{ColMac1} where it is noted that some LV modifications of Dirac matrices can be defined as well. However, an open question is whether one can, starting from some LV superfield gauge theory, obtain the action of the gauge field coupled to the new geometry consistently, so that not only derivatives, but also vector component fields would be contracted to a new metric.  Another possible line of studies could consist in a generalization of this approach for the case of a space-time with a non-zero curvature through suggesting the tensor $k_{\mu\nu}$ to be a fixed function of space-time coordinates rather than a constant. However, in this case the deformed supersymmetry algebra will clearly be much more complicated, and other difficulties, intrinsic for studies of Lorentz symmetry breaking in a curved space-time, which will be discussed in the next chapter of the book, will arise as well.

Third, the problem of consistent supersymmetric extension of Horava-Lifshitz-like theories continues to be open. The reason is that usual superfield models always involve lower (second) spatial derivatives in a kinetic term, therefore, one cannot arrive at a theory involving only terms with two time derivatives and just $2z$ spatial derivatives, while implementing of higher spatial derivatives into a spinor supercovariant derivative clearly breaks the Leibnitz rule. Some attempts to proceed in this case are presented in \cite{Quei}, however, this methodology is still under development.

It must be mentioned that there is one more manner to implement Lorentz symmetry breaking specific for supersymmetric field theories, that is, the fermionic noncommutativity \cite{Sei}. Following this approach, the anticommutation relations between fermionic coordinates are deformed as $\{\theta^{\alpha},\theta^{\beta}\}=\Sigma^{\alpha\beta}$ where $\Sigma^{\alpha\beta}$ is a constant symmetric matrix clearly breaking the Lorentz symmetry (e.g., in a three-dimensional space-time this matrix is equivalent to the constant vector $\Sigma^{\mu}=\frac{1}{2}(\gamma^{\mu})_{\alpha\beta}\Sigma^{\alpha\beta}$). However, this methodology is known to meet its own difficulties. We close this section with a conclusion that the problem of construction of a consistent supersymmetric extension for LV theories is still open.

\chapter{Lorentz and CPT symmetry breaking in gravity}

In this chapter we describe some first steps in study of CPT and/or Lorentz breaking extensions of gravity. 
Implementation of the Lorentz symmetry breaking within the gravity context is certainly one of the most complicated issues within general studies of LV theories, facing many difficulties. The most important reason for these difficulties is the following one. In the curved space-time, the symmetry group is that one of general coordinate transformations like $x^{\mu}=x^{\mu}(x')$, which, at the same time, represents itself as the extension both of the Lorentz group and of the gauge group. As we noted in previous chapters, in many cases the Lorentz symmetry breaking implies the CPT symmetry breaking as well, whereas the gauge symmetry is not broken, the paradigmatic example is the CFJ term. Therefore, it is natural to require for the most interesting CPT, and in certain cases Lorentz breaking extensions of gravity to be consistent with the general coordinate (that is, gauge) invariance, and there is only a few known examples where this consistency is possible. The most important of such examples, is the four-dimensional Chern-Simons modified gravity which we discuss in this chapter. Another possible approach could consist in consideration of the \index{weak (linearized) gravity} weak (linearized) gravity limit, where we consider only the dynamics of the symmetric metric fluctuation tensor $h_{\mu\nu}$, and we can apply the methods similar to those ones used for studies of LV extensions of QED.  In this chapter we consider both these approaches within the Riemannian framework. 

Besides of this, we note that introducing Lorentz symmetry breaking in a curved space-time faces one difficulty more. As we noted in previous chapters, in a flat space-time the Lorentz symmetry can be violated explicitly through introducing new terms proportional to constant vectors (tensors) which cannot be introduced consistently in a non-zero curvature case. Indeed, let us consider for example a constant vector $k^{\mu}$. In the flat space-time it satisfies the condition $\partial_{\nu}k^{\mu}=0$, but this condition evidently breaks general covariance. A straightforward covariant generalization of this condition, looking like $\nabla_{\nu}k^{\mu}=0$, imposes additional restrictions on space-time geometry (so-called no-go constraints), which are hard to satisfy in general (see the discussion in \cite{KosLiGrav}). And if no such conditions are imposed, one faces an infinite tower of terms proportional to $\nabla_{\nu_1}\ldots\nabla_{\nu_n}k^{\mu}$ which makes all calculations to be extremely complicated (some lower terms of this form are discussed in \cite{ShaBerr}, where renormalization of LV QED in a curved space-time is performed). Therefore, the most promising way of breaking Lorentz symmetry in the presence of gravity is based on spontaneous Lorentz symmetry breaking, instead of the explicit one. The Einstein-aether and bumblebee models allowing to break Lorentz symmetry spontaneously on a curved background, will also be discussed in this chapter.

\section[Motivations for 4D gravitational CS term]{Motivations for $4D$ gravitational Chern-Simons term}

We start our study of LV modifications in gravity with the paradigmatic example of the additive CPT-, and in certain cases Lorentz-breaking term, consistent at the same time with the gauge symmetry, that is,  the four-dimensional gravitational CS term. It represents itself as a natural generalization of the three-dimensional Lorentz-invariant gravitational CS term introduced in \cite{DJT} and looking like
\bea
\label{acsg}
S_{CS}&=&\frac{1}{2\mu\kappa^2}\int d^3x\epsilon^{\mu\nu\alpha}\Big[(\pa_{\mu}\omega_{\nu ab})\omega_{\alpha}^{\phantom{\alpha}ab}+\frac{2}{3}\omega_{\mu a}^{\phantom{\alpha a}c}\omega_{\nu cb}\omega_{\alpha}^{\phantom{\alpha}ab}
\Big],
\eea
where $\mu$ is a mass, $\kappa^2$ is a gravitational constant (of mass dimension $-1$, in $3D$), and $\omega_{\mu}^{\phantom{\mu}ab}$ is a connection. Here and further in this section, we follow notations adopted in \cite{ptime}, so, the Greek indices are for the curved space-time, while the Latin indices are for the tangent one. As our geometry is assumed to be Riemannian, the connection can be expressed in terms of vielbeins or metrics through well-known expressions. One can straightforwardly verify that the action (\ref{acsg}) is gauge invariant. We note that there is no $\sqrt{|g|}$ factor in this integral since the invariant Levi-Civita tensor in a curved space is $\frac{\epsilon^{\mu\nu\alpha}}{\sqrt{|g|}}$ rather than $\epsilon^{\mu\nu\alpha}$.

Introducing a constant (pseudo)vector $b_{\mu}$, it is easy to generalize the CS action to the four-dimensional case. From the formal viewpoint, this generalization is similar to promoting the usual CS term to the CFJ term through the replacement $\epsilon^{\mu\nu\lambda}\to b_{\rho}\epsilon^{\rho\mu\nu\lambda}$ corresponding to adding of one more dimension.
As a result, treating $\omega_{\mu}^{\phantom{\mu}ab}$ as a Riemannian connection, we arrive at the $4D$ CS action (we note that in four dimensions, no $\mu$ multiplier is needed):
\bea
\label{acsg1}
S_{CS}&=&\int d^4x\epsilon^{\rho\mu\nu\alpha}b_{\rho}\Big[(\pa_{\mu}\omega_{\nu ab})\omega_{\alpha}^{ab}+\frac{2}{3}\omega_{\mu a}^{c}\omega_{\nu c}^b\omega_{\alpha b}^a
\Big].
\eea
Originally, this term has been introduced in \cite{JaPi}, and in this section we review some results obtained in \cite{JaPi}. To prove gauge invariance of this term, that is, its invariance under general coordinate transformations, one can write down its equivalent form: indeed, we can define the vector
\bea
K^{\rho}=\epsilon^{\rho\mu\nu\alpha}\Big[(\pa_{\mu}\omega_{\nu ab})\omega_{\alpha}^{ab}+\frac{2}{3}\omega_{\mu a}^{c}\omega_{\nu c}^b\omega_{\alpha b}^a
\Big], 
\eea
so that $\pa_{\mu}K^{\mu}=\frac{1}{2}{}^*R R$, where
\bea
{}^*R R=\frac{1}{2}\epsilon^{\alpha\beta\gamma\delta}R_{\alpha\beta}^{\phantom{\alpha\beta}\mu\nu}R_{\gamma\delta\mu\nu}
\eea
is \index{Pontryagin term} a pseudoscalar Pontryagin density. So, suggesting that the vector $b_{\mu}$ is expressed as $b_{\mu}=\pa_{\mu}\vartheta$, where $\vartheta$ is a pseudoscalar called the CS coefficient, which, for the first step, is treated as an external function, one can rewrite the $4D$ gravitational CS term as
\bea
S_{CS}&=&\frac{1}{4}\int d^4x\vartheta {}^*R R.
\eea
We see immediately that this term is both Lorentz and gauge invariant, but parity-breaking. If we require the $\vartheta$ to be linear in coordinates, just as we did in the section 3.3 during the study of three-dimensional CS terms, that is, $\vartheta=b_{\mu}x^{\mu}$, with $b_{\mu}$ be constant, after a simple integration by parts we arrive at the particular Lorentz-breaking CS term (\ref{acsg1}).  We note that this is the lower gauge-invariant CPT-Lorentz breaking term in gravity consistent with the gauge invariance requirement, and other possible CPT-Lorentz breaking gauge-invariant terms for the gravity would necessarily involve higher orders in derivatives.

The complete action of the CS modified gravity involving both the usual Einstein-Hilbert term and the CS term, looks like \cite{JaPi}:
\bea
\label{CSMG}
S_{CSMG}=\frac{1}{16\pi G}\int d^4x(\sqrt{|g|}R-\frac{1}{2}b_{\mu}K^{\mu}).
\eea
Comparing the theory (\ref{CSMG}) and the LV electrodynamics with the CFJ term (\ref{lvqed}), we can conclude that the $4D$ gravitational CS term is related with the gravitational anomalies originally introduced in \cite{AlvWitt}, without any concerning of the Lorentz symmetry breaking, in the manner similar to that one relating the CFJ term with the Adler-Bell-Jackiw anomaly (see the discussion of this correspondence in \cite{JackAmb}).

The equations of motion for this theory also were obtained originally in \cite{JaPi}, they have the form
\bea
\label{emot}
G_{\mu\nu}+C_{\mu\nu}=-8\pi G T_{\mu\nu},
\eea
where $T_{\mu\nu}$ is the energy-momentum tensor of a matter, $G_{\mu\nu}=R_{\mu\nu}-\frac{1}{2}Rg_{\mu\nu}$ is the usual Einstein tensor, and $C_{\mu\nu}$ is a Cotton tensor \index{Cotton tensor} defined as
\bea
C^{\mu\nu}=-\frac{1}{2\sqrt{|g|}}\epsilon^{\sigma\mu\alpha\beta}\Big[b_{\sigma}\nabla_{\alpha}R^{\nu}_{\beta}+b_{\sigma\tau}R^{\tau\nu}_{\phantom{\tau\nu}\alpha\beta}
\Big]+(\mu\leftrightarrow\nu).
\eea
Here $\nabla_{\alpha}$ is a covariant derivative, and $b_{\mu}=\nabla_{\mu}\vartheta$, $b_{\mu\nu}=\nabla_{\mu}\nabla_{\nu}\vartheta$ are constructed on the base of the CS coefficient.

We can find the divergence of the modified Einstein equations (\ref{emot}). In the r.h.s. we obtain zero due to the conservation of the energy-momentum tensor of the matter, then, the identity $\nabla_{\mu}G^{\mu\nu}=0$ continues to be valid because of the  Bianchi identities. Finally we arrive at
\bea
\nabla_{\mu}C^{\mu\nu}=\frac{1}{8\sqrt{|g|}}b^{\nu}{}^*R R.
\eea
This is a constraint for the possible solutions of Eq. (\ref{emot}). In many relevant cases, e.g. spherically symmetric metrics, one has ${}^*R R=0$, so, the Cotton tensor will be conserved (see e.g. \cite{Grumiller} for a general discussion).

One can suggest the CS coefficient to be not an external object but a dynamical field. In this case one starts with the action of the dynamical CS modified gravity \cite{Grumiller}:
\bea
S_{DCSMG}=\frac{1}{16\pi G}\int d^4x(\sqrt{|g|}R+\frac{1}{4}\vartheta{}^*RR+\frac{1}{2}\sqrt{|g|}\nabla_{\mu}\vartheta\nabla^{\mu}\vartheta).
\eea
In this case, one finds an additional contribution to the energy-momentum tensor of the matter, that is, the energy-momentum tensor of the scalar field $\vartheta$. 

It remains to write down the linearized form of the gravitational CS term. Suggesting that the metric looks like $g_{\mu\nu}=\eta_{\mu\nu}+h_{\mu\nu}$, where $h_{\mu\nu}$ is a small metric fluctuation, and relabeling $b_{\mu}$ by $v_{\mu}$, we have \cite{JaPi,gravCS}:
\bea
\label{gravcsterm}
S_{CS}=\frac{1}{4}\int d^4x h^{\mu\nu}v^{\lambda}\epsilon_{\alpha\mu\lambda\rho}\pa^{\rho}(\Box h^{\alpha}_{\nu}-\pa_{\nu}\pa_{\gamma}h^{\gamma\alpha}).
\eea
We see that the gravitational CS term involves higher derivatives. However, it was proved in \cite{JaPi} that for physical degrees of freedom there is no higher time derivatives in the corresponding equation of motion, hence, neither unitarity nor causality are violated.

So, let us discuss some aspects relating to this term, namely, its perturbative generation and the ambiguity of the result.

\section{Perturbative generation of the gravitational CS term}

Now, let us present some schemes to generate the gravitational CS term in four dimensions. The first of these schemes has been proposed in \cite{gravCS}.
The starting point of the calculation is the following classical action of the spinor field $\psi$ on a curved background:
\bea
\label{full}
S=\int d^4x e e^{\mu}_a\bar{\psi}\Big(\frac{1}{2}i\gamma^a\stackrel{\leftrightarrow}{\pa}_{\mu}+
\frac{1}{4}i\omega_{\mu cd}\Gamma^{acd}-b_{\mu}\gamma^a\gamma_5-m
\Big)\psi.
\eea
Here $\Gamma^{acd}=\frac{1}{6}\gamma^a\gamma^c\gamma^d+\ldots$ is the antisymmetrized product of three Dirac matrices. 
The corresponding one-loop effective action of the gravitational field can be cast as
\bea
\label{full1}
\Gamma^{(1)}=-i{\rm Tr}\ln(i\ds-m-\bs\gamma_5+\os).
\eea
To consider the weak field approximation, we write $g_{\mu\nu}=\eta_{\mu\nu}+h_{\mu\nu}$, so, the vielbein is $e^{\mu}_a=\delta^{\mu}_a+\frac{1}{2}h^{\mu}_a$, and $e=1+\frac{1}{2}h^a_a$ (again, we note that Greek indices are for the curved space-time, and Latin ones -- for the tangent one). So, omitting the irrelevant terms, in particular, those ones proportional to $h^{\mu}_{\mu}$, we get the action of $\psi$ coupled to $h_{\mu\nu}$:
\bea
\label{gravlin}
S=\int d^4x \bar{\psi}\Big(\frac{1}{2}i\Gamma^{\mu}\stackrel{\leftrightarrow}{\pa}_{\mu}+
h_{\mu\nu}\Gamma^{\mu\nu}-b_{\mu}\gamma^{\mu}\gamma_5-m
\Big)\psi,
\eea
where $\Gamma^{\mu}=\gamma^{\mu}-\frac{1}{2}h^{\mu\nu}\gamma_{\nu}$, and $\Gamma^{\mu\nu}=\frac{1}{2}b^{\mu}\gamma^{\nu}\gamma_5-\frac{i}{16}(\pa_{\rho}h_{\alpha\beta})\eta^{\beta\nu}\Gamma^{\rho\mu\alpha}$.

The nontrivial contributions to the two-point function of the metric fluctuation $h_{\mu\nu}$ are given by the Feynman diagrams depicted at Fig. 3.4, the only difference is that now the external lines are for the metric fluctuation. It is clear that the quartic vertex and that one involving the $b^{\mu}h_{\mu\nu}$ contraction evidently will not yield CS-like results. 

These Feynman diagrams are superficially divergent. However, after long and very involved calculations described in \cite{gravCS}, adopting a special calculation scheme based on 't Hooft-Veltman prescription \cite{HofVel}, we find that all divergences cancel out, and, in the zero mass limit, where possible divergent contributions of the above mentioned Feynman diagrams vanish, arrive at the finite result
\bea
\label{lincs}
S_{CS}=\frac{1}{192\pi^2}\int d^4x h^{\mu\nu}b^{\lambda}\epsilon_{\alpha\mu\lambda\rho}\pa^{\rho}(\Box h^{\alpha}_{\nu}-\pa_{\nu}\pa_{\gamma}h^{\gamma\alpha}),
\eea
that is, the vector $v^{\mu}$ (\ref{gravcsterm}) is expressed as $v^{\mu}=\frac{1}{48\pi^2}b^{\mu}$. 
So, this result appears to be unique, at least within this approach. However, this is not sufficient to conclude whether the gravitational CS term we obtained is indeed unambiguous in general. First of all, we should emphasize that the 't Hooft-Veltman prescription adopted within \cite{gravCS} is nothing more that a fixing of the regularization scheme so that in this case one obtains a definite result. Besides, as we observed in the Chapter 3, the finiteness of a superficially divergent result typically signalizes about its ambiguity.

To demonstrate that our result is indeed ambiguous, we use another manner of calculating the gravitational CS term based on the proper-time method which was for the first time carried out in \cite{ptime}. In this case, we can consider the complete action of the spinor on a curved background (\ref{full}), thus avoiding use of the weak field approximation, so, we can obtain the full-fledged CS term.

To compute the CS term, we can put $e\simeq 1$. This approximation is sufficient for obtaining the gravitational CS term since all higher terms in expansion of $e$ will be irrelevant for our calculations. Then, in order to complete the operator whose trace we study, up to the quadratic one, as it is required by the proper-time technique (cf. Section 3.3), we add to the one-loop effective action (\ref{full1}) the constant of the form 
\bea
C_0=-i{\rm Tr}\ln(i\ds+m+\bs\gamma_5).
\eea
As a result, after multiplication of determinants with use of the fact that $\det AB=\det A\det B$, we arrive at the following expression for the one-loop effective action \cite{ptime}:
\bea
\Gamma^{(1)}&=&-i {\rm Tr}\ln\Big[
-\Box+i\os\ds+m\omega-m^2+(\os-2m)\bs\gamma_5+\nonumber\\ &+& 2i(b\cdot\pa)\gamma_5-b^2
\Big].
\eea
Now, we expand this expression up to the first order in the LV vector $b^{\mu}$ and employ the Schwinger proper-time representation $T^{-1}=\int_0^{\infty}ds e^{-sT}$.

Afterward, the divergent parts easily can be found to cancel. Then, the finite contribution of the second order in connections $\omega$ looks like
\bea
\label{SG2fin}
S^{(2)}_{fin} &=& {\rm tr} \int d^4x\int_0^{\infty}ds\,e^{-sm^2}\left[ s^2m^2\SLASH{\omega}(\SLASH{\partial}\SLASH{\omega})\Slash{b}\gamma_5+2m^2s^3\SLASH{\omega}\SLASH{\partial}(\partial_\alpha\SLASH{\omega})\partial^\alpha\Slash{b}\gamma_5\right. \nonumber\\
&&\left.+2m^2s^3\SLASH{\omega}(\partial_\alpha\SLASH{\omega})\partial^\alpha\SLASH{\partial}\Slash{b}\gamma_5\right]e^{-s\Box}\delta(x-x')\big|_{x'=x}.
\eea
To perform calculations, we employ the following representation of the geodesic bi-scalar $\sigma(x,x')$ \cite{Ojima}:
\bea
\delta(x-x') = \int\frac{d^4k}{(2\pi)^4} e^{ik_\alpha \nabla^\alpha\sigma(x,x')}.
\eea
Calculating the trace (see details in \cite{ptime}), we arrive at
\bea\label{SG2fin2}
S^{(2)}_{CS} &=& \frac i4\int d^4x\int_0^{\infty}ds\,e^{-sm^2} b_\mu\,\omega_{\nu ab}\,\partial_\lambda\omega_{\rho cd}
\int\frac{d^4k}{(2\pi)^4}e^{sk^2}s^2m^2\epsilon^{\mu\nu\lambda\rho}(g^{ac}g^{bd}-g^{ad}g^{bc}) \nonumber\\
&=& \frac1{32\pi^2}\int d^4x \epsilon^{\mu\nu\lambda\rho}b_\mu\partial_\nu\omega_{\lambda ab}\,{\omega_\rho}^{ba},
\eea
with $\epsilon^{\mu\nu\lambda\rho}=e\,{e^\mu}_a{e^\nu}_b{e^\lambda}_c{e^\rho}_d\epsilon^{abcd}$. 

The relevant term of third order in connections $\omega$ is
\bea
S^{(3)}_{fin} &=& i\,{\rm tr} \int d^4x \int_0^{\infty}ds\,e^{-sm^2}\left[\frac{s^2}{2}m^2\SLASH{\omega}\SLASH{\omega}\SLASH{\omega}\SLASH{b}\gamma^5+\right.\nonumber\\
&+&\left.\frac{s^3}{3}m^2\left(\SLASH{\omega}\SSLASH{\nabla}\SLASH{\omega}\SSLASH{\nabla}\SLASH{\omega}+\SLASH{\omega}\SSLASH{\nabla}\SLASH{\omega}\SLASH{\omega}\SSLASH{\nabla}+\SLASH{\omega}\SLASH{\omega}\SSLASH{\nabla}\SLASH{\omega}\SSLASH{\nabla}\right)\Slash{b}\gamma_5\right.\\
&-&\left.\frac{s^3}{3}m^2\left(\SLASH{\omega}\SSLASH{\nabla}\SLASH{\omega}\SLASH{\omega}+\SLASH{\omega}\SLASH{\omega}\SSLASH{\nabla}\SLASH{\omega}+\SLASH{\omega}\SLASH{\omega}\SLASH{\omega}\SSLASH{\nabla}\right)(b\cdot \nabla)\gamma_5-\frac{s^3}{3}m^4\SLASH{\omega}\SLASH{\omega}\SLASH{\omega}\SLASH{b}\gamma^5\right]\times\nonumber\\&\times& e^{-s\Box}\delta(x-x')\big|_{x'=x}.\nonumber
\eea
After calculating the trace we arrive at
\bea
S^{(3)}_{CS} &=& -\frac i{16} \int d^4x\int_0^{\infty}ds\,e^{-sm^2} b_\mu\,\omega_{\nu ab}\,\omega_{\lambda cd}\,\omega_{\rho ef} \int\frac{d^4k}{(2\pi)^4}e^{sk^2}\left(\frac{s^2}2m^2\epsilon^{\mu\nu\lambda\rho}+
\right.\nonumber\\&+&\left.
\frac{s^3}3m^2\epsilon^{\mu\nu\lambda\rho}k^2+
\frac{s^3}3m^2\epsilon^{\alpha\nu\lambda\rho}k_\alpha k^\mu-\frac{s^3}3m^4\epsilon^{\mu\nu\lambda\rho}\right)\times \nonumber\\
&\times&\left[g^{af}(g^{bc}g^{de}\!-\!g^{bd}g^{ce})\!+\!g^{ae}(g^{bd}g^{cf}\!-\!g^{bc}g^{df})\!+
\right.\nonumber\\ &+&\left.
\!g^{ad}(g^{bf}g^{ce}\!-\!g^{be}g^{cf})\!+\!g^{ac}(g^{be}g^{df}\!-\!	g^{bf}g^{de})\right]. \nonumber
\eea
By integrating over the momenta $k$ and the Schwinger parameter $s$, we find
\bea
S^{(3)}_{CS} = -\frac1{48\pi^2}\int d^4x \epsilon^{\mu\nu\lambda\rho} b_\mu\,\omega_{\nu ab}\,{\omega_\lambda}^{bc}\,{\omega_{\rho c}}^a,
\eea
so that, summing this expression with (\ref{SG2fin2}), we find that our complete result yields the gravitational CS term \cite{JaPi}: 
\bea
\label{csfull}
S_{CS} = \frac{1}{32\pi^2} \int d^4x \epsilon^{\mu\nu\lambda\rho}b_\mu\left(\partial_\nu\omega_{\lambda ab}\,{\omega_\rho}^{ba} - \frac23\omega_{\nu ab}\,{\omega_\lambda}^{bc}\,{\omega_{\rho c}}^a\right).
\eea
It is easy to verify that in the weak field approximation this term does not reproduce the value of the numerical coefficient  presented in (\ref{lincs}). We conclude that the four-dimensional gravitational CS term is ambiguous. Unlike the approach developed in \cite{gravCS}, in this case there is no need to use the zero mass limit to obtain our result. 

At the same time, using the methodology based on transformations of the functional integral, developed in \cite{Chung}, we can show that  the gravitational CS term (\ref{csfull}) is accompanied by a completely undetermined constant $C$:
\bea
\label{csfull1}
S_{CS} = C \int d^4x \epsilon^{\mu\nu\lambda\rho}b_\mu\left(\partial_\nu\omega_{\lambda ab}\,{\omega_\rho}^{ba} - \frac23\omega_{\nu ab}\,{\omega_\lambda}^{bc}\,{\omega_{\rho c}}^a\right).
\eea
This constant arises from the modification of a CS conserved current which is totally arbitrary \cite{Chung}. Some more details of this calculation can be found in \cite{ambgrav}.
At the same time, it has been argued in \cite{AltKarki,AltKarki2}, that the result $C=0$, i.e. vanishing of the one-loop contribution to the gravitational CS term, is preferable in a certain sense. Following these arguments, if one assumes $b_{\mu}$ to be a dynamical field, instead of a given vector or a gradient of a given scalar, the gravitational CS term loses its gauge invariance, therefore, the consistent manner of calculations should imply a zero result for it. As we noted in the section 3.4.1, the CFJ term displays the analogous behaviour.

Another interesting discussion of ambiguity of the four-dimensional gravitational CS term has been presented in \cite{Scarpgrav}. In this paper, the methodology of the implicit regularization  \cite{ScarpReg} is applied to generation of the gravitational CS term, where the starting point is the action of the spinors on a weak gravity background (\ref{gravlin}) taken in the zero mass case, $m=0$. Within this scheme of calculations, the self-energy tensor $\Pi^{\mu\nu\alpha\beta}$ is found to be (see \cite{Scarpgrav} for details):
\bea
\Pi^{\mu\nu\alpha\beta}&=&-\frac{i}{8}\epsilon^{\lambda\rho\beta\mu}b_{\lambda}p_{\rho}\Big[
(\frac{i}{48\pi^2}-64\sigma_0-4v_0+4\xi_0)p^{\alpha}p^{\nu}-\\&-&(\frac{i}{48\pi^2}+32\sigma_0)\eta^{\alpha\nu}p^2
\Big]+(\alpha\leftrightarrow\beta)+(\mu\leftrightarrow\nu)+(\alpha\leftrightarrow\beta,\mu\leftrightarrow\nu).\nonumber
\eea
Here $\sigma_0$, $v_0$ and $\xi_0$ are finite but yet undetermined parameters of implicit regularization. The next step consists in imposing the requirement of transversality for this self-energy tensor, $p_{\mu}\Pi^{\mu\nu\alpha\beta}=0$, which implies a requirement for these parameters to satisfy the relation $\xi_0-v_0=24\sigma_0$, so, one has
\bea
\Pi^{\mu\nu\alpha\beta}&=&-\frac{i}{24}\epsilon^{\lambda\rho\beta\mu}b_{\lambda}p_{\rho}
(\frac{i}{16\pi^2}+96\sigma_0)[p^{\alpha}p^{\nu}-\eta^{\alpha\nu}p^2]+\\ &+& 
(\alpha\leftrightarrow\beta)+(\mu\leftrightarrow\nu)+(\alpha\leftrightarrow\beta,\mu\leftrightarrow\nu).\nonumber
\eea
Therefore, we recover the explicit form of the gravitational CS term given by (\ref{gravcsterm}), where $v^{\mu}=(\frac{1}{24\pi^2}-64i\sigma_0)b^{\mu}$. We see that the result unavoidably depends on an arbitrary regularization parameter $\sigma_0$, thus, this calculation is consistent with the above-mentioned fact that the constant accompanying the gravitational CS term is completely undetermined, cf. \cite{ambgrav}. 

To close the discussion of the four-dimensional gravitational CS term, we note again that this term is related with the gravitational anomaly \cite{AlvWitt} just in the same manner as the CFJ term is related with the Adler-Bell-Jackiw anomaly (see \cite{JackAmb} and the Section 3.4 of this book). 

\section{Problem of spontaneous Lorentz symmetry breaking in gravity}

As we noted in the beginning of this chapter, constant vectors (tensors) used to construct LV terms in a flat space-time, in general cannot be defined for a curved one -- in \cite{KosLiGrav}, possible restrictions for geometry arising from requirement for such vectors and tensors to be constant, called no-go constraints, have been discussed in great details, and it was argued that in general, these restrictions cannot be satisfied, perhaps except of some fine-tuned situations. Moreover, these restrictions typically contradict to Bianchi identities and to conservation of the energy-momentum tensor. Hence, the explicit Lorentz symmetry breaking, in general, does not appear to be an appropriate mechanism in gravity (nevertheless, some interesting results have been obtained within this approach as well, see f.e. \cite{Sehic} where conservation laws for nondynamical backgrounds were studied, and \cite{Wen} where some schemes allowing to maintain Bianchi identities in the explicit Lorentz symmetry breaking case were proposed), and one must employ the spontaneous Lorentz symmetry breaking mechanism, within which, the vector (tensor) introducing a privileged space-time direction, is not required to be constant. 
	
	So, let us discuss models used to break the Lorentz symmetry spontaneously in a curved space-time. There are two known theories considered in this context, the Einstein-aether gravity and the bumblebee gravity.
	
	Let us discuss first the \index{Einstein-aether gravity} Einstein-aether gravity. Originally, it was introduced in \cite{Jacobson0}. The action of this theory is \cite{Jacobson}:
	\bea
	\label{eae}
	S=-\frac{1}{16\pi G}\int d^4x\sqrt{|g|}\Big[R+\lambda(u^{\mu}u_{\mu}-1)+K^{\alpha\beta}_{\mu\nu}\nabla_{\alpha}u^{\mu}\nabla_{\beta}u^{\nu}
	\Big],
	\eea
	where $u^{\mu}$ is a dynamical vector satisfying the constraint $u^{\mu}u_{\mu}=1$, so, choosing of a possible value of this vector breaks the Lorentz symmetry spontaneously, the $\lambda$ is the Lagrange multiplier field, the tensor $K^{\alpha\beta}_{\mu\nu}$ looks like
	\bea
	K^{\alpha\beta}_{\mu\nu}=c_1g^{\alpha\beta}g_{\mu\nu}+c_2\delta^{\alpha}_{\mu}\delta^{\beta}_{\nu}+c_3\delta^{\alpha}_{\nu}\delta^{\beta}_{\mu}+c_4u^{\alpha}u^{\beta}g_{\mu\nu},
	\eea
	and $c_1,c_2,c_3,c_4$ are constants.

The corresponding equations of motion are:
\bea
&&g_{\alpha\beta}u^{\alpha}u^{\beta}=1; \quad\, \nabla_{\alpha}J^{\alpha}_{\phantom{a}\mu}-c_4\dot{u}_{\alpha}\nabla_{\mu} u^{\alpha}=\lambda u^{\mu};\nonumber\\
T_{\alpha\beta}&=&-\frac{1}{2}g_{\alpha\beta}{\cal L}_u+\nabla_{\mu}\Big(J^{\alpha}_{\phantom{a}(\mu}u_{\beta)}-J^{\mu}_{\phantom{m}(\alpha}u_{\beta)}-J_{(\alpha\beta)}u^{\mu}
\Big)+\\
&+&c_1[(\nabla_{\mu} u_{\alpha})(\nabla^{\mu}u_{\nu})-(\nabla_{\alpha}u_{mu})(\nabla_{\beta}u^{\mu})]+
c_4\dot{u}_{\alpha}\dot{u}_{\beta}+[u_{\nu}\nabla_{\mu} J^{\mu\nu}-c_4\dot{u}^2]u_{\alpha}u_{\beta}.\nonumber
\eea
Here $\dot{u}^{\mu}=u^{\alpha}\nabla_{\alpha}u^{\mu}$, $J^{\alpha}_{\phantom{a}\mu}=K^{\alpha\beta}_{\mu\nu}\nabla_{\beta}u^{\nu}$, and ${\cal L}_u$ is $u$-dependent part of the Lagrangian. 

Consistency of some known metrics, in particular, cosmological and black hole ones, in this theory has been verified in a number of papers (see e.g. \cite{JacStat} and references therein). It is clear that this theory can be extended through adding different terms where the various degrees of the $u^{\mu}$ vector are coupled to different fields, for example, scalar, spinor and gauge ones, for example, it is possible to implement aether terms discussed in the Chapter 3, like that one for the gauge field, $u^{\alpha}F_{\alpha\mu}u_{\beta}F^{\beta\mu}$. In the flat space limit, the $u^{\mu}$ vectors can (but are not restricted to) be constant ones.
	
In the Einstein-aether theory, the spontaneous Lorentz symmetry breaking is introduced with use of the constraint. At the same time, it is known that presence of constraints makes perturbative studies of such theories more complicated requiring special methodologies like $1/N$ expansion, which, moreover, cannot be applied in our case as the $u^{\mu}$ field possesses only four components. The bumblebee gravity, where the spontaneous Lorentz symmetry breaking is implemented in other manner, that is, through choosing a minimum for some potential, proposed in \cite{KosGra}, is free of this problem. 
The action of the bumblebee gravity looks like
\bea
\label{bumbact}
S=-\int d^4x\sqrt{|g|}\left( \frac{1}{16\pi G}(R+\xi B^{\mu}B^{\nu}R_{\mu\nu}) -\frac{1}{4}B_{\mu\nu}B^{\mu\nu}-V(B^{\mu}B_{\mu}\pm b^2)\right).
\eea
Here $V$ is the potential, typically one can use $V=\frac{\lambda}{4}(B^{\mu}B_{\mu}\pm b^2)^2$, and $b^2>0$ is a constant. The $B_{\mu\nu}=\partial_{\mu}B_{\nu}-\partial_{\nu}B_{\mu}$ is the stress tensor, and $\xi$ is a coupling constant characterizing the magnitude of the non-minimal coupling between the bumblebee field and the Ricci tensor.

Let us briefly review the most important results obtained within studies of the bumblebee gravity. As we have already noted, the main direction of research in various modified gravity theories, including the bumblebee gravity, is study of consistency of known results found within general relativity, especially, cosmological solutions and black holes, within the new modified gravity. First we discuss the solutions originally treated within the bumblebee context in \cite{Bertolami}. In this case, the action of the model is given by (\ref{bumbact}). In this theory, the static spherically symmetric metric was considered and shown to imply a modification of the \index{Schwarzschild metric} Schwarzschild solution so that its $00$ and $11$ components behave as $-g_{00}=g_{11}^{-1}=1-2\frac{G_Lm}{r^{1-L}}$, with $G_L$ is a modified gravitational constant,  $L\simeq b^2_0/2$, and the dimensionless $b_0$ characterizes the radial component of the vector implementing the spontaneous Lorentz symmetry breaking. Another important solution is \index{G\"{o}del metric} the G\"{o}del one, which, as it is known, closed timelike curves (CTCs) can arise. In the bumblebee gravity this metric was shown to be consistent for certain vacua \cite{Ferrbumb}, so, this Lorentz-breaking scenario does not exclude CTCs. The third solution usually tested for various modified gravity models is the \index{Friedmann-Robertson-Walker metric} Friedmann-Robertson-Walker cosmological metric, and it has been proved in \cite{Capelo} that within the bumblebee gravity, cosmic acceleration, including late-time de Sitter expansion, can occur.

\section{Possible LV terms in gravity}

The four-dimensional gravitational CS term we have considered above is certainly a paradigmatic example of the CPT-Lorentz breaking term in gravity. Nevertheless, other LV additive terms in gravity, although perhaps less advantageous, can be considered as well. The list of possible terms with dimensions up to 8 can be found in \cite{KosLiGrav}. Let us briefly characterize their general features.

As an example of a Lorentz-breaking extension for the Einstein gravity, we can discuss the aether-like term for the gravity \cite{Carroll} in a space-time of an arbitrary dimension $D$, looking like
\bea
\label{aethergra}
S_{aether}^{grav}=\alpha \int d^Dx \sqrt{|g|}u^{\mu}u^{\nu}R_{\mu\nu},
\eea
with $\alpha$ is a small constant. This term is CPT-even. Actually, this is a particular form of the CPT-even term $S_{even}=\int d^4x \sqrt{|g|}s^{\mu\nu}R_{\mu\nu}$, whose different aspects, including Hamiltonian formulation and cosmological issues, were studied in various papers, see f.e. \cite{ReShre1,ReShre2,ReShre3}. As we noted in the previous section, to avoid the explicit breaking of the diffeomorphism invariance, one should suggest that the vector $u^{\mu}$ is not a constant but, instead of this, arises as a result of the spontaneous Lorentz symmetry breaking occurring for some potential. As an example, we can choose the bumblebee-like one, see e.g. \cite{KosGra}:
\bea
V(u_{\mu})=\lambda_n(u^{\mu}u_{\mu}\pm v^2)^n
\eea
where $v^2$ is some positive constant, so, we should add this potential to the action (\ref{aethergra}), therefore, the complete action would be
\bea
S=M_*^{D-4}\int d^Dx\sqrt{|g|}(\frac{1}{\kappa}R+\alpha u^{\mu}u^{\nu}R_{\mu\nu}+{\cal L}^{kin}_{bumb}[u]+\lambda_n(u^{\mu}u_{\mu}\pm v^2)^n).
\eea
Here, ${\cal L}^{kin}_{bumb}[u]=-\frac{1}{4}F_{\mu\nu}[u]F^{\mu\nu}[u]$ is the Maxwell kinetic term for the bumblebee field, denoted here as $u^{\mu}$ instead of $B^{\mu}$ used in the previous section, and the $M_*$ is a parameter with the mass dimension 1 introduced to ensure a correct dimension of the action. We note that this action essentially differs from the Enstein-aether theory (\ref{eae}) discussed in the previous section. The term (\ref{aethergra}), proposed originally in \cite{Carroll} to couple the aether term with gravity is, first, more convenient for studies within the weak gravity framework since it allows, in particular, to treat the vector $u^{\mu}$ as constant one, second, is more similar with other aether-like terms defined in \cite{Carroll,aether} for scalar, spinor and gauge fields. One can check that the gauge symmetry in this theory can be maintained for special restrictions on $g_{\mu\nu}$ (fot the weak gravity case, $h_{\mu\nu}$) and $u_{\mu}$. For example, if we consider the five-dimensional space-time choosing a vector $u^{\mu}$ directed along the extra (fifth) dimension, and remind that, for the weak gravity, the gauge transformations for the metric fluctuation and the LV vector are $\delta h_{\mu\nu}=\pa_{\mu}\xi_{\nu}+\pa_{\nu}\xi_{\mu}$ and $\delta u^{\mu}=\pa_5\xi^{\nu}$, we can impose restrictions $h_{a 5}=0$ (with $a=0,1,2,3$) and $\pa_5\xi^a=0$ \cite{Carroll}. We note that for this term, as well as for many other Lorentz and/or CPT-breaking terms in gravity proposed in \cite{KosGra,ShaBerr}, with the only known exception in a full-fledged gravity is the gravitational CS term, the gauge invariance requires imposing of special conditions (no-go constraints), e.g., the terms like $t^{\mu\nu\lambda\rho}R_{\mu\nu\lambda\rho}$ are in general not gauge invariant by the same reasons as (\ref{aethergra}). Moreover, the term of the first order in derivatives,  being generated as a quantum correction in the linearized theory described by the action (\ref{gravlin}), is essentially divergent \cite{ncqugra}. The same is valid for terms linear in Riemann and Ricci tensors in a full-fledged gravity with LV terms (i.e. the terms of second order in derivatives of $h_{\mu\nu}$ in the weak gravity case, cf. \cite{ShaBerr}}. This emphasizes the importance of the gravitational CS term which, as we noted above, is one-loop finite. In principle, one can abandon as well the restriction for the geometry to be Riemannian, through introduction of the torsion, but up to now, neither perturbative generation of any LV term involving torsion nor studies of impact of CPT-Lorentz breaking terms involving torsion were carried out. 

Besides of the aether-like term (\ref{aethergra}), other terms involving couplings of the LV vector $u^{\mu}$ with various geometrical objects can be introduced, such as, e.g., $s^{\mu\nu}R_{\mu\nu}$, $t^{\mu\nu\lambda\rho}C_{\mu\nu\lambda\rho}$, with $s^{\mu\nu}$ and $t^{\mu\nu\lambda\rho}$ are constructed on the base of the $u^{\mu}$, and $C_{\mu\nu\lambda\rho}$ is the Weyl tensor, see f.e. \cite{BaileyKost}. In that paper, the lower contributions from these terms to the parametrized post-newtonian (PPN) expansion are found, and classical dynamics of the vector field in this theory is studied in \cite{SeifVect}.
	
These terms are the examples of dimension-4 LV terms in the gravitational sector. A list of possible LV terms in the gravitational sector with dimensions up to 8 is presented in \cite{KosLiGrav}. Among most important terms, we can emphasize the following ones: first, the gravitational Chern-Simons term, second, the terms proportional to various orders of covariant derivatives of the Riemann tensor, like e.g. $(\stackrel{\smile}{k}^{(n)}_D)^{\alpha\beta\gamma\delta\mu_1\ldots\mu_n}D_{(\mu_1}\ldots D_{\mu_n)}R_{\alpha\beta\gamma\delta}$, third, those ones proportional to direct products of various degrees of Riemann tensor, like e.g. $(\stackrel{\smile}{k}^{(6)}_R)^{\alpha_1\beta_1\gamma_1\delta_1\alpha_2\beta_2\gamma_2\delta_2}R_{\alpha_1\beta_1\gamma_1\delta_1}R_{\alpha_2\beta_2\gamma_2\delta_2}$, fourth, those ones involving various non-zero degrees both of the Riemann tensor and of its covariant derivatives. Except of the gravitational Chern-Simons term, all these terms, being considered in the weak field limit, with the LV vectores (tensors) assumed to be constant, break the gauge symmetry (the importance of gauge symmetry breaking within the gravity context has been discussed in \cite{KosMew2017}). The dispersion relations for linearized gravity models whose action is given by the sum of the Einstein-Hilbert action and each of these terms, for a simplified case when the constant tensors, e.g. $(\stackrel{\smile}{k}^{(n)}_D)^{\alpha\beta\gamma\delta\mu_1\ldots\mu_n}$, are completely characterized by only one constant vector, have been studied in \cite{Vieira}, and were shown to display a usual form $E^2=\vec{p}^2$, being unaffected by the Lorentz-breaking vectors. Studies of dispersion relations for a more generic form of Lorentz-breaking tensors are presented in  \cite{Mewes2019}.

 To illustrate the problem with the gauge (non)invariance we  turn again to the weak gravity limit, where the dynamical field is the metric fluctuation, and consider the term initially introduced in \cite{ncqugra}:
\bea
\label{lingrav}
{\cal L}_{lingrav}=-2\epsilon^{\lambda\mu\nu\rho}b_{\rho}h_{\nu\sigma}\pa_{\lambda}h_{\mu}^{\sigma}.
\eea
The main motivation to define this term in \cite{ncqugra} was the generalization of the noncommutative field method discussed in the section 4.4 and based on deformation of the canonical commutation algebra between fields and their canonically conjugated momenta applied in \cite{Gamboa} for the gauge fields, to the case of the (linearized) gravity, and namely this structure rests when we impose a special gauge allowing to rule out the non-local contributions, whose presence is a price we should pay for the gauge symmetry. If we consider the usual gauge transformations for this term, we will see that it is not invariant. Among other properties of this term we should emphasize as well the birefringence of the gravitational waves. Also, if one would try to generate this term from the usual theory of spinors on a curved background (\ref{gravlin}), the result will be divergent, which, however, is natural since the theory of spinors on a curved background (\ref{full}) is non-renormalizable.

Nevertheless, many gauge invariant Lorentz-breaking terms in the linearized gravity case are possible as well. Besides of the gravitational Chern-Simons term discussed in the section 6.2, there are other gauge invariant CPT-odd LV terms and CPT-even LV terms presented in \cite{Vieira}. However, all these terms are higher-derivative ones. Some examples are
\bea
{\cal L}_{even}=\frac{1}{2}K_{\alpha}\Pi^{\alpha\beta}K_{\beta},
\eea
with $K_{\alpha}=b_{\mu}\Pi^{\mu\nu}h_{\nu\alpha}$, where $\Pi^{\mu\nu}=\eta^{\mu\nu}\Box-\pa^{\mu}\pa^{\nu}$ is the projection-like operator, and
	\bea
\label{odd0}
{\cal L}_{odd}=\epsilon^{\alpha\beta\gamma\delta}b_{\alpha}K_{\beta}\pa_{\gamma}K_{\delta}.
\eea
The dispersion relations in both these cases are $E^2=\vec{p}^2$, being unaffected by the Lorentz-breaking vectors.
Other higher-derivative gauge invariant LV terms for the linearized gravity can be constructed as well, they involve more derivatives (e.g. one can consider linear combinations of terms $\epsilon^{\alpha\beta\gamma\delta}b_{\alpha}K_{\beta}\pa_{\gamma}\Box^nK_{\delta}$, $\frac{1}{2}K_{\alpha}\Pi^{\alpha\beta}\Box^nK_{\beta}$ for various $n\geq 1$, being straightforward generalizations of  above terms). However, it is clear that more studies of higher-derivatives LV additive terms in gravity are still to be done.

\section{Conclusions}

In this chapter, we described some first steps carried out to consider the CPT-Lorentz breaking extensions in the gravity. We demonstrated that the four-dimensional gravitational CS term is apparently the best possible CPT-Lorentz breaking extension of the Einstein gravity, since it, first, can be perturbatively generated in a consistent manner being finite, second, does not display any problems with unitarity and/or causality, third, is gauge invariant.  Clearly, it calls the interest to studying of possible classical solutions of the CS modified gravity, that is, first of all, verification of consistency of known general relativity solutions within the CSMG, either in dynamical or non-dynamical cases.

For the first time, this study was carried out already in the seminal work \cite{JaPi} where the consistency of the Schwarzschild metric within the CSMG was proved. As a continuation, in \cite{Grumiller} it was proved that the spherically symmetric solutions of general relativity, whose most important example are clearly the cosmological solutions, and the static solutions with axial symmetry, as well as their generalizations, so-called stationary solutions where the metric depends on time only through the factor $t-\alpha\phi$, with $\alpha$ is some constant, continue to be consistent within the CSMG, where the Kerr metric is no more consistent. Further, it was shown in \cite{Konno} that to achieve the consistency within the DCSMG, the Kerr metric should be deformed by additive terms proportional to the CS coefficient $\vartheta$, with the desired consistency can be shown order by order. A great review on different exact solutions in CSMG is presented in \cite{AlexYunes}. Other solutions, which do not match these examples, are the G\"{o}del-type solutions characterized by the possibility of the \index{closed time-like curves} closed time-like curves \cite{Godel,RT}. Their consistency within the CSMG was checked both in non-dynamical and dynamical cases in \cite{ourG1,ourG2}, where it was shown that, in the CSMG, both in non-dynamical and dynamical cases, besides of the known G\"{o}del-type solutions, the new solutions, e.g., completely causal ones, can arise. 

However, it is clear that the problem of possible manners to implement the Lorentz symmetry breaking within the gravity, and, further, to study the resulting theories at the quantum level, is still open. As we already noted, one important issue is the gauge symmetry, that is, the general covariance, whose maintaining in an expected LV extension of a full-fledged gravity  requires special efforts. For example, as we already noted above, one of the main difficulties is related to the development of an appropriate manner for introduction of privileged directions in a curved space-time without breaking the general covariance; it was claimed in \cite{Bertolami} that a consistent manner to do it is based on a spontaneous Lorentz symmetry breaking introduced within the bumblebee model in a curved space-time, however, up to now, only some first studies in this direction were carried out. Another problem is the search for the renormalizable model of the gravity itself. Certainly, constructing of Lorentz-breaking non-Riemannian gravity models, which could involve torsion and nonmetricity, as well as development of LV massive gravity theories, is also an important direction of studies (some first steps along this way are discussed in \cite{KosLiNR,KosPot,ReShre2}). Therefore we conclude that constructing of consistent LV extension of gravity certainly requires more active studies.

\chapter[Experimental studies]{Experimental studies of the Lorentz symmetry breaking}

It is clear that all new theories should be verified through some experiments. So, this book would be clearly incomplete without a review of different experimental studies of possible LV phenomena and measurements of LV parameters characterizing various LV extensions of known field theory models. Therefore, in this last chapter we present this review, in which we discuss both macroscopic studies of LV effects, especially gravitational ones, and the microscopic studies, especially quantum ones.   Also, one of the main reasons for experimental studies of the Lorentz symmetry breaking 
is the determination of possible limits of applicability for special relativity, which, as well as those ones for any physical theory, require a careful study. 
Besides of this, there are strong cosmological motivations for these studies, such as, e.g., the hypothetic possibility for cosmic radiation anisotropy discussed in \cite{MagAxEv} where it was called the "axis of evil". While in \cite{MagAxEv1} it was argued that it is more reasonable to attribute this anisotropy to inappropriate methods of statistical analysis rather than to fundamental physical effects, the possibility of this anisotropy certainly requires further studies. 

One more (and historically, the first) effect treated as a possible manifestation of the Lorentz symmetry breaking was the Greisen-Zatsepin-Kuzmin (GZK) effect \index{Greisen-Zatsepin-Kuzmin (GZK) effect} \cite{Gre,ZaKu} which showed that the flux of the cosmic rays strongly decreases with the energy more than about $10^{20}$ eV, which naturally called the interest to the idea of deformed dispersion relations which essentially involve a certain energy scale as it is suggested within the double special relativity \cite{rainbow} (see also the discussion in the section 2.2 of this book and \cite{MagSmo}). Other important effects within this context are the possible birefringence of a light in a vacuum and rotation of a plane of polarization of light in a vacuum which can occur in certain LV extensions of electrodynamics (see discussions in Section 2.2) and the Cherenkov radiation \index{Cherenkov radiation} which in the LV case is possible even in a vacuum, see f.e. \cite{Lehn,Lehn1} (various aspects of Cherenkov radiation in different LV extensions of QED are discussed also, e.g.  in \cite{Alt,Alt1,Alt2,KliShr}, see also references therein, in \cite{PotCher}, the Cherenkov radiation for massive photons is treated, and in \cite{ShreCher}, it is studied for LV fermions).

The most appropriate theory to treat all these issues related with Lorentz symmetry breaking, is, clearly, the minimal LV Standard Model extension (LV SME) \cite{ColKost1,ColKost2} which has been discussed throughout this book and includes as ingredients almost all terms considered here -- the CFJ term (both in Abelian and non-Abelian forms), the aether term for scalar, gauge and spinor fields, and different LV spinor-scalar and spinor-vector couplings. In \cite{KosGra}, this model was generalized to include the gravity. This model clearly allows various manners for experimental measurements of different LV effects.  Treating numerical estimations of the parameters of LV SME, we, first of all, should refer  to the Data Tables \cite{datatables} which collect the results for estimated values of all LV parameters known up to now.  Certainly, non-minimal extensions of this model discussed in \cite{KosLi,KosLiGrav} also must be studied experimentally.

We note that, within verifying the Lorentz symmetry breaking and/or measuring its parameters,  two directions of studies can be emphasized: (i)  non-gravitational studies, such as, first, the determinations of unusual behavior of electromagnetic waves which can originate from the Lorentz symmetry breaking, such as birefringence and rotation of the plane of polarization, second, the possible modifications of dispersion relations, third, measurements of the parameters of the Lorentz-violating extension of the Standard Model; (ii) gravitational and cosmological studies.

\section{Non-gravitational studies}

One of the most known experiments aimed to study possible unusual behavior of electromagnetic waves in a vacuum is the PVLAS experiment \index{PVLAS} (see e.g. \cite{PVLAS}) in which the rotation of plane of polarization of light in an external magnetic field was measured. In the paper \cite{ChaShe} it was claimed that this rotation can be attributed to space-time noncommutativity, which, as we noted in the Chapter 4, represents itself as one of the known forms of Lorentz symmetry breaking, with the noncommutativity parameter is estimated as $\Theta\simeq (30\, GeV)^{-2}$. Nevertheless, further \cite{Pinzul} it was argued that this rotation should be attributed to the axion-photon coupling, whereas the impact of the noncommutativity must be unobservable.    

Another important line of experimental studies for the possible Lorentz symmetry breaking is based on studies of cosmic rays. Indeed, as we already noted, one of interpretations of the GZK effect was based on the idea that it can be explained by a strong Lorentz symmetry breaking at some energy scale, as is suggested by the double special relativity, therefore the dispersion relations should be modified by extra terms becoming important at very large energies. In the paper \cite{Bietenholz},  one of the first reviews of studies of cosmic rays within this context was presented. Following \cite{Bietenholz}, the \index{ultra-high energy cosmic rays (UHECRs)} ultra-high energy cosmic rays (UHECRs)  with the energy $E\geq 10^{18.5}$ eV,  mostly have the extragalactic origin. Actually they are frequently related with the $\gamma$-ray bursts, \index{$\gamma$-ray bursts} see e.g. \cite{CWEP}. The main experimental studies of UHECRs presented in \cite{Bietenholz} are performed by the Pierre Auger Collaboration. While the experiments discussed in \cite{Bietenholz} showed that the Lorentz invariance is valid up to the value of the usual Lorentz $\gamma$ factor up to $10^{11}$, we can conclude that these experiments impose strong restrictions on the parameters of Lorentz symmetry breaking but do not rule it out, and, as it is claimed in \cite{Bietenholz}, new studies in this direction clearly will establish new questions, and, among the suggestions given in that paper, one should emphasize the need to study, first,  $\gamma$-ray bursts, second, high-energy neutrinos. 

Within this context, the discussion of the $\gamma$-ray bursts, first time observed in 1967 and representing themselves as processes characterized by extremely high energy scale, is especially important. In \cite{Shore}, it was suggested that namely by this reason, it is natural to expect the LV effects within the $\gamma$-ray bursts observations, with the main line of study could consist in determination of polarization of the electromagnetic waves emitted within the bursts. Also, in \cite{Shore}, some spectroscopic experiments in order to determine the LV parameters $k_{\mu}$ and $\kappa_{\mu\nu\lambda\rho}$ are suggested; the results of some these experiments are presented in \cite{datatables} where the typical bound for $k_{\mu}$ obtained from different experiments is about $10^{-43}$ GeV, and for the dimensionless $\kappa_{\mu\nu\lambda\rho}$ -- about $10^{-15}\div 10^{-17}$. More recent results for study of the $\gamma$-ray bursts in this context are presented in \cite{Vasi}, where the results are obtained with use of the {\it Fermi} satellite, where it is shown on the base of these observations that the characteristic energy of Lorentz symmetry breaking should be no less than about $7,6M_{Pl}$, that is, larger than the Planck energy. Besides of this, one should mention as well the paper \cite{KosMewGRB}, where the constraints for the LV coefficients, including the higher-dimensional coefficients, obtained on the base of the $\gamma$-ray bursts studies are given. Also, in \cite{Laurent}, on the base of studies of $\gamma$-ray bursts some estimations for upper limits for the vacuum birefringence effect were obtained, however, these limits are too large, e.g., for $\xi$ characterizing the Myers-Pospelov term (\ref{MP}) they give $\xi<2.6\times 10^8$.  Besides of this, in \cite{CWEP} estimations for modification of dispersion relations and for rotation of the polarization plane of the light in a vacuum based on study of the $\gamma$-ray burst GRB 041219A were carried out, and the possibility of Horava-Lifshitz-like deformations of the dispersion relations, with additive $\alpha_z\vec{k}^{2z}$ terms, was estimated, and it was shown that time delay expected from suggestions of existence of additive terms characterized by the critical exponent $z=2$, calculated on the base of results obtained for this $\gamma$-ray burst, is bounded by $10^{-21}$ s.

The LV parameters can be estimated as well on the base of astrophysical observations. In this context, it is important to mention results obtained from observations of astrophysical $\gamma$-ray sources which allowed to estimate the bounds for $c^{\mu\nu}$ coefficients from the spinor sector of LV QED to be of the order of $10^{-16}\div 10^{-17}$ \cite{AltAstro}.

It should be noted that the LV modifications of dispersion relations for the electromagnetic field can imply as well in the Cherenkov radiation, see e.g. \cite{Alt,Matti}. For example, in \cite{Matti} it was argued that the Cherenkov radiation can occur for the Myers-Pospelov-like dispersion relation for the photon $E^2=p^2\pm\frac{\xi p^3}{M}$, and for the electron $E^2=p^2+m^2+\frac{\eta_{R,L}p^3}{M}$, with $M$ is a Planck-order mass scale, $\eta_{R,L}$ correspond to different electron helicities, and the sign $\pm$ is for two photon helicities. It was argued in \cite{Matti} that it follows from observations of the Crab nebula that $|\xi|\le 10^{-3}$, and $|\eta_L-\eta_R|<4$.

Various experiments with elementary particles are used for tests and measurements of parameters of Lorentz symmetry breaking. In this context, the experiments with \index{Penning traps} Penning traps, allowing to detect a possible particle-antiparticle asymmetry, play the special role. It was proved in these studies \cite{KostDing}, that the bounds for components of the LV axial vector $b_{\mu}$ from the minimal LV extended QED (\ref{genrenmod}) are about $10^{-24}$ GeV. In the same paper, some estimations for non-minimal LV parameters are also presented. More results for non-minimal LV parameters can be found in \cite{Ding1}.
In \cite{Vasi}, some other experiments within the particle physics allowing to measure the LV parameters were proposed. 

Many other studies of elementary particles aimed to detect the impacts of Lorentz-CPT symmetry breaking were performed as well. Among them, experiments with neutrinos have a special role. First of all, it should be noted that the neutrino oscillations \index{neutrino oscillations} also can be explained with use of the Lorentz symmetry breaking, as it was argued in \cite{neutosc,neutosc2}. It is interesting to notice that despite the famous ``discovery of superluminal neutrinos" within the OPERA experiment \index{OPERA} was proved to be an experimental error, it called much attention namely to studying of possible LV effects in neutrino physics. Further, studies of neutrino oscillations were used to obtain estimations for LV parameters in \cite{Rahaman}, where it was shown that the bound for the parameter $a_{\mu}$ defined in (\ref{genrenmod}), estimated on the base of these oscillations, is about $10^{-23}$ GeV. Various estimations for the Lorentz-breaking parameters were obtained also within the framework of the \index{DUNE} DUNE experiment, see f.e. \cite{DUNE}, and the IceCube experiments, \index{IceCube} see f.e. \cite{IceCube}. As another application of neutrinos, in \cite{neut3} it was argued that the neutrinos are the very convenient objects for measurements allowing for imposing very strong bounds on LV parameters from gravity experiments which we consider in the next section.

\section{Gravitational effects}

Purely gravitational studies presented by an important line of experiments have been described in \cite{Tasson1}. Within these studies, the LV extensions for theory of fermions and bumblebee model coupled to the gravity were considered. Different tests, representing themselves either as free-fall gravimeter tests, where falling corner cubes or matter interferometers were used as free-fall gravimeters, or as the tests of the weak equivalence principle (WEP) were carried out. Other gravimeters used within the study performed in \cite{Tasson1} were the force-comparison gravimeters based on comparing of the gravitational force with an appropriate electromagnetic force. As a result, e.g., for the force-comparison gravimeter tests, the experimental bounds were showed to be: about $10^{-8}$ GeV for the coefficient $a_{\mu}$ characterizing the term $a_{\mu}ee^{\mu}_a\bar{\psi}\gamma^a\psi$, and about $10^{-8}$ for the coefficients $c^{\mu\nu}$ entering the term $-\frac{1}{2}ic_{\lambda\nu}ee^{\mu}_ae^{\nu a}e^{\lambda}_b\bar{\psi}\stackrel{\leftrightarrow}{D}_{\mu}\psi$.  In \cite{Tasson1}, also other tests for the WEP were presented, e.g., the satellite-based ones, the Solar system ones based on the precession of the perihelion, which yield the results: $|a_{\mu}|\leq 10^{-6}$ GeV, and $|c_{\mu\nu}|\leq 10^{-7}$, tests with antimatter, and a great number of photon tests. Some possible satellite tests have been discussed as well in \cite{Alttest}.

One of the most important experimental discoveries of recent years is, certainly, the detection of gravitational waves \cite{gravwaves}. Therefore, it is natural to suggest that the LV impacts can be studied within the context of gravitational waves as well. In particular, as we already mentioned in the Chapter 6, some LV extensions of gravity can allow for birefringence of gravitational waves (see e.g. \cite{ncqugra}). The results of first studies in this direction based on study of the gravitational-wave event  GW150914 are presented in \cite{KosGW},  for a more detailed discussion see \cite{Mewes2019}. Later experimental studies of birefringence of gravitational waves, for a certain higher-derivative extension of the Einstein gravity, can be found in \cite{GWbir}. It is interesting to note that some studies in that paper suggest to consider higher-derivative extensions of gravity, while up to now, except of the Chern-Simons modified gravity, no other examples of higher-derivative LV gravity models were sufficiently studied. It worth mentioning that within this context, the gravitational Cherenkov radiation can exist as well and will produce the energy losses, see the discussion in \cite{Koscosrays}, where a typical extension of gravity implying in the modified Einstein equations
$$
G_{\mu\nu}=8\pi GT_{\mu\nu}+\hat{\bar{s}}^{\alpha\beta}\tilde{R}_{\alpha\mu\beta\nu},
$$
with $\tilde{R}_{\alpha\mu\beta\nu}$ is a double dual of the Riemann tensor obtained through its contraction with two Levi-Civita symbols, and $\hat{\bar{s}}^{\alpha\beta}$ is a differential operator involving LV parameters of dimensions $d=4, 6, 8, \ldots$ like $\hat{\bar{s}}_{\mu\nu}=\sum\limits_d
(\bar{s}^{(d)})_{\mu\nu}^{\phantom{\mu\nu}\alpha_1\ldots\alpha_{d-4}}\pa_{\alpha_1}\ldots\pa_{\alpha_{d-4}}$, was considered, and observations of cosmic rays were used to estimate the LV parameters $(\bar{s}^{(d)})_{\mu\nu}^{\phantom{\mu\nu}\alpha_1\ldots\alpha_{d-4}}$, e.g., it was found that the bound for the dimensionless $(\bar{s}^{(4)})_{\mu\nu}$ is of the order $10^{-14}$. In \cite{Bourgoin:2017fpo}, it was estimated with use of the lunar laser range, to be about $10^{-9}$ (the lunar laser range was used to estimate LV parameters also in \cite{Bourgoin:2020ckq}), and in \cite{Bailey:2017lbo} this parameter was determined by gravity tests of the Solar system objects, where the typical bound was found to be about $10^{-9}\ldots 10^{-11}$. Some aspects of gravitational Cherenkov radiation are discussed also in \cite{CherRadGrav}. And in \cite{TasCher}, estimations for the coefficients up to $(\bar{s}^{(10)})$ are done on the base of different cosmic ray observations, so that the typical bound of  $(\bar{s}^{(10)})_{ij}$ turns out to be about $10^{-61}$ GeV$^{-6}$. As an aside observation, we once more see that the studies of cosmic rays and $\gamma$-ray bursts allow to obtain the most important information about the scale of LV parameters. Further, in \cite{Bailey2021}, a methodology allowing for estimation of LV parameters from observations of gravitational waves was presented, and it was claimed in this paper that this analysis is currently on the way. Among other experimental studies of LV effects in gravity, the short-range gravity tests \cite{BailKost} are worth mentioning. The most recent results of various experiments on  Lorentz and CPT breaking, both  in flat and curved spacetime, can be found in \cite{snowmass}.

To close this book, we note that there are numerous experiments performed in order to study different LV extensions of various field theories, either of gravity, QED or other models, and apparently, more new experiments are to be performed. And in principle, we note that the complete picture of a possible LV extension of the Standard Model is still very far from its conclusion, and many problems in this context continue to be open and will be open during many years, and the main attention in nearest years certainly should be paid to studying LV extensions of gravity.

\newpage

{\bf Acknowledgements.} Authors are grateful to J. F. Assun\c{c}\~{a}o, A. P. Baeta Scarpelli, H. Belich, L. H. C. Borges, F. A. Brito, L. C. T. Brito,  M. Cvetic, A. G. Dias, J. C. C. Felipe, A. F. Ferrari, C. Furtado, M. S. Guimaraes, H. O. Girotti, M. Gomes,  A. C. Lehum, R. V. Maluf, C. Marat Reyes, E. Passos, P. Porfirio, A. A. Ribeiro, R. F. Ribeiro, V. O. Rivelles, A. F. Santos, M. Schreck, A. J. da Silva, A. R. Vieira, C. Wotzasek, C. A. D. Zarro,  for fruitful collaboration and interesting discussions. The work has been partially supported by CNPq.

\printindex
\end{document}